\documentclass[aip,amsmath,amssymb,reprint]{revtex4-1}

\usepackage{graphicx}
\usepackage{dcolumn}
\usepackage{bm}

\usepackage[utf8]{inputenc}
\usepackage[T1]{fontenc}
\usepackage{mathptmx}
\usepackage{multirow}
\usepackage{xcolor}
\usepackage{footnote}
\usepackage{braket}
\usepackage{siunitx}
\usepackage{hyperref}
\usepackage{booktabs}

\begin{document}

\title{Quasi-relativistic approach to analytical gradients of parity
violating potentials}
\thanks{Parts of this work were reported in preliminary form in S.
Br\"uck, Diploma thesis, University of Frankfurt, 2011.}

\author{Sascha A Br\"{u}ck}
\affiliation{Frankfurt Institute for Advanced Studies,
Ruth-Moufang-Stra{\ss}e 1, 60438 Frankfurt am Main, Germany}
\author{Nityananda Sahu}
\affiliation{Fachbereich Chemie, Philipps--Universit{\"a}t Marburg,
Hans-Meerwein-Stra{\ss}e 4, 35032 Marburg, Germany}
\author{Konstantin Gaul}
\affiliation{Fachbereich Chemie, Philipps--Universit{\"a}t Marburg,
Hans-Meerwein-Stra{\ss}e 4, 35032 Marburg, Germany}
\author{Robert Berger}
\affiliation{Frankfurt Institute for Advanced Studies,
Ruth-Moufang-Stra{\ss}e 1, 60438 Frankfurt am Main, Germany}
\affiliation{Fachbereich Chemie, Philipps--Universit{\"a}t Marburg,
Hans-Meerwein-Stra{\ss}e 4, 35032 Marburg, Germany}
\affiliation{Clemens-Sch{\"o}pf-Institut, Technische Universit{\"a}t
Darmstadt, Alarich-Weiss-Stra{\ss}e 4, 64287, Darmstadt, Germany}

\date{\today}

\begin{abstract}
An analytic gradient approach for the computation of derivatives of
parity-violating (PV) potentials with respect to displacements of
the nuclei in chiral molecules is described and implemented within
a quasirelativistic mean-field framework. Calculated PV potential
gradients are utilised for estimating PV frequency splittings
between enantiomers in rotational and vibrational spectra of four
chiral polyhalomethanes, i.e. CHBrClF, CHClFI, CHBrFI and CHAtFI.
Values calculated within the single-mode approximation for the
frequency shifts agree well with previously reported theoretical
values. The influence of non-separable anharmonic effects
(multi-mode effects) on the vibrational frequency shifts, which are
readily accessible with the present analytic derivative approach,
are estimated for the C--F stretching fundamental of all four
molecules and computed for each of the fundamentals in CHBrClF and 
CHAtFI. Multi-mode effects are found to be significant, in particular
for the C--F stretching modes, being for some modes and cases of similar
size as the single-mode contribution.
\end{abstract}

\maketitle

\section{\label{sec:intro} Introduction}
Early after the proof of parity violation (PV) was provided in 1957 by the 
famous experiment of $\beta^{-}$ disintegration of Co nuclei \cite{wu:1957},
which realised a proposal by the theoreticians Lee and Yang \cite{lee:1956},
it was suggested by Yamagata \cite{yamagata:1966} that PV weak interactions 
can induce a tiny energy difference between a chiral
molecule and its non-identical
mirror-image.\cite{gajzago:1974,letokhov:1975,zeldovich:1977,zeldovich:1977a,hegstrom:1980,khriplovich:1980,gorshkov:1982a,gorshkov:1982,quack:1986}
Diverse experimental schemes have been
proposed to detect PV effects in chiral molecules, ranging from
M\"ossbauer spectroscopy\cite{compton:2002} to vibrational
spectroscopy,\cite{kompanets:1976,arimondo:1977,bauder:1997,daussy:1999}
electron paramagnetic resonance spectroscopy,\cite{harris:1980} 
rotational spectroscopy \cite{bauder:1997,schnell:2011} and nuclear
magnetic resonance spectroscopy
\cite{barra:1986,barra:1988a,barra:1996,eills:2017} to time-dependent
approaches and quantum-beat experiments
\cite{harris:1978,harris:1981,quack:1986,berger:2003}. We refer the reader
for more details and an extended overview to a collections of reviews
on this subject \cite{quack:1989,quack:2002,berger:2004a,crassous:2005,quack:2008,schwerdtfeger:2010,berger:2019}.
The most accurate experimental attempts reported so far were in the high-resolution infrared spectroscopy of
bromochlorofluoromethane (CHBrClF). An upper bound of $\Delta 
\nu_\mathrm{PV}/\nu \approx 10^{-13} $ was obtained for the relative PV
difference of the C-F stretching frequency between (\textit{R})- and
(\textit{S})-enantiomers.\cite{daussy:1999} The
theoretically \cite{quack:2000,laerdahl:2000a,viglione:2000,quack:2000a,schwerdtfeger:2002,quack:2005,berger:2007,thierfelder:2010} predicted value for this
relative frequency splitting is of the order of $10^{-17}$, however. Later, with an
improved experimental set-up, a measurement of the same compound with a
resolution of $5\times 10^{-14}$ has been reported in
2002\cite{ziskind:2002} and
with a new set-up, it may be hoped that a precision of $10^{-16}$ can
be reached.\cite{darquie:2010,cournol:2019} As nuclear spin-independent electroweak PV effects in chiral compounds scale approximately
with nuclear charge $Z$ to the power of five (in the presence of
spin-orbit coupling), compounds containing heavy metal nuclei could
be of greater experimental value than the originally used organic
molecules. Theoretical searches in this direction have been already
made on molecules containing for instance bismuth, rhenium, mercury and
astatine.\cite{faglioni:2003,schwerdtfeger:2003,bast:2003,berger:2007} 

In general, a measurement of the PV energy as a difference between electronic
energies of separated enantiomers would be very difficult since it comes on top of the
rest mass energy of the molecule, \cite{letokhov:1975} but the scheme
proposed by Quack \cite{quack:1986}, for instance, circumvents this by directly
measuring the PV energy via the PV induced time-dependent interconversion
between states of opposite parity \cite{quack:1989}. An alternative for
detecting molecular parity violation is to
measure the difference between PV energy differences arising due
to the PV potential [$E_\mathrm{PV}(\vec{q})$] in the vibrational and
rotational transitions in chiral compounds, \cite{letokhov:1975} with $\vec{q}$
denoting the vector of dimensionless reduced normal coordinates. Often
also $V_\mathrm{PV}$ is used as symbol for the PV potential.  
In first order perturbation theory, $E_\mathrm{PV}(\vec{q})$ gives rise to a PV shift in the
$n^\mathrm{th}$ vibrational energy level of a given enantiomer
(\textit{R} or \textit{S}) according to\cite{quack:2000a} 
\begin {equation}
E_{n,\mathrm{PV}}^{R,S}\approx \langle
\Psi_{n}^{R,S}|E_\mathrm{PV}(\vec{q})| \Psi_{n}^{R,S}\rangle,
\end {equation}
where $|\Psi_{n}^{R,S}\rangle$ denote the $n^\mathrm{th}$
vibrational state of the \textit{R}-  and \textit{S}-enantiomer, which
are obtained by solving the parity-conserving (PC) rovibrational
Schr{\"o}dinger equation for each enantiomer. Since,
$E_\mathrm{PV}(\vec{q})$ is parity odd, $E_{n,\mathrm{PV}}^{R}$
and $E_{n,\mathrm{PV}}^{S}$ values are numerically of equal magnitude but have an opposite sign. Thus, the
corresponding PV energy difference between the $n^\mathrm{th}$
vibrational levels of the two enantiomers is 
\begin{equation}
\label {EPV}
\Delta E_{n,\mathrm{PV}} =
E_{n,\mathrm{PV}}^{S}-E_{n,\mathrm{PV}}^{R}\approx 2\langle
\Psi_{n}^{S}|E_\mathrm{PV}(\vec{q})| \Psi_{n}^{S}\rangle.
\end {equation}
The relative change in vibrational ($\Delta E_\mathrm{vib,PV} = \Delta
E_{m,\mathrm{PV}} - \Delta E_{n,\mathrm{PV}}$) and
rotational ($\Delta E_\mathrm{rot,PV}$) transition energies between right- and
left-handed molecules is expected to scale to same order of magnitude
as that of the relative change in electronic ($\Delta E_\mathrm{el,PV}$)
transition energy.\cite{messiah} 
\begin{equation}
\frac{\Delta E_\mathrm{el,PV}}{E_\mathrm{el}} \approx  \frac{\Delta E_\mathrm{vib,PV}}{E_\mathrm{vib}} \approx \frac{\Delta E_\mathrm{rot,PV}}{E_\mathrm{rot}} 
\end{equation}

The introduction of PV within electroweak quantum chemistry not only affects the
energy of an enantiomer, but also its equilibrium structure if the
PC vibrational potential gets modified by the PV
potential.\cite{quack:2000a} In general, the PV potential induces a
minute change in the equilibrium structure of a molecule compared to
the equilibrium structure without PV contribution, if the PV energy
gradient ($\vec \nabla E_\mathrm{PV}$) does not vanish there. 
Thus, the weak interaction leads to a change of the equilibrium 
structure of a chiral molecule, which is different for both
enantiomers due to PV effects. 

This effect could in principle be measured by microwave spectroscopy,
since it leads to a shift of the rotational
constants.\cite{quack:2000} In practice, the change of the structure
due to $\vec \nabla E_\mathrm{PV}$ at the minimum of the parity
conserving potential can be estimated with the help of the vibrational
Hessian $\mathbf{F}_\mathrm{MW}$,\cite{quack:2000a} of the PC
potential, given in mass-weighted Cartesian displacement coordinates.
From this $3N_\mathrm{nuc}\times3N_\mathrm{nuc}$-dimensional
vibrational Hessian, with $N_\mathrm{nuc}$ being the number of nuclei
in the system, contributions of infinitesimal translational and
rotational displacements can be eliminated with the
$3N_\mathrm{nuc}\times(3N_\mathrm{nuc}-6)$ dimensional projection
matrix $\mathbf{A}$ by forming $(\mathbf{A}^\mathsf{T}
\mathbf{F}_\mathrm{MW} \mathbf{A})^{-1}$, so that the corresponding
Hessian in internal Cartesian displacement coordinates results, which
can be inverted. When $\mathbf{M}$ is a $(3N_\mathrm{nuc}\times
3N_\mathrm{nuc})$ diagonal matrix containing the
masses of the various atoms, the final displacements
($\delta_\mathrm{PV} \vec R$)
of the atoms from their equilibrium position due to $\vec \nabla
E_\mathrm{PV}$ read
\begin{equation}
\label{dispequation}
\delta_\mathrm{PV} \vec R=- \mathbf{M}^{-1/2}\mathbf{A} (\mathbf{A}^\mathsf{T} \mathbf{F}_\mathrm{MW} \mathbf{A})^{-1} \mathbf{A}^\mathsf{T} \mathbf{M}^{-1/2} \vec \nabla E_\mathrm{PV}.
\end{equation}
Coordinates and displacements are transformed subsequently to
the principal axis system of the PC equilibrium
structure and the splittings of the diagonal elements ($\Delta \mathbf{I}$) of
the moment of inertia tensor due to PV effects are approximated to first order
in $\delta_\mathrm{PV} \vec R$. For the $xx$ component of $\Delta
\mathbf{I}$ we have for instance:
\begin{align}
\Delta I_x = 2 m_A (y_A \delta_\mathrm{PV} y_A + z_A
\delta_\mathrm{PV} z_A ) 
\end{align} 
and analogous for the other diagonal elements.
Then, the changes in diagonal elements are used to estimate 
the splittings of rotational constants ($\Delta X_\mathrm R$) between
two enantiomers within the approximate expression \cite{quack:2000a}  
\begin{equation}
\label{dispequation1}
\frac{\Delta X_\mathrm R}{X_\mathrm R} \approx -\frac{\Delta I_X}{I_X}
\end{equation}
with $X_\mathrm R$ being one of the rotational constants ($A$, $B$ or $C$) and 
$I_X$ being the corresponding eigenvalues of the moment of inertia tensor. 
From Eq.~\ref{dispequation} the importance of the gradient of the parity
violating potential with respect to displacements of the nuclei becomes 
particularly evident. The numerical calculation of this term, however, is
tedious, in particular within a (quasi)relativistic electronic structure 
framework, which is why we present in this work an approach for calculating
this gradient analytically within a quasirelativistic mean-field framework.

Another promising experiment for detecting parity violation in chiral
molecules is vibrational
spectroscopy. Most of the effort has been put into the measurement of
vibrational frequency shifts due to PV
effects.\cite{kompanets:1976,bauder:1997,daussy:1999,marrel:2001,ziskind:2002,quack:2000,laerdahl:2000,viglione:2000,quack:2001,schwerdtfeger:2002,schwerdtfeger:2005,
berger:2007}
The relative vibrational frequency shift for a
transition from state $n$ to $m$ is determined by 
taking the difference of the vibrationally averaged PV 
potentials of the enantiomers and dividing by the
corresponding transition energy $h\nu_{mn}$ \cite{quack:2000a,berger:2007}
\begin{equation}
\frac{\Delta \nu_{mn,\mathrm{PV}}}{\nu_{mn}}=2\frac{ \left( \langle
\Psi_m|E_\mathrm{PV}(\vec{q})| \Psi_m\rangle-\langle
\Psi_n|E_\mathrm{PV}(\vec{q})| \Psi_n\rangle \right)}{h\nu_{mn}}
\end{equation}
where the factor two enters into the equation since the difference of
the $R$ and $S$ enantiomer is twice (\textit {cf.}  Eq. $\ref{EPV}$ )
the difference between one enantiomer and the parity conserving case,
where no shift occurs.\cite{berger:2007} The vibrationally averaged
potentials depend on the multi-dimensional PV energy surface and not
only on a single point energy at the equilibrium structure. In
addition, the complete rovibrational wavefunction would be needed 
to compute the expectation value. As this problem is extremely
complex even for relatively small molecules, the PV effects of
electronic, vibrational and rotational degrees of freedom are typically
separated in a first step. Still the computational effort is in
general too high, so as a second step, the multi-dimensional problem is usually
split into one-dimensional problems, where the movements along the normal
coordinates are treated separately as if they were independent of each
other.\cite{quack:2000a,berger:2007} This can be augmented by adding contributions from
the potential depending on a smaller number of modes order by
order.\cite{quack:2003a,rauhut:2004} In practice, the PC potential
($V_\mathrm{BO}$) and PV potential [$E_\mathrm{PV}(\vec q)$] 
can be evaluated at one-dimensional cuts along the dimensionless
reduced normal coordinates $\vec{q}$.\cite{berger:2007,quack:2000a} This
approach has already been used to estimate the vibrational frequency
splitting of the C-F stretching mode in chiral halogenated methane
derivatives.\cite{quack:2000,quack:2000a,laerdahl:2000a,schwerdtfeger:2002,schwerdtfeger:2005,berger:2007} 

A different approach comes from perturbation theory,\cite{buckingham:1975}
where the PC and PV potentials are expanded in a
Taylor series. In this, the influence of multi-mode effects are estimated 
by the calculations of the derivatives of the PV potentials 
with respect to all normal coordinates. 
The contributions of lowest non-vanishing order to the 
$n^\mathrm{th}$ vibrational energy levels of a normal mode $\nu_r$ 
within the perturbative treatment are
\begin{equation}
\begin{split}
\label{vib2D}
&\langle \Psi_{n_r}|E_\mathrm{PV}(\vec{q})| \Psi_{n_r}\rangle \approx E_\mathrm{PV}(\vec R_0)\\
&+\frac 1 2 \left( n_r +\frac 1 2 \right) \left(
\frac{\partial^2 E_\mathrm{PV}}{\partial q_r^2}-\sum_{s} \frac 1
{\hbar \omega_s} \frac{\partial^3 V_\mathrm{BO}}{\partial q_r^2
\partial q_s} \frac{\partial E_\mathrm{PV}}{\partial q_s} \right)
\end {split}
\end{equation}
where $E_\mathrm{PV}(\vec{R}_0)$ is the PV energy at the equilibrium
structure and $\omega_s$ denotes the harmonic vibrational angular frequency of 
normal mode $\nu_s$.
The terms $\frac{\partial E_\mathrm{PV}}{\partial q_s}$ and
$\frac{\partial^2 E_\mathrm{PV}}{\partial q_r^2}$ 
are the first (gradient) and second partial derivatives of the PV energy along 
normal coordinates ($q_s$, $q_r$), respectively. Ignoring all the 
$r \not= s$ terms in the second
sum of Eq. $\ref{vib2D}$ gives the one-dimensional (1D)
perturbative estimate of the vibrational energy for normal mode $\nu_r$ as
\begin{multline}
\label{vib}
\langle \Psi_{n_r}|E_\mathrm{PV}(q_r)|
\Psi_{n_r}\rangle ^\mathrm{1D} = E_\mathrm{PV}(\vec R_0)\\
 + \frac 1 2 \left( n_r +\frac 1 2 \right) \left( \frac{\partial^2
E_\mathrm{PV}}{\partial q_r^2}-\frac 1 {\hbar \omega_r}
\frac{\partial^3 V_\mathrm{BO}}{\partial q_r^3} \frac{\partial
E_\mathrm{PV}}{\partial q_r} \right).
\end{multline}
The drawback of this method is that the one-dimensional terms of the PC and
PV potentials are not
included to infinite order. But the advantage is that, once the
Cartesian gradient of the PV energy at the equilibrium structure is known, the two-dimensional
coupling term can be included without too much effort, because only
the semidiagonal cubic force constants are needed, whereas the gradient of PV
energy along all the normal mode can conveniently be determined by
projecting the Cartesian PV energy gradient onto displacements
along to the normal coordinates. Hence, contributions from other modes
(\textit {cf.}  Eq. $\ref{vibD}$) in addition to the single-mode terms
described by Eq. $\ref{vib}$ 
can easily be accounted for by considering first order multi-mode (MM) effects on the vibrational 
energy levels as shown in Eq. $\ref{vib2D}$ 
\begin{equation}
\begin{split}
\label{vibD}
\langle \Psi_{n_r}|E_\mathrm{PV}(\vec{q})|
\Psi_{n_r}\rangle ^\mathrm{MM} = -\frac 1 2 \left( n_r +\frac 1 2 \right) \sum_{s
\ne r}\frac 1 {\hbar \omega_s} \frac{\partial^3
V_\mathrm{BO}}{\partial q_r^2 \partial q_s} \frac{\partial
E_\mathrm{PV}}{\partial q_s}.
\end {split}
\end{equation}
We define with these MM terms the perturbative approximation of the 
vibrational energy shifts with 2D coupling terms as
\begin{equation}
\begin{split}
\langle \Psi_{n_r}|E_\mathrm{PV}(\vec{q})|\Psi_{n_r}\rangle ^\mathrm{2D} =&
\langle \Psi_{n_r}|E_\mathrm{PV}(\vec{q})|\Psi_{n_r}\rangle ^\mathrm{1D}\\ &+
\langle \Psi_{n_r}|E_\mathrm{PV}(\vec{q})|\Psi_{n_r}\rangle ^\mathrm{MM}.
\end{split}
\end{equation}

Thus, for the calculation of the influence of PV on the vibrational
and rotational spectra, which is one of the main goals of the current
manuscript, the knowledge of the gradient of the PV potential along
all the normal modes is essential. The present article describes an
analytical approach for calculating the gradient of the PV energy
($\vec \nabla E_\mathrm{PV}$) at HF and LDA level of theory (see the
following section \ref{sec:theory}). These
values obtained from this analytic gradient approach, with corresponding 
computational details being described in section \ref{sec:compdetails}, are 
then utilised in section \ref{sec:results} for determining the relative shifts
in rotational constants and vibrational frequencies, due to PV
effects in chiral methane derivatives. Besides this, the influence of
non-separable anharmonic effects (multi-mode effects) on PV induced
vibrational frequency shifts for all the vibrational normal modes 
is calculated for CHBrClF and CHAtFI molecules.

\section{\label{sec:theory}Theory}
For the description of the  vibrational movement of nuclei in the
Born-Oppenheimer approximation, it is
often sufficient to evaluate the electronic structure for a single
fixed arrangement of the nuclei described by the coordinates
$\vec{R}_0$ and to treat the displacement of
nuclei perturbatively (see also
Refs.~\onlinecite{jayatilaka:1992,bast:2011a,helgaker:2012}). 

The electroweak parity-violating potential is very small compared to
the PC potential due to the appearance of the Fermi
coupling constant which is $2.22249\times10^{-14}\,E_\mathrm{h}a_0^3$
in atomic units. Therefore, we can treat the parity-violating
electroweak Hamiltonian $\hat{H}_\mathrm{PV}$ as a small addition to
the parity conserving molecular Hamiltonian $\hat{H}_0$:
\begin{equation}
\hat{H}=\hat{H}_0+\lambda_\mathrm{PV}
\hat{H}_\mathrm{PV},
\end{equation}
where we have introduced $\lambda_\mathrm{PV}$ as a formal perturbation 
parameter.
 
In a mean-field approach the leading order parity
violating contribution $E_\mathrm{PV}(\vec{R}_0+\vec{\eta},\lambda_\mathrm{PV}) =
E(\vec{R}_0+\vec{\eta},\lambda_\mathrm{PV})-E(\vec{R}_0)$ to the
gradient of the variational energy with respect to
nuclear displacements is:
\begin{multline}
\left.\vec{\nabla}_{\vec{\eta}}
E_\mathrm{PV}(\vec{R}_0+\vec{\eta},\lambda_\mathrm{PV})\right|_{\vec{\eta}=\vec{0}}
\approx \lambda_\mathrm{PV} \left. \frac{\partial\vec{\nabla}_{\vec{\eta}}
E(\vec{R}_0+\vec{\eta},
\lambda_\mathrm{PV})}{\partial\lambda_\mathrm{PV}}\right|_{\substack{\vec{\eta}=\vec{0}\\\lambda_\mathrm{PV}=0}}
\label{eq: gradientPV}
\end{multline}
For details on the specific case of electroweak parity
violation, see Appendix \ref{variational_pt}, whereas 
for the general case of variational perturbation theory,
see e.g. Ref.~\onlinecite{sellers:1988}.

In this paper, we want to focus on the molecular Hamiltonian
approximated within a (quasi-relativistic) two-component zeroth-order
regular approximation (ZORA) framework\cite{chang:1986,lenthe:1996}
on the level of Generalised Hartree-Fock (GHF) or Generalised
Kohn-Sham (GKS) density functional theory (DFT) (see
Ref.~\onlinecite{wullen:1998,wullen:2010}):
\begin{equation}
\hat{H}_0=\hat{H}_\mathrm{ZORA}=\mathbf{V}_{\mathrm{GHF,GKS}} +
\underbrace{\hat{V}_\mathrm{nuc}(\vec{r})+
c^2\vec{\bm{\sigma}}\cdot\hat{\vec{p}}\omega(\vec{r})\vec{\bm{\sigma}}\cdot\hat{\vec{p}}}_{\hat{h}_\mathrm{ZORA}}
\end{equation}
Here, $\mathbf{V}_{\mathrm{GHF,GKS}}$ is the effective electron
repulsion potential within the GHF or GKS framework, $\hat{\vec{p}}$
is the linear momentum operator, $\vec{\bm{\sigma}}$ is the vector of
Pauli spin matrices and
$\omega(\vec{r})=\frac{1}{2c^2-\tilde{V}(\vec{r})}$ is the ZORA factor
with the ZORA model potential $\tilde{V}$ as proposed by van W\"ullen
to alleviate the gauge dependence of ZORA.~\cite{wullen:1998} The ZORA
one-electron operator has an electron spin-independent and an electron spin-dependent
contribution:
\begin{equation}
\hat{h}_\mathrm{ZORA} = \hat{V}_\mathrm{nuc} + 
\underbrace{c^2\hat{\vec{p}}\cdot\left(\omega(\vec{r})\hat{\vec{p}}\right)}_{\hat{h}_\mathrm{ZORA}^{(0)}}
+\underbrace{c^2\imath\hat{\vec{p}}\times\left(\omega(\vec{r})\hat{\vec{p}}\right)}_{\hat{h}_\mathrm{ZORA}^{(1,2,3)}}
\cdot\vec{\bm{\sigma}}
\end{equation}

Within ZORA, the nuclear-spin independent parity-violating electroweak
one-electron Hamiltonian appears as\cite{berger:2005}
\begin{equation}
\hat{h}_\mathrm{PV}=\underbrace{\frac{G_\mathrm{F}}{2\sqrt{2}}\sum\limits_{A=1}^{N_\mathrm{nuc}}
\left\{\omega(\vec{r})Q_{\mathrm{W},A}\rho_A(\vec{r}),c\hat{\vec{p}}\right\}_+}_{\hat{h}_\mathrm{PV}^{(1,2,3)}}\cdot\vec{\bm{\sigma}}, 
\end{equation}
where $\left\{A,B\right\}_+=AB+BA$ is the anti-commutator,
$Q_{\mathrm{W},A}$ is the weak charge of nucleus $A$ and $\rho_A$ is
the normalised nuclear density distribution. This operator has only
electron spin-dependent contributions.

We expand the ZORA two-component HF or KS molecular orbitals (MOs) $\phi_i$
in a linear combination of real one-component basis functions $\chi_\mu$ and
complex two-component coefficients
$\vec{C}_{\mu i}=\begin{pmatrix}C^{(\alpha)}_{\mu i}\\C^{(\beta)}_{\mu
i}\end{pmatrix}$ as
\begin{equation}
\phi_i = \sum\limits_{\mu}\vec{C}_{\mu i}\chi_\mu
\end{equation}
In this two-component framework, we can define four complex
one-component density matrices ($\kappa=0,1,2,3$): 
\begin{equation}
D^{(\kappa)}_{\mu\nu}=\sum\limits_{i=1}^{N_\mathrm{orb}}n_i
\underbrace{\vec{C}_{\mu i}^\dagger \bm{\sigma}^{\kappa}
\vec{C}_{\nu i}}_{D^{(\kappa)}_{i\mu\nu}}\,,
\label{eq: density_matrix}
\end{equation}
where the 0$^\mathrm{th}$ component of the Pauli spin matrices is the
$2\times2$ identity matrix.  The two-component density matrix can be
written in terms of these one-component density matrices as
\begin{equation}
\mathbf{D}=\frac{1}{2}\sum\limits_{\kappa=0}^3(\bm{\sigma}^{\kappa})^*\otimes\mathbf{D}^{(\kappa)}\,.
\end{equation}

We can write the expectation value of the one electron ZORA operator as
\begin{align}
h_{\mathrm{ZORA}} &= \sum\limits_{\mu\nu}
\mathfrak{R} \left\{\sum\limits_{\kappa=0}^{3}D^{(\kappa)}_{\mu\nu}
h_{\mathrm{ZORA},\mu\nu}^{(\kappa)}\right\}\,,
\end{align}
and the energy contribution of the parity-violating electroweak potential as
\begin{equation}
h_{\mathrm{PV}} = \sum\limits_{\mu\nu}
\mathfrak{R} \left\{\sum\limits_{\kappa=1}^{3}D^{(\kappa)}_{\mu\nu}
h_{\mathrm{PV},\mu\nu}^{(\kappa)}\right\}\,.
\end{equation}

In the following, the effective potential
$\mathbf{V}_{\mathrm{GHF,GKS}}$ will be represented in the space of
basis functions in terms of the matrix of contracted two-electron
integrals $\mathbf{G}$.  In the general case of hybrid DFT, $\mathbf{G}^{(\kappa)}$ is
constructed as
\begin{widetext}
\begin{equation}
G_{\mu\nu}^{(\kappa)}(\mathbf{D})=\sum\limits_{\rho\sigma}\left[\delta_{k0}D^{(\kappa)}_{\rho\sigma}(\mu\nu|\rho\sigma)-a_{\mathrm{X}}\frac{1}{2}D^{(\kappa)}_{\rho\sigma}(\mu\sigma|\rho\nu)+a_\mathrm{DFT}\Braket{\chi_\mu|V_\mathrm{XC}^{(\kappa)}(\mathbf{D})|\chi_\nu}\right],
\label{eq: gmatrix_aobasis}
\end{equation}
\end{widetext}
where the Mulliken notation for two electron integrals is employed:
$(\mu\nu|\rho\sigma)=\iint\mathrm{d}^3r_1\mathrm{d}^3r_2\chi_\mu(\vec{r}_1)\chi_\rho(\vec{r}_2)\frac{1}{\left|\vec{r}_{1}-\vec{r}_2\right|}\chi_\nu(\vec{r}_1)\chi_\sigma(\vec{r}_2)$.
In case of pure DFT (non-hybrid) we have $a_\mathrm{X}=0$ and in case
of pure HF we have $a_\mathrm{X}=1$ and $a_\mathrm{DFT}=0$.

We consider non-relativistic density functionals which do not depend
on the current density, which are commonly employed even in
relativistic electronic structure theory.  In this case, the matrix
elements of the exchange-correlation potential
$\Braket{\chi_\mu|V_\mathrm{XC}^{(\kappa)}|\chi_\nu}$ are always real.
In this paper, we restrict the discussion to the spin-unpolarised
local density approximation (LDA), in which the exchange-correlation
potential has the form
\begin{equation}
V_\mathrm{XC,LDA}^{(0)}=\frac{\delta
F_\mathrm{XC,LDA}[\rho_\mathrm{e}(\vec{r};\mathbf{D}^{(0)})]}{\delta
\rho_\mathrm{e}(\vec{r};\mathbf{D}^{(0)})},
\end{equation}
with $F_\mathrm{XC,LDA}$ being the LDA density functional and the
electronic number density function being 
\begin{equation}
\rho_\mathrm{e}(\vec{r};\mathbf{D}^{(0)})=\sum_{\mu\nu}
\mathfrak{R} \left\{D^{(0)}_{\mu\nu}\right\}\chi_\mu(\vec{r})
\chi_\nu(\vec{r})\,.
\end{equation}
For the general form of the exchange-correlation potential see
Appendix \ref{xcfun}.
 
The expectation value of the electron repulsion potential is
\begin{equation}
V_{\mathrm{GHF,GKS}}= 
\mathfrak{R} \left\{\sum\limits_{\kappa=0}^3D^{(\kappa)}_{\mu\nu}G_{\mu\nu}^{(\kappa)}(\mathbf{D})\right\}.
\end{equation}

The total ZORA energy in presence of the full perturbation $\hat{H}_\mathrm{PV}$, which implies 
$\lambda_\mathrm{PV}=1$, is given by
\begin{widetext}
\begin{equation}
\begin{aligned}
E_\infty &= h_{\mathrm{ZORA}}+ \frac{1}{2}
V_{\mathrm{GHF,GKS}} + h_\mathrm{PV} \\
&= \mathfrak{R}\left\{\sum\limits_{\mu\nu}\left[
\sum\limits_{\kappa=0}^{3}
D^{\infty,(\kappa)}_{\mu\nu}
h_{\mathrm{ZORA},\mu\nu}^{(\kappa)}
+ \frac{1}{2}\sum\limits_{\kappa=0}^3 D^{\infty,(\kappa)}_{\mu\nu}G_{\mu\nu}^{(\kappa)}(\mathbf{D})
+ \sum\limits_{\kappa=1}^{3} D^{\infty,(\kappa)}_{\mu\nu} h_{\mathrm{PV},\mu\nu}^{(\kappa)}
\right]\right\}\,, 
\end{aligned}
\end{equation}
\end{widetext}
where, $\infty$ refers to density matrices obtained variationally in
the presence of the perturbation, which is analogous to an infinite
order perturbation theory treatment of $\hat{H}_\mathrm{PV}$ (for details see Eq.
(\ref{eq: parityviolationPT}) in the Appendix), provided that this converges. Thus, the 
gradient with respect to nuclear displacements can be written as
\begin{widetext}
\begin{equation}
\begin{aligned}
\vec{\nabla}_{\vec{\eta}}E_\infty &=
\mathfrak{R}\left\{\sum\limits_{\mu\nu}
\left[
\sum\limits_{\kappa=0}^{3}
\left(\vec{\nabla}_{\vec{\eta}}D^{\infty,(\kappa)}_{\mu\nu}\right)h_{\mathrm{ZORA},\mu\nu}^{(\kappa)} 
+\sum\limits_{\kappa=0}^{3} 
D^{\infty,(\kappa)}_{\mu\nu}\left(\vec{\nabla}_{\vec{\eta}}h_{\mathrm{ZORA},\mu\nu}^{(\kappa)}\right)
\right.\right.\\&\left.\left.
+ \frac{1}{2}\sum\limits_{\kappa=0}^3 
\left(\vec{\nabla}_{\vec{\eta}}D^{\infty,(\kappa)}_{\mu\nu}\right)G_{\mu\nu}^{(\kappa)}(\mathbf{D}^{\infty})
+\frac{1}{2}\sum\limits_{\kappa=0}^3
D^{\infty,(\kappa)}_{\mu\nu} \left(\vec{\nabla}_{\vec{\eta}}G_{\mu\nu}^{(\kappa)}(\mathbf{D}^{\infty})\right)
\right.\right.\\&\left.\left.
+ \sum\limits_{\kappa=1}^{3} 
\left(\vec{\nabla}_{\vec{\eta}}D^{\infty,(\kappa)}_{\mu\nu}\right) h_{\mathrm{PV},\mu\nu}^{(\kappa)}
+\sum\limits_{\kappa=1}^{3} 
D^{\infty,(\kappa)}_{\mu\nu} \left(\vec{\nabla}_{\vec{\eta}}h_{\mathrm{PV},\mu\nu}^{(\kappa)}\right)
\right]\right\} 
\end{aligned}
\end{equation}
\end{widetext}
For a first order property that does not depend on the basis functions
such as the parity-violating potential, this expression can be
simplified by using the orthonormality condition of the HF equations
and with Eq. (\ref{eq: gradientPV}) we receive the leading order
parity-violating energy gradient (see Appendix
\ref{nucdisplace_property}): 
\begin{widetext}
\begin{equation}
\begin{aligned}
\label{eq: gradientPVfinal}
\vec{\nabla}_{\vec{\eta}}E_\mathrm{PV} &\approx
\mathfrak{R}\left\{\sum\limits_{\mu\nu}
\left[\sum\limits_{\kappa=0}^{3}
D'^{(\kappa)}_{\mu\nu}\left(\vec{\nabla}_{\vec{\eta}}h_{\mathrm{ZORA},\mu\nu}^{(\kappa)}\right)
+ 
\sum\limits_{\kappa=0}^3 
D^{(\kappa)}_{\mu\nu}
\vec{G}^{(\kappa)}_{\mathrm{grad},\mu\nu}(\mathbf{D}, \mathbf{D}')
+ \sum\limits_{\kappa=1}^{3} 
D^{(\kappa)}_{\mu\nu} \left(\vec{\nabla}_{\vec{\eta}}h_{\mathrm{PV},\mu\nu}^{(\kappa)}\right)
-W'^{(0)}_{\mu\nu}\left(\vec{\nabla}_{\vec{\eta}}S_{\mu\nu}\right)
\right]\right\}\,. 
\end{aligned}
\end{equation}
where we have introduced the energy weighted density matrix
(EWDM) $\mathbf W$ as
\begin{equation}
\mathbf{W}^{(\kappa)} = \sum\limits_{i=1}^{N_\mathrm{orb}}n_i\varepsilon_i
\mathbf{D}^{(\kappa)}_i\,,
\end{equation} 
and the matrix $\vec{\mathbf{G}}_\mathrm{grad}^{(\kappa)}$ of
contracted gradients of two-electron integrals with elements:
\begin{equation}
\vec{G}^{(\kappa)}_{\mathrm{grad},\mu\nu}(\mathbf{D},\mathbf{D}')=\sum\limits_{\rho\sigma}\left[\delta_{k0}D'^{(\kappa)}_{\rho\sigma}\vec{\nabla}_{\vec{\eta}}(\mu\nu|\rho\sigma)-a_{\mathrm{X}}\frac{1}{2}D'^{(\kappa)}_{\rho\sigma}\vec{\nabla}_{\vec{\eta}}(\mu\sigma|\rho\nu)+2a_\mathrm{DFT}\Braket{\chi_\mu|\hat{V}'^{(\kappa)}_\mathrm{XC}(\mathbf{D},\mathbf{D}')|\vec{\nabla}_{\vec{\eta}}\chi_\nu}\right].
\end{equation}
\end{widetext}
Note, that in the DFT case, derivatives of the exchange-correlation
potential have to be computed for the calculation of the perturbed
two-electron gradients, which in the spin-unpolarised LDA
case are
\begin{equation}
V'_\mathrm{XC,LDA}(\mathbf{D}^{(0)}, \mathbf{D}'^{(0)})=\frac{\delta
V_\mathrm{XC,LDA}(\mathbf{D}^{(0)})}{\delta\rho(\vec{r};\mathbf{D}^{(0)})}\rho(\vec{r};\mathbf{D}'^{(0)})\,.
\label{eq: perturbed_fun}
\end{equation}
For other functionals see Appendix \ref{xcfun}.  $\mathbf{D}$ is the
SCF density matrix of the unperturbed system, i.e.  received with the
Hamiltonian $\hat{H}_0$ only, and $\mathbf{D}'$ and $\mathbf{W}'$ are
the perturbed density matrix and EWDM of first order in
$\lambda_\mathrm{PV}$.

The nuclear displacement gradients of the full unperturbed ZORA
Hamiltonian, i.e. the gradients of one-electron ZORA integrals
$\vec{\nabla}_{\vec{\eta}}h_{\mathrm{ZORA},\mu\nu}^{(\kappa)}$, as well as
gradients of two-electron integrals and the exchange-correlation
potentials needed for
$\vec{G}^{(\kappa)}_{\mathrm{grad},\mu\nu}(\mathbf{D})$ have
been implemented in Ref.~\onlinecite{wullen:2007}. 
In this work, we have
altered the previous one-electron gradient part to account also for a finite 
nucleus model when computing the derivatives of the one-electron
integrals.
What remains then to be obtained
are the perturbed density matrices $\mathbf{D}'^{(\kappa)}$, the
perturbed spin-independent EWDM $\mathbf{W}'^{(0)}$ and the gradient
of the PV integrals
$\vec{\nabla}_{\vec{\eta}}h_{\mathrm{PV},\mu\nu}^{(\kappa)}$. We describe
the scheme for computing the latter in the following subsection
\ref{subsec:pvints}, before we discuss in subsection \ref{subsec:response} 
the linear response scheme that is used for computation of
perturbed density matrices.

\subsection{Gradient of the PV integrals\label{subsec:pvints}}
The matrix elements of the one electron PV operator in basis set
representation is
\begin{equation}
\begin{aligned}
h_{\mathrm{PV},\mu\nu}^{(\kappa)}&=\Braket{\chi_{\mu,B}|\hat{h}_{\mathrm{PV}}^{(\kappa)}|\chi_{\nu,C}}
\\&=\iota \hbar \frac {G_\mathrm F}{2\sqrt2}\sum\limits_{A=1}^{N_\mathrm{nuc}}
\Braket{\chi_{\mu,B}|\hat{h}_{\mathrm{PV},A}^{(\kappa)} |\chi_{\nu,C}}
\\&=\iota \hbar \frac {G_\mathrm F}{2\sqrt2}\sum\limits_{A=1}^{N_\mathrm{nuc}}
\left(\Braket{\chi_{\mu,B}|
\omega(\vec{r}) Q_{\mathrm{W},A}\rho_{\mathrm{nuc},A}(\vec{r})
|\partial_{\kappa}\chi_{\nu,C}} 
\right.\\&\qquad\left.
- \Braket{\partial_{\kappa}\chi_{\mu,B}|
\omega(\vec{r}) Q_{\mathrm{W},A}\rho_{\mathrm{nuc},A}(\vec{r})
|\chi_{\nu,C}}\right)
\end{aligned}
\label{eq: integral_pv}
\end{equation}
with $\partial_\kappa=\frac{\partial}{\partial x^\kappa}$ denoting a component of the electronic four
derivative, with the four vector being
$x^{(0,1,2,3)}=(c t, x, y, z)^\mathsf{T}$. Indices $A,B,C$
denote the nuclei at which the basis
functions $\chi$ or nuclear density distribution $\rho_\mathrm{nuc}$ are hooked. As the
basis functions depend on the nuclear coordinates, the geometry
gradient of the PV integrals is
\begin{equation}
\begin{aligned}
\vec{\nabla}_{\vec{\eta}}h_{\mathrm{PV},A,\mu\nu}^{(\kappa)}
&=
\Braket{\vec{\nabla}_{\vec{\eta}}\chi_{\mu,B}|\hat{h}_{\mathrm{PV},A}^{(\kappa)}|\chi_{\nu,C}}\\
&+\Braket{\chi_{\mu,B}|\vec{\nabla}_{\vec{\eta}}\hat{h}_{\mathrm{PV},A}^{(\kappa)}|\chi_{\nu,C}}\\
&+\Braket{\chi_{\mu,B}|\hat{h}_{\mathrm{PV},A}^{(\kappa)}|\vec{\nabla}_{\vec{\eta}}\chi_{\nu,C}}\,
\end{aligned}
\label{eq: integral_pvgrad}
\end{equation}
where, the second term is the Hellmann-Feynman term and the first and
third term are the basis function contributions. The integrals of the
PV operator are purely imaginary and due to Hermiticity of the PV
operator, the basis function contributions are connected by
\begin{equation}
\Braket{\vec{\nabla}_{\vec{\eta}}\chi_{\mu,B}|\hat{h}_{\mathrm{PV},A}^{(\kappa)}|\chi_{\nu,C}}
=
-\Braket{\chi_{\nu,C}|\hat{h}_{\mathrm{PV},A}^{(\kappa)}|\vec{\nabla}_{\vec{\eta}}\chi_{\mu,B}}\,.
\end {equation}

We split the nuclear displacements $\vec{\eta}$ into the separate
contributions from different nuclei $A$.  The basis functions depend
on $\vec{r} - \vec{R}_A$. Therefore, only the subset of basis
functions centered at $A$ contributes to the corresponding integrals.
Furthermore, as we deal with Gaussian basis functions, derivatives with
respect to nuclear coordinates can be represented by derivatives with
respect to electronic coordinates as
$\partial_{A,\kappa}\chi_B=-\delta_{AB}\partial_{\kappa}\chi_B$, where
$\partial_{A,\kappa}$ denotes a component of the four derivative of
nucleus $A$. For the basis function contribution to the PV gradient
integrals, we arrive at 
\begin{widetext}
\begin{multline}
\Braket{\chi_{\mu,B}|\hat{h}_{\mathrm{PV},A}^{(\kappa)}|\vec{\nabla}_D\chi_{\nu,C}}=
\imath \hbar \frac {G_\mathrm F}{2\sqrt2}\delta_{DC}\sum\limits_{A=1}^{N_\mathrm{nuc}}
\left(
-\Braket{\chi_{\mu,B}|
\omega(\vec{r}) Q_{\mathrm{W},A}\rho_{\mathrm{nuc},A}(\vec{r})
|\vec{\nabla}\partial_{\kappa}\chi_{\nu,C}} 
+ \Braket{\partial_\kappa\chi_{\mu,B}|
\omega(\vec{r}) Q_{\mathrm{W},A}\rho_A(\vec{r})
|\vec{\nabla}\chi_{\nu,C}}
\right).
\end{multline}
For a Gaussian shaped nuclear density distribution, the Hellmann-Feynman term reads
\begin{multline}
\Braket{\chi_{\mu,B}|\vec{\nabla}_{D}\hat{h}_{\mathrm{PV}}^{(\kappa)}|\chi_{\nu,C}}
=\imath \hbar \frac {G_\mathrm F}{2\sqrt2}
\left(
-\Braket{\chi_{\mu,B}|
-(\vec{\nabla}_D\omega(\vec{r}))\sum\limits_A Q_{\mathrm{W},A}\rho_A(\vec{r}) 
+ \omega(\vec{r})Q_{\mathrm{W},D}(\vec{\nabla}\rho_D(\vec{r}))
|\partial_{\kappa}\chi_{\nu,C}} 
\right.\\\left.
+ \Braket{\partial_\kappa\chi_{\mu,B}|
(-\vec{\nabla}_D\omega(\vec{r}))\sum\limits_A  Q_{\mathrm{W},A}\rho_A(\vec{r}) 
+ \omega(\vec{r})Q_{\mathrm{W},D}(\vec{\nabla}\rho_D(\vec{r}))
|\chi_{\nu,C}}\right)\,,
\end{multline}
\end{widetext}
with the gradient of a Gaussian normalised nuclear density distribution being
\begin{equation}
\vec{\nabla}\rho_A(\vec{r}) = -2\zeta_A\left(\frac{\zeta_A}{\pi}\right)^{3/2}(\vec{r}-\vec{R}_A)
\exp\left\{-\zeta_A\left|\vec{r}-\vec{R}_A\right|^2\right\}\,.
\end{equation}
The geometry gradient of the ZORA factor $\vec{\nabla}_A\omega$ is
discussed in Ref.~\onlinecite{wullen:1998}. The construction of ZORA, PV
operators and its derivatives is done by means of numerical
integration.

The above integrals are evaluated numerically on a grid.  In the
present work, an atom centered grids is used employing the
Treuter-Ahlrichs\cite{treutler:1995} version of a Becke
grid,\cite{becke:1988}
where the partitioning is done by weight functions $w_i$ for each grid
point $i$. As these weight functions depend on the nuclear coordinates
they give an additional contribution to the PV gradient which is
calculated as
\begin{equation}
\sum\limits_{i}^{N_\mathrm{grid}}
\left(\chi_{\mu,B}(\vec{r}_i)\hat{h}_{\mathrm{PV},A}^{(\kappa)}(\vec{r}_i)\chi_{\nu,C}(\vec{r}_i)\right)
\vec{\nabla}_{\vec{\eta}}w_i(\vec{R})  
\label{eq: grid_deriv}
\end{equation}

The grid points move when the nuclei are slightly displaced.  To
account for this effect, we employ that translational invariance of
the molecule holds and, therewith, the net force on the molecule has
to be zero. We, therefore, subtract the net force that results from
the numerical integration procedure from the numerically integrated
gradient contribution
$(\vec{\nabla}_{D}h_{\mathrm{PV},A,\mu\nu}^{(\kappa)})_\mathrm{num}$:
\begin{equation}
\vec{\nabla}_{D}h_{\mathrm{PV},A,\mu\nu}^{(\kappa)}
= (\vec{\nabla}_{D}h_{\mathrm{PV},A,\mu\nu}^{(\kappa)})_\mathrm{num} -
\sum\limits_{B}
(\vec{\nabla}_{B}h_{\mathrm{PV},A,\mu\nu}^{(\kappa)})_\mathrm{num}\,.
\end{equation}

\subsection{Linear response computation of perturbed density
matrices\label{subsec:response}}
In first order the perturbed density matrix can be written in terms of
the unoccupied-occupied block $\mathbf{T}_\mathrm{uo}$ of an
anti-Hermitian transformation matrix
$\mathbf{T}$ (see Appendix \ref{first_order_pt_density} for details):
\begin{align}
D'^{(\kappa)}_{\mu\nu}(\mathbf{T}_\mathrm{uo}) &=
\sum\limits_{i}^{\mathrm{occ}}\sum\limits_{a}^{\mathrm{unocc}}\left[
\vec{C}^\dagger_{\mu a}\bm{\sigma}^{(\kappa)}\vec{C}_{\nu i} T_{ai}^*  
+
\vec{C}_{\nu i}^\dagger\bm{\sigma}^{(\kappa)}\vec{C}_{\mu a} T_{ai}
\right]
\label{eq: perturbed_dm}
\,.
\end{align}
For how to compute perturbed density matrices to arbitrary order see
Ref.~\onlinecite{ringholm:2014}.

We introduce the MO transformed PV operator as
\begin{equation}
\begin{aligned}
H_{\mathrm{PV,MO},ij}&=\sum\limits_{\kappa=1}^{3}\sum\limits_{\mu\nu}\vec{C}^\dagger_{\mu
i}\bm{\sigma}^\kappa\vec{C}_{\nu j}h^{(\kappa)}_{\mathrm{PV},\mu\nu}.
\label{eq: hpv_mo}
\end{aligned}
\end{equation}

If we would assume that the first order perturbed wave function would
be calculated by transformation of the orbital coefficients as 
\begin{equation}
\vec{C}'_{\mu i} \approx \sum\limits_a^{\mathrm{unocc}}
\vec{C}_{\mu a}(\varepsilon_a-\varepsilon_i)^{-1} H_{\mathrm{PV,MO},ai}
\,
\end{equation}
then the perturbed density matrix would simply calculated as
$D'^{(\kappa)}_{\mu\nu}(\bm{\Delta^{-1}}\circ\mathbf{H}^{\mathrm{uo}}_{\mathrm{PV,MO}})$
with $\circ$ being the Hadamard product and the matrices
$\bm{\Delta}$, $\bm{\Delta^{-1}}$ having the elements
$\Delta_{ai}=(\epsilon_a-\epsilon_i)$ and
$\Delta^{-1}_{ai}=(\epsilon_a-\epsilon_i)^{-1}$, respectively. Here
$a$ and $i$ are defined in the space of unoccupied and occupied
orbitals, respectively.

However, this does not account for the response of the orbitals to the
perturbation but correspond to a simple sum over states approach in
which it is assumed that the electronic Hessian is diagonal and
therefore the perturbed SCF equations are uncoupled.  Within HF and
KS, however, the electronic Hessian is not diagonal as the
two-electron matrix $\mathbf{G}$ is a function of the orbitals and one
has to solve the coupled perturbed HF (CPHF) or coupled perturbed KS
(CPKS) equations (see Appendix \ref{cphf_ks} and for a detailed
derivation e.g. Ref.\onlinecite{olsen:1989}).
To solve the response equations, we use the reduced form (see
Refs.~\onlinecite{salek:2002,saue:2003} and Appendix \ref{cphf_ks} for
details): 
\begin{equation}
\sum\limits_{bj}\mathbf{A}_{bj}T_{bj} +\mathbf{B}_{bj}T_{bj}^* =
-\mathbf{H}^\mathrm{uo}_\mathrm{PV,MO}\,
\label{eq: response_reduced}
\end{equation}
where indices $b$ run over unoccupied orbitals and indices $j$ over
occupied orbitals. The
elements of the electronic Hessian are 
\begin{align}
A_{ai,bj}
&=\left(\epsilon_a-\epsilon_i\right)\delta_{ab}\delta_{ij}+\tilde{G}_{ai,jb}
\\
B_{ai,bj}&=\tilde{G}_{ai,bj}\,,
\end{align}
with $\tilde{G}_{ai,bj}$ being an element of the four-index MO
transformed two-electron tensor:
\begin{equation}
\tilde{G}_{ai,bj} =
\sum\limits_{\kappa=0}^{3}\sum\limits_{\mu\nu}D^{(\kappa)}_{ai\mu\nu}G_{\mu\nu}^{(\kappa)}(\mathbf{D}_{bj}), 
\end{equation}
with the transition density matrix $\mathbf{D}_{ai}$ having the elements
\begin{equation}
D^{(\kappa)}_{ai,\mu\nu} = \vec{C}_{\mu a}^\dagger \bm{\sigma}^{\kappa}
\vec{C}_{\nu i}\,.
\end{equation}
Eq.~(\ref{eq: response_reduced}) is solved iteratively within a preconditioned 
conjugate gradient algorithm.
Thereby, as initial guess $(0)$, we employ trial vectors $\tilde{\mathbf{T}}$
that represent the uncoupled solutions:
\begin{align}
\label{eq: init_guess}
\tilde{\mathbf{T}}^{(0)}&=-\mathbf{H}^{\mathrm{uo}}_\mathrm{PV,MO}\circ\bm{\Delta^{-1}}\\
\mathbf{R}^{(0)}&=\tilde{\mathbf{T}}^{(0)}\circ\bm{\Delta}\\
\mathbf{P}^{(0)}&=\tilde{\mathbf{T}}^{(0)}.
\end{align}
  
In an iterative procedure in each step $i$ from the trial vector
$\tilde{\mathbf{T}}^{(i-1)}$, we construct the perturbed density
matrices $(\mathbf{D}'^{(\kappa)})^{(i)}$ following Eq.
(\ref{eq: perturbed_dm}) and calculate the contracted electronic Hessian
as
\begin{equation}
\begin{aligned}
\tilde{\mathbf{E}}^{[2],(i)}&=\tilde{\mathbf{T}}^{(i-1)}\circ\bm{\Delta}
+
\mathbf{G}^\mathrm{uo}_{\mathrm{MO}}(\mathbf{D}'(\tilde{\mathbf{T}}^{(i-1)}))\,,
\end{aligned}
\end{equation}
with the two-index MO transformed two-electron matrix:
\begin{equation}
G_{\mathrm{MO},ai}(\mathbf{D}') =
\sum\limits_{\kappa=0}^{3}\sum\limits_{\mu\nu}\vec{C}^\dagger_{\mu
a}\bm{\sigma}^\kappa\vec{C}_{\nu
i}G_{\mu\nu}^{(\kappa)}(\mathbf{D}'), 
\end{equation}
From this we update 
\begin{align}
\alpha^{(i)} &=
\frac{r^{(i-1)}}{\mathrm{Tr}\left(\left(\tilde{\mathbf{T}}^{(i-1)}\right)^\dagger\tilde{\mathbf{E}}^{[2],(i)}\right)}\\
\mathbf{R}^{(i)}&=\mathbf{R}^{(i-1)}-\alpha^{(i)}
\tilde{\mathbf{E}}^{[2],(i)}\\
\mathbf{P}^{(i)}&=-\mathbf{R}^{(i)}\circ\bm{\Delta^{-1}}\\
r^{(i)}&=\left|\mathrm{Tr}\left(\left(\mathbf{R}^{(i)}\right)^\dagger\mathbf{P}^{(i)}\right)\right|
\end{align}
For all calculations in this paper, the Fletcher-Reeves weight of the precondition was employed, which is calculated as
$\beta_\mathrm{FR}=\frac{r^{(i)}}{r^{(i-1)}}$ and the algorithm is
followed until convergence of the norm
$\sqrt{\frac{r^{(i)}}{r^{(0)}}}$ to a given threshold.

The trial matrix is updated as
\begin{equation}
\tilde{\mathbf{T}}^{(i)}=\mathbf{P}^{(i)}+\beta\tilde{\mathbf{T}}^{(i-1)}.
\end{equation}
Upon convergence, the solution is received as the sum over all weighted trial matrices from the $N_\mathrm{iter}$ iterations: 
\begin{equation}
\mathbf{T}_\mathrm{uo}=\sum\limits_{i=1}^{N_\mathrm{iter}}\alpha^{(i)}\tilde{\mathbf{T}}^{(i-1)}
\label{eq: response_solution}
\end{equation}
From the solution vectors of the linear response equations $\mathbf{T}$, new perturbed density matrices are calculated following
Eq. (\ref{eq: perturbed_dm}) and the perturbed EWDM can be calculated as
\begin{equation}
\begin{aligned}
W'^{(\kappa)}_{\mu\nu}(\mathbf{T}_\mathrm{uo}) =&
\sum\limits_{i}^{\mathrm{occ}}\sum\limits_{a}^{\mathrm{unocc}}\left[\vec{C}^\dagger_{\mu
a}\bm{\sigma}^{(\kappa)}\vec{C}_{\nu i}
\left(\varepsilon_iT_{ai}^*+ (F'_{ai})^*\right)
\right.\\&\left.
\qquad+
\vec{C}_{\nu i}^\dagger\bm{\sigma}^{(\kappa)}\vec{C}_{\mu a}
\left(\varepsilon_iT_{ai} + F'_{ai}\right)  
\right]\,,
\end{aligned}
\label{eq: perturbed_ewdm}
\end{equation}
with the MO transformed Fock matrix being 
\begin{equation}
\mathbf{F}' = \mathbf{H}^{\mathrm{uo}}_{\mathrm{PV,MO}} +
\mathbf{G}^{\mathrm{uo}}_{\mathrm{MO}}(\mathbf{D}'(\mathbf{T}_\mathrm{uo})).
\end{equation}

In case of DFT contributions from the exchange correlation potential
to the perturbed $\mathbf{G}$ matrix have to be calculated via
derivatives of the exchange correlation potential as shown as in
\ref{eq: perturbed_fun} for the case of spin unpolarised LDA; for a
more general discussion of exchange-correlation functionals see
Appendix \ref{xcfun} and Ref.~\onlinecite{salek:2002}).

\section{Computational Details\label{sec:compdetails}}
In the pilot implementation of the PV energy gradient, unperturbed LCAO coefficients ($\mathbf{C}$),
orbital energies ($\epsilon$), PV operators as well as two electron
integrals ($\mathbf{G}$) were computed with a modified
version\cite{wullen:1998,berger:2005,nahrwold:09,isaev:2012} of the
Turbomole program package.\cite{haser:1989,ahlrichs:1989} 
Therein, the conjugate gradient algorithm for solving the linear
response equations [Eqs. (\ref{eq: init_guess}-\ref{eq: response_solution})]
was implemented within MATLAB\cite{MATLAB:2018} and the modified
Turbomole program was called to compute the perturbed $\mathbf{G}$
matrix in AO basis following Eq. (\ref{eq: gmatrix_aobasis}). The
resulting perturbed density matrices [Eqs. (\ref{eq: perturbed_dm}) and
(\ref{eq: perturbed_ewdm})] were used in a modified version
of the gradient implementation of Ref.~\onlinecite{wullen:2007} to assemble
the analytic gradient of the parity-violating potential.

The calculation of the two-electron part (HF case and Coulomb
contribution in LDA) of the perturbed contracted two-electron matrix
$\mathbf{G}$ is carried out via contraction of the two electron tensor
$(\mu\nu|\rho\sigma)$ with the two-particle density matrix
$\bm{\Gamma}^{(\kappa,\lambda)}$ which can be constructed
from the one-particle density matrices as 
\begin{equation}
\Gamma_{\mu\nu\rho\sigma}^{(\kappa,\lambda)}(\mathbf{D}^{(\kappa)},\mathbf{D}^{(\lambda)})
=a_\mathrm{C}^{\kappa,\lambda}D^{(\kappa)}_{\mu\nu}D^{(\lambda)}_{\rho\sigma}-a_\mathrm{X}^{\kappa,\lambda}\frac{1}{2}D^{(\kappa)}_{\mu\sigma}D^{(\lambda)}_{\rho\nu}
\label{eq: 2edens}
\end{equation}
Note, that for the employed ZORA operator without two-electron
spin-orbit or spin-spin coupling terms, only two-electron densities
with $\kappa=\lambda$ are required at the HF or DFT level.
Furthermore, we introduced the scaling factors for the direct Coulomb
contribution $a_\mathrm{C}^{\kappa,\lambda}$, which is in all present
calculations set as
$a_\mathrm{C}^{\kappa,\lambda}=\delta_{0\kappa}\delta_{0\lambda}$ and
a more flexible scaling parameter for the exchange contribution
$a_\mathrm{X}^{\kappa,\lambda}$, which is in the present
implementation set to be constant
$a_\mathrm{X}^{\kappa,\lambda}=a_\mathrm{X}$.

For the calculation of the two-electron contribution to the PV energy
gradient, the perturbed two particle density matrices are needed:
\begin{equation}
{\Gamma'}_{\mu\nu\rho\sigma}^{(\kappa,\lambda)}
=\Gamma_{\mu\nu\rho\sigma}^{(\kappa,\lambda)}(\mathbf{D'}^{(\kappa)},\mathbf{D}^{(\lambda)})
=a_\mathrm{C}^{\kappa,\lambda}{D'}^{(\kappa)}_{\mu\nu}D^{(\lambda)}_{\rho\sigma}-a_\mathrm{X}^{\kappa,\lambda}\frac{1}{2}D'^{(\kappa)}_{\mu\sigma}D^{(\lambda)}_{\rho\nu}
\label{eq: perturbed_2edens}
\end{equation}
Whereas in the present implementation Eq. (\ref{eq: perturbed_2edens}) is
employed directly, in the pilot implementation the perturbed two-electron
density matrices were calculated for practical reasons via
\begin{widetext}
\begin{equation}
\label{eq: perturbed_2edens_approx}
\bm{\Gamma'}^{(\kappa,\lambda)}\simeq \frac{1}{2}\left [ \bm{\Gamma}^{(\kappa,\lambda)}\left ( \mathbf{D}^{(\kappa)}+\mathbf{D'}^{(\kappa)},\mathbf{D}^{(\lambda)}+\mathbf{D'}^{(\lambda)} \right )
-\bm{\Gamma}^{(\kappa,\lambda)}\left (  \mathbf{D}^{(\kappa)},\mathbf{D}^{(\lambda)}\right
)-\bm{\Gamma}^{(\kappa,\lambda)}\left (  \mathbf{D'}^{(\kappa)},\mathbf{D'}^{(\lambda)}\right )
\right ],
\end{equation}
\end{widetext}
where a constant, numerical scaling factor of $2\sqrt{2}/[\alpha G_\mathrm{F}
/(E_\mathrm{h} a_0^3)] \approx 1.74 \times 10^{16}$ was used to scale up
artificially the very small numerical values in the perturbed density 
matrices to avoid substractive cancellation in 
Eq.~\ref{eq: perturbed_2edens_approx}.

The present implementation of 
$\vec \nabla E_\mathrm{PV}$ as defined in Eq.  (\ref{eq:
gradientPVfinal}) as well
as the response equations [Eqs. (\ref{eq: init_guess}-\ref{eq: response_solution})]
were included in a modified 
version\cite{wullen:2010,gaul:2020}
of the Turbomole program.\cite{haser:1989,ahlrichs:1989} The results
from the current production-level implementation are identical to those of the pilot
implementation. The two different implementations provided an internal
test of our results.
The implementation proceeds as follows:
\begin{enumerate}
\item For a given molecular structure, unperturbed LCAO coefficients 
and orbital energies are received from a SCF computation
at the two-component ZORA level with the
modified version\cite{wullen:1998,berger:2005,nahrwold:09,isaev:2012}
of Turbomole\cite{haser:1989,ahlrichs:1989} and written on disk. 
\item Integrals of the PV operator [Eq. (\ref{eq: integral_pv})] are computed and from this
the initial guess for the response equations is formed [Eq. (\ref{eq: hpv_mo})] with
the program described in Ref.~\onlinecite{gaul:2020}.
\item The linear response equations are solved in an iterative manner
following Eqs. (\ref{eq: init_guess}-\ref{eq: response_solution})
within the program of Ref.~\onlinecite{gaul:2020} and the perturbed density
matrices and perturbed energy weighted density matrix are written on
disk.
\item Within a modified version of the program described in
Ref.~\onlinecite{wullen:2007} the gradient integrals of the PV
operator  [Eq. (\ref{eq: integral_pvgrad}-\ref{eq: grid_deriv})], of
the ZORA operator
$\vec{\nabla}_{\vec{\eta}}h_{\mathrm{ZORA},\mu\nu}^{(\kappa)}$ and of
the two-electron integrals
$\vec{\nabla}_{\vec{\eta}}(\mu\nu|\rho\sigma)$ are computed and
combined with the unperturbed and perturbed density matrices following
Eq. (\ref{eq: gradientPVfinal}).  For two-electron contributions the
perturbed two electron matrix is formed [Eq. (\ref{eq:
perturbed_2edens})] and contracted with the gradient of two-electron
integrals $\vec{\nabla}_{\vec{\eta}}(\mu\nu|\rho\sigma)$.
\end{enumerate}

In the following, the scheme described above for computing $\vec \nabla
E_\mathrm{PV}$ is used to study electroweak parity-violating effects in 
halogenated methane derivatives; CHBrClF,
CHClFI, CHBrFI and CHAtFI.  As mentioned in the Introduction, their
vibrational spectra, in particular for CHBrClF have been extensively
studied theoretically
\cite{berger:2007,quack:2000,laerdahl:2000a,schwerdtfeger:2005,viglione:2000,quack:2001,schwerdtfeger:2002}
and
experimentally\cite{kompanets:1976,bauder:1997,daussy:1999,marrel:2001,ziskind:2002}
to detect PV effects. A high resolution of 5$\times 10^{-14}$ has been
achieved for the C-F stretching mode of CHBrClF with CO$_{2}$ laser
spectroscopy,\cite{ziskind:2002} which is nevertheless about three or
four orders of magnitude larger than the theoretical predictions of
the size of the effect for the molecule under investigation.
Therefore, the vibrational and rotational spectra of heavier
homologues will be studied herein, as has been done previously for the
vibrational frequency shifts in a one-dimensional anharmonic
approximation.\cite{berger:2007} Whereas in the previous work
single-mode anharmonic effects were included variationally
by solving the one-dimensional anharmonic vibrational Schr{\"o}dinger 
equation, we will use in the present work vibrational perturbation 
theory to account for anharmonic effects. Most importantly,
we can include now also multi-mode contributions to 
parity-violating frequency shifts efficiently, which were neglected 
in essentially all previous studies on parity-violating frequency shifts 
in chiral molecules with the notable exceptions of Ref.~\onlinecite{quack:2003a},
where CDBrClF was studied in a four-dimensional anharmonic model that
span the C--F stretching as well as C-D stretching and the two C-D
bending modes and of Ref.~\onlinecite{barone:2005}, where up to third
order vibrational effects in chiral
arsenic and lead compounds were studied.

For all the methane derivatives mentioned above, we use the same molecular
structures, electronic structure methods and basis sets from earlier
work \cite{berger:2007} to allow for direct comparison. As in 
Ref.~\onlinecite{berger:2007}, the $S$-enantiomers of a given chiral compound 
is considered when parity-violating potentials, and in the current work also 
gradients of the parity-violating potentials, are reported. Splittings of
a property $A$ between enantiomers are given as $\Delta A = A^{S}-A^{R}$.
The equilibrium structures, harmonic vibrational
frequencies and potential energy surfaces (PES) were computed
on the CCSD(T) level with the cc-pVDZ basis set for 
the first to third row elements and quasi-relativistic Stuttgart
pseudopotentials in the neutral atom reference system together with
energy optimised valence basis sets for Br, I\cite{bergner:1993} and
At.\cite{kuchle:1991} Equilibriums structures, harmonic vibrational
frequencies, corresponding normal coordinates and displaced structures
as reported in Ref.~\onlinecite{berger:2007} were reused in the present work. 
The PES have been determined by the SURF module\cite{rauhut:2004} of
MOLPRO.\cite{molpro2012a,qianli:2018,werner:2020}. The double zeta basis
is expected to behave comparatively poorly, but is kept herein to speed up
the calculations and to obtain results that are directly comparable to
previous work. An improved description of the anharmonic PES, in particular
for the promising astatine containing methane derivative will be left for a
later study.

In computations of the PV operator a Weinberg parameter of $\sin^2
\theta_\mathrm{W}=0.2319$ has been used to determine the weak
nuclear charge $Q_{\mathrm{W},A} = (1-4\sin^2\theta_\mathrm{W})Z_A - N_A$,
with $Z_A$ being the number of protons and $N_A$ being the number of neutrons
in nucleus $A$. In all two-component ZORA calculations an even tempered basis set, with the
exception of an uncontracted aug-cc-pVDZ basis set for hydrogen, has
been used in order to compare the results to previous
works.\cite{berger:2007,laerdahl:1999} The parameters of the even tempered
series are $\alpha_i= \gamma \beta_N^{N-i}, \, i=1,\dots,N$, with
$N=26$, $\gamma=0.02$ and $\alpha_1=500000000$. For s and p functions,
the exponents $\alpha_{1-25}$ and $\alpha_{2-26}$ have been used,
while in the case of d functions, $\alpha_{20-24}$ has been chosen for
elements of the second and third row of the periodic table of
elements, $\alpha_{15-25}$ for the fourth and fifth row and
$\alpha_{12-25}$ for the sixth row, which is the only one to contain f
functions as well with exponents $\alpha_{15-22}$.

For LDA, Dirac exchange\cite{dirac:1930} and VWN5
correlation\cite{vosko:1980} potentials have been used. For LDA a
standard DFT integration grid was used, whereas matrix elements of the ZORA and PV
operators were computed on a very dense grid. MOs have been
converged until the change of the SCF energy and relative change of spin-orbit
energy (except CHClFI at HF level) between two-successive iterations
dropped below at least $10^{-6}$~$E_{\textrm{h}}$ and $10^{-12}$
respectively. MOs for CHClFI at HF level have been converged to less 
than $10^{-15}$ for the relative spin-orbit energy change as the 
corresponding PV energy for the equilibrium structure was not converged 
to the desired accuracy with the $10^{-12}$ criterion. In practice, the spin-orbit energy criterion 
was by far the more restrictive one, such that at the end of the iterative process, the change in SCF energy between two cycles typically dropped below
$10^{-9}$~$E_{\textrm{h}}$. 
The threshold for negelection of
gradients of two-electron integrals was set to
$10^{-15}$~$E_{\textrm{h}}\,a_{0}^{-1}$. 

Following the regular spectroscopic notation, the normal modes of the
methane derivatives are named in descending order of the frequency
values. 

\section{Results and Discussion\label{sec:results}}
We first make a comparison of the directional derivatives of the 
PV energy along the C-F stretching mode ($\nu_4$) as obtained 
from the analytical PV energy gradients with those of the numerical ones 
from an earlier study.\cite{berger:2007} The PV energy 
gradients are utilised for estimating the shifts of 
the rotational constants corresponding to the equilibrium 
structures. The vibrational frequency shifts in vibrational 
transitions for the C-F stretching mode are calculated within 
a perturbative treatment using the PV energy gradients.
At the end, the multi-mode effects in the vibrational 
transitions for all the normal modes in CHBrClF 
and CHAtFI molecules are discussed.

\subsection{PV energy gradients}
All the chiral methane derivatives studied here show a characteristic
C-F stretching mode ($\nu_4$) which is amenable to high-resolution
CO$_{2}$ laser spectroscopy. The directional derivative of the PV energy 
along the C-F stretching normal coordinate ($q_4$) is obtained herein
by projecting the analytical Cartesian PV gradient $\vec \nabla
E_\mathrm{PV}$ onto the Cartesian displacement vector $\hat{u}_{q_4}$ corresponding 
to a unit shift along the dimensionless reduced normal coordinate $q_4$.
Table \ref{table:EqGrad} depicts the HF and LDA level directional
derivative along C-F stretching mode at the equilibrium structure.
See the Supporting Information for the corresponding Cartesian 
HF and LDA level $\vec \nabla E_\mathrm{PV}$ for each of the equilibrium structures. 

\begin{table}[]
\caption{Analytical directional derivative ($\vec \nabla
E_\mathrm{PV}(\vec{R}_0) \cdot \hat{u}_{q_4} $ in 
$10^{-12}$ $\mathrm{cm}^{-1}$) along the
dimensionless reduced normal coordinates corresponding to the C-F
stretching mode ($\nu_4$) of the chiral halogenated methane derivatives as
computed at the equilibrium structure.}
\begin{center}
\begin{tabular}{llSS}
\toprule
molecules &  $\vec \nabla E_\mathrm{PV}(\vec{R}_0)\cdot\hat{u}_{q_4}$
                                    &\multicolumn{1}{c}{HF}&\multicolumn{1}{c}{LDA}\\ 
\midrule
\multirow{ 2}{*}{CHBrClF} &analytical                           &+0.4205    &+0.4330  \\
                          &numerical\footnote{Numerical derivatives obtained in Ref.~\onlinecite{berger:2007} from a
polynomial fit of a one-dimensional cut through the parity-violating potential along the dimensionless reduced normal coordinates $q_4$. 
For the equilibrium structure of CHClFI, however, the HF value was slightly less tightly converged, so that this value was recomputed 
herein (see supplement) and the potential refitted, leading to only slightly improved fit values.\label{F1}}       &+0.42148(5) &+0.4333(2) \\
\\
\multirow{ 2}{*}{CHClFI}  &analytical                           &+4.1407    &+2.9285    \\
                          &numerical\textsuperscript{\ref{F1}}  &+4.1452(9)  &+2.92880(4) \\
\\
\multirow{ 2}{*}{CHBrFI}  &analytical                            &+7.0813    &+7.6584    \\
                          &numerical\textsuperscript{\ref{F1}}   &+7.0842(5) &+7.65824(5) \\
\\
\multirow{ 2}{*}{CHAtFI}  &analytical                            &+60.1744 &-28.0288 \\
                          &numerical\textsuperscript{\ref{F1}}   &+59.91(5) &-27.63(7) \\
\bottomrule
\end{tabular}
\end{center}
\label{table:EqGrad}
\end{table}

In general, the analytically computed gradient is expected
to be more accurate than a numerical counterpart obtained by a finite 
difference scheme, provided that the underlying self-consistent field 
solutions are well converged. 
We note in reference to Table \ref{table:EqGrad}, that a direct comparison
of the analytical and numerical directional derivative of the parity
violating potential is partially hampered by the fact that the analytical
gradient is calculated at the equilibrium structure of the molecule,
whereas the numerical directional derivative was determined in
Ref.~\onlinecite{berger:2007} from the linear term of a polynomial fit of the
parity-violating potential as calculated along a one-dimensional cut along
the C-F stretching normal coordinate $q_4$ in the range from $-3$ to $3$.
The two become better comparable, if one computes also the analytical
directional derivatives at several points along the same normal mode
displacement and performs a polynomial fit. Therefore,
as a next step, the coordinate dependence of the PV gradient is
examined, checking the higher order terms as well, by fitting the
$\vec \nabla E_\mathrm{PV}\cdot\hat{u}_{q_r}$ to the third order polynomial.  
\begin{equation}
\label{eq:gradfit}
\vec \nabla E_\mathrm{PV}\left (q_r\right )\cdot \hat{u}_{q_r}
\approx\sum_{i=0}^{3}a_{i}q_r^{i}
\end{equation}
where $\hat{u}_{q_r}$ is the Cartesian displacement vector corresponding to
a unit displacement along the normal coordinate
$q_r$
and $a_{i}$ would correspond to $\frac{1}{i!} \left.\frac{\partial^{i}
\vec \nabla E_\mathrm{PV}\cdot\hat{u}_{q_r}}{\partial q_r^{i}}\right|_{q_r=0}$ in a
Taylor series expansion of $\vec\nabla E_\mathrm{PV}\cdot \hat{u}_{q_r}$.
Thus, $a_{0}$ is similar to the value of the $\vec \nabla
E_\mathrm{PV}\cdot \hat{u}_{q_r}$ at the equilibrium structure along the normal
mode $\nu_r$. The other fitting coefficients $a_{1}$, $a_{2}$ and $a_{3}$
are related to the first, second and third order derivatives, respectively, 
of $\vec \nabla E_\mathrm{PV}\cdot \hat{u}_{q_r}$ with respect to the dimensionless
reduced normal coordinate $q_r$ for $\nu_r$ mode. 

\begin{table}[]
\caption{Fitting coefficients of the HF PV energy  ($E_\mathrm{PV}$)
and the HF PV energy gradients ($\vec \nabla E_\mathrm{PV}$) along the
C-F stretching mode ($\nu_4$) of the chiral halogenated methane derivatives in $10^{-12}$ cm$^{-1}$.}
\begin{tabular}{lS[table-format=+4.5(2)]S[table-format=+4.5(2)]S[table-format=+4.5(2)]l}
\toprule
molecules &                       &\multicolumn{1}{c}{$E_\mathrm{PV}$\footnote{Ref. \onlinecite{berger:2007}; for CHClFI
data see also footnote to Tab.~\ref{table:EqGrad}.}}&         \multicolumn{1}{c}{$\vec \nabla E_\mathrm{PV}$} &    \\ 
\midrule
\multirow{ 5}{*}{CHBrClF} &b$_{0}$       &   -1.4536(1)  &                  &                \\
                          &b$_{1}$       &   +0.42148(5) &     +0.4210(2)   &      a$_{0}$   \\
                          &b$_{2}$       &   -0.04242(9) &     -0.0424(1) &      a$_{1}$/2 \\ 
                          &b$_{4}$       &   +0.00019(1) &     +0.00019(1) &      a$_{3}$/4 \\ 
\\
\multirow{ 5}{*}{CHClFI} &b$_{0}$        &  -13.7228(9) &                                     &                    \\
                         &b$_{1}$        &   +4.1452(9)  &      +4.145(2)    &      a$_{0}$  \\
                         &b$_{2}$        &   -0.3157(7)  &      -0.316(1)    &      a$_{1}$/2  \\
                         &b$_{3}$        &   +0.0022(1)  &      +0.0017(2)   &      a$_{2}$/3  \\
                         &b$_{4}$        &   +0.00104(8) &      +0.0011(1)   &      a$_{3}$/4  \\
\\
\multirow{ 5}{*}{CHBrFI} &b$_{0}$        & -38.4858(5)  &                   &               \\
                         &b$_{1}$        &  +7.0842(5)  &      +7.084(2)    &      a$_{0}$  \\
                         &b$_{2}$        &  -0.4819(4)  &      -0.4818(9)   &      a$_{1}$/2  \\
                         &b$_{3}$        &  +0.00874(8) &      +0.0085(1)   &      a$_{2}$/3  \\
                         &b$_{4}$        &  +0.00060(5) &      +0.00060(7)  &      a$_{3}$/4  \\
\\
\multirow{ 5}{*}{CHAtFI} &b$_{0}$        & -2314.06(5)  &                   &                \\
                         &b$_{1}$        &   +59.91(5)  &     +59.9(2)      &      a$_{0}$  \\
                         &b$_{2}$        &   -69.48(4)  &     -69.46(9)     &      a$_{1}$/2  \\
                         &b$_{3}$        &    +7.946(8) &      +7.97(1)   &      a$_{2}$/3  \\
                         &b$_{4}$        &    -0.350(4) &      -0.353(6)    &      a$_{3}$/4  \\
\bottomrule
\end{tabular}
\label{table:TableFit}
\end{table}

\begin{figure}[]
\includegraphics[width=\linewidth]{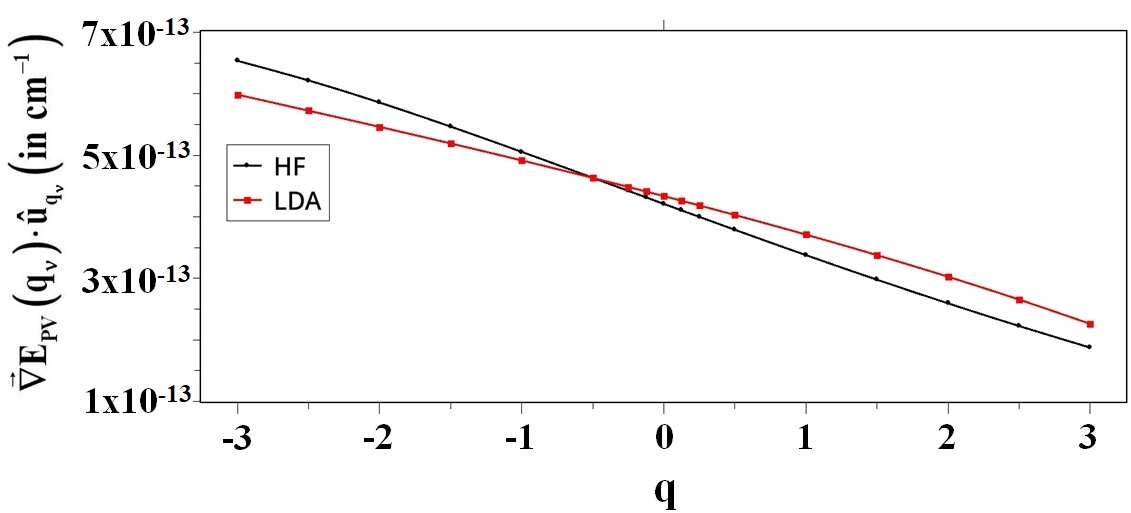}
\caption{HF and LDA level PV energy gradient (fitted to
$\sum_{i=0}^{3}a_{i}q_{r}^{i}$) along the C-F stretching mode of
CHBrClF. See Table \ref{table:TableFit} for the fitting coefficients ($a_{i}$) due to HF
PV energy gradient. For the fitting coefficients of LDA PV energy gradient, see Table S23 in the Supporting Information.}
\label{fig:chbrclf-pvgrad}
\end{figure}

\begin{figure}[]
\includegraphics[width=\linewidth]{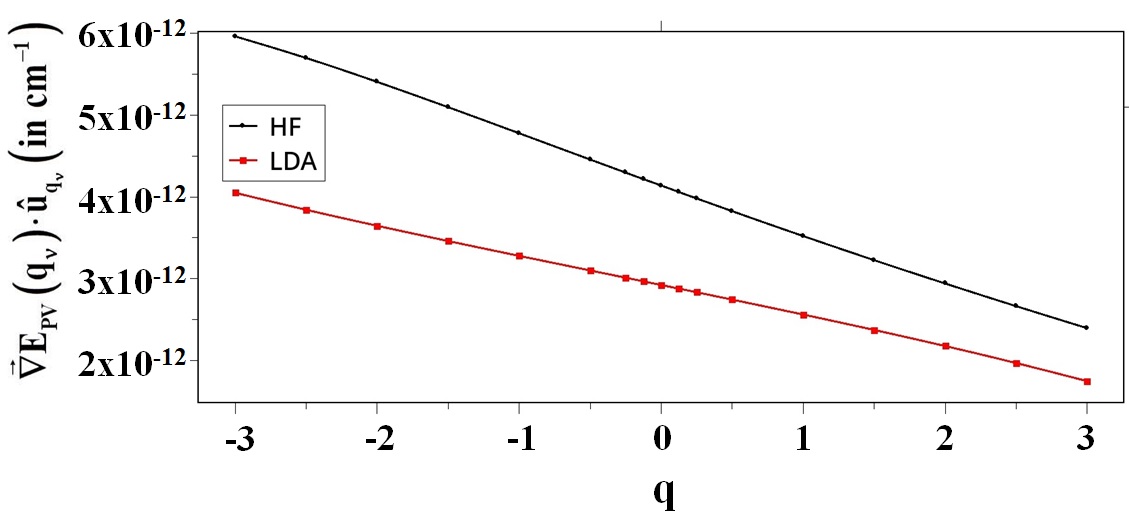}
\caption{HF and LDA level PV energy gradient (fitted to
$\sum_{i=0}^{3}a_{i}q_{r}^{i}$) along the C-F stretching mode of
CHClFI. See Table \ref{table:TableFit} for the fitting coefficients ($a_{i}$) due to HF PV
energy gradient. For the fitting coefficients of LDA PV energy gradient, see Table S23 in the Supporting Information.}  
\label{fig:chclfi-pvgrad}
\end{figure}

\begin{figure}[]
\includegraphics[width=\linewidth]{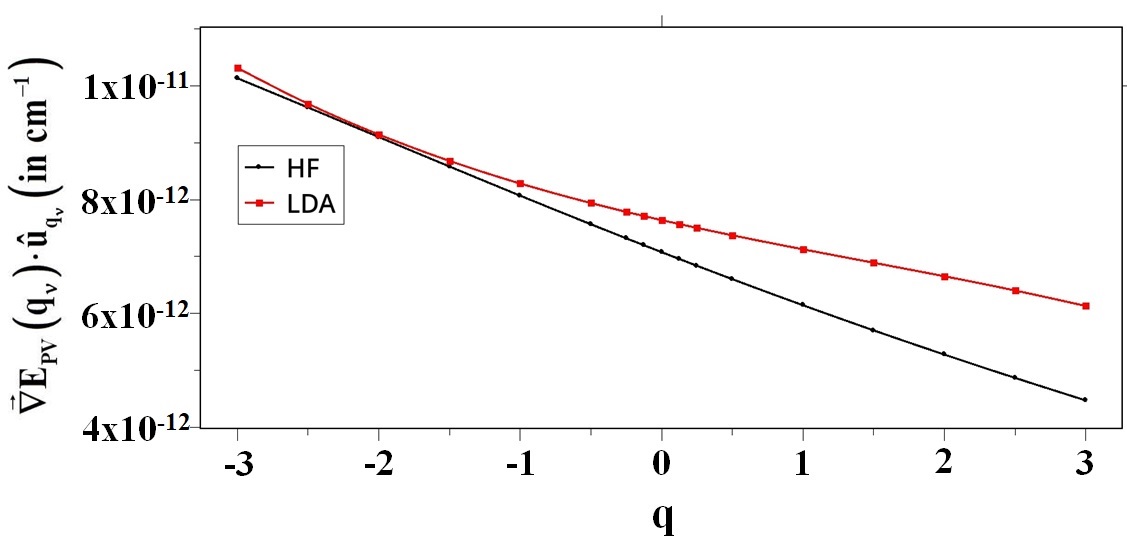}
\caption{HF and LDA level PV energy gradient (fitted to
$\sum_{i=0}^{3}a_{i}q_{r}^{i}$) along the C-F stretching mode of
CHBrFI. See Table \ref{table:TableFit} for the fitting coefficients ($a_{i}$) due to HF PV
energy gradient. For the fitting coefficients of LDA PV energy gradient, see Table S23 in the Supporting Information.} 
\label{fig:chbrfi-pvgrad}
\end{figure}

\begin{figure}[]
\includegraphics[width=\linewidth]{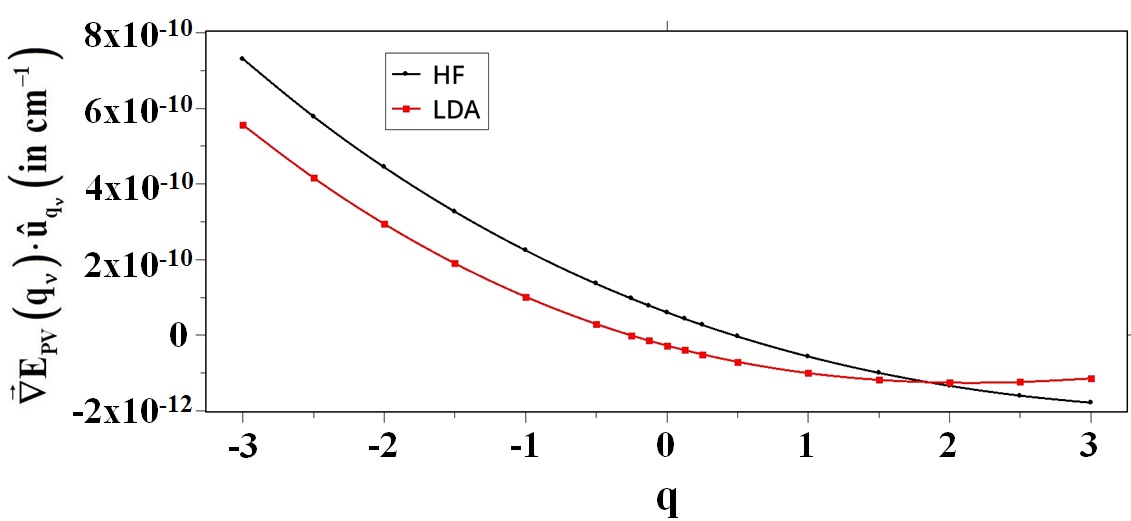}
\caption{HF and LDA level PV energy gradient (fitted to
$\sum_{i=0}^{3}a_{i}q_{r}^{i}$) along the C-F stretching mode of
CHAtFI. See Table \ref{table:TableFit} for the fitting coefficients ($a_{i}$) due to HF PV
energy gradient. For the fitting coefficients of LDA PV energy gradient, see Table S23 in the Supporting Information.}
\label{fig:chatfi-pvgrad}
\end{figure}

For this purpose, structures that are displaced along the normal
coordinates of a vibrational mode are employed to generate a
one-dimensional cut through the PV potential energy surface
along $q$ that varies from $-3$ to $3$ through the equilibrium structure
($q =0$). A total of 17 points is taken into consideration. Here,
for each of the methane derivatives, we consider the one-dimensional
cut for the C-F stretching mode ($\nu_{4}$).  Figures
\ref{fig:chbrclf-pvgrad}--\ref{fig:chatfi-pvgrad} display the HF and
LDA level $\vec \nabla E_\mathrm{PV}(q_4)\cdot\hat{u}_{q_4}$ along the 
dimensionless reduced normal coordinates corresponding to the C-F stretching mode of the
methane derivatives. The corresponding numerical values of $\vec \nabla E_\mathrm{PV}(q_4)\cdot\hat{u}_{q_4}$
are given in Tables S15-S18 in the Supporting Information file. The fitting coefficients need to be adapted to
compare the structure dependence of the PV energy $E_\mathrm{PV}(\vec q_{r})$ 
with the directional derivative of the PV energy 
$\vec \nabla E_\mathrm{PV}\cdot\hat{u}_{q_4}$ along the
dimensionless reduced normal coordinates $q_4$ corresponding to the C-F
stretching mode, because the former can be fitted to a polynomial
expansion, too, namely
\begin{equation}
\label{eq:potfit}
E_\mathrm{PV} (q_{r}) \approx \sum_{i=0}^{4}b_{i}q_{r}^{i},
\end{equation}
where $b_{i}$ would correspond to $\frac{1}{i!} \left.\frac{\partial^i
E_\mathrm{PV}(q_r)}{\partial q_r^i}\right|_{q_r=0}$ in a Taylor series
expansion of $E_\mathrm{PV}$.  Thus, one has to
contrast $E_\mathrm{PV}(\vec q_{r})$ with $\int \vec \nabla
E_\mathrm{PV}(q_{r}) \cdot \hat{u}_{q_{r}}\,\mathrm{d}q_{r}$ and hence
the values of $a_{0}$, $a_{1}/2$, $a_{2}/3$ and $a_{3}/4$ are similar
to the values of $b_{1}$, $b_{2}$, $b_{3}$ and $b_{4}$ respectively.
The term $b_{0}$ corresponds in this approximation to the PV energy at the 
equilibrium structure.

The comparison of these fitting coefficients for PV energy (taken from Ref.~\onlinecite{berger:2007}) and directional
derivative of the PV energy along the C-F stretching mode for the HF
level is presented in Table \ref{table:TableFit}. 
Both numerical and analytical $\vec \nabla
E_\mathrm{PV}\cdot\hat{u}_{q_4}$ values for all the systems are in
good agreement, with the numbers in
parentheses denoting the error due to the fit procedure, which involved one
free parameter less for the directional derivative
(eq.~\ref{eq:gradfit}) than the PV potential
(eq.~\ref{eq:potfit}). The
corresponding comparisons of these fitting PV energy and PV energy
gradient coefficients along the C-F stretching mode at the LDA level are
given in the Supporting Information.

For the determination of the second partial derivatives $\frac{\partial ^{2}E_{PV}}{\partial q_r^{2}}$ 
(same as $a_{1}$ in Table \ref{table:TableFit}),
which are needed for calculating the vibrational energy levels as shown in Eqs.
$\ref{vib2D}$ and $\ref{vib}$, one can expect the availability of the analytical
gradient to simplify the calculations, because fewer points are
needed from the PV energy surface. As a consistency check, a
linear fit of the analytically calculated directional derivatives
computed at $q = -0.125$, $q = 0$ and $q = 0.125$ 
has been opted for in all of the molecules, where the fitting values
($a_{1}$) are found to be in good agreement with the values presented
in Table \ref{table:TableFit}. These linear fit values for the C-F stretching mode
are reported [Table S22] in the Supporting Information. 

\subsection{Shifts of the rotational constants}
The values shown in Tables I and II reveal that the equilibrium
structure of these halogenated methane derivatives possess non-zero
$\vec \nabla E_\mathrm{PV}$.  Due to these non-zero $\vec \nabla
E_\mathrm{PV}$ values, the PV potential can induces a minute change in
the equilibrium structure. This leads to a shift of the rotational
constants, which could in principle be measured by microwave
spectroscopy. As already been discussed in the Introduction, the
change of the structure and hence the shifts in the rotational
constants due to the existence of non-zero $\vec \nabla E_\mathrm{PV}$
at the minimum of the parity conserving potential is calculated with
the help of the vibrational Hessian \textbf{F} (see Eqs.
$\ref{dispequation}$ and  $\ref{dispequation1}$). The change of the
inertia tensor ($\Delta I_{x}$) in the principal axis system and
subsequently the change of the rotational constants ($\Delta X_{R}$)
can be approximated assuming that the result depends linearly on the
displacements.\cite{quack:2000a} 

\begin{table*}[]
\caption{Relative shift of the rotational constants in chiral halogenated methane derivatives.}
\begin{center}
\begin{tabular}{lcS[table-format=-1.3e-2]S[table-format=-1.3e-2]S[table-format=-1.3e-2]cS[table-format=-1.3e-2]S[table-format=-1.3e-2]S[table-format=-1.3e-2]}
\toprule
\multirow{ 2}{*} {} & &\multicolumn{3}{c}{HF}&&\multicolumn{3}{c}{LDA}\\ 
\cline{3-5}\cline{7-9}
          &     & $\Delta A/A$   & $\Delta B/B$ & $\Delta C/C$ && $\Delta A/A$ & $\Delta B/B$ & $\Delta C/C$ \\ 
\midrule
CHBrClF  &  & 2.093e-17   & 1.663e-17& 2.111e-17 & & 7.361e-18 & 4.291e-18 & 5.833e-18 \\
CHClFI   &  & 1.438e-16   & 2.004e-16& 2.171e-16 & & 7.868e-17 & 8.091e-18 & 2.162e-17 \\
CHBrFI   &  & 1.604e-15   & 1.818e-16& 4.388e-16 & & 1.472e-15 &-1.374e-16 & 1.033e-16 \\
CHAtFI   &  & 6.882e-14   & 1.300e-15& 9.137e-15 & & 1.271e-14 & 1.383e-14 & 1.395e-14 \\
\bottomrule
\end{tabular}
\end{center}
\label{table:Rotshift}
\end{table*}

Shifts in the rotational constants as reported in 
Table \ref{table:Rotshift} are highly sensitive to the PV energy 
gradients $\vec \nabla E_\mathrm{PV}$. The numbers obtained 
on the LDA levels are found to be about one quarter of
the HF results for the CHBrClF molecule. The calculated relative effect in
the microwave spectrum ($\Delta X/X \approx 10^{-17}$) of CHBrClF, although
it agrees well with the earlier 
finding by Quack and Stohner\cite{quack:2000}, is still far from the
current experimental resolution. 
Similarly, the corresponding relative shifts in CHClFI and CHBrFI
are about one order of magnitude larger in absolute value than for CHBrClF, but still far below the present
experimental resolution (see Table \ref{table:Rotshift}). Considering the current
scenario, the PV induced shifts of the rotational constants in
CHBrClF, CHClFI and CHBrFI are not expected to be measurable with present
experimental setups. On the other hand, due to the heavy mass of
astatine, a larger shift in the rotational constants is expected for
CHAtFI. For this heavier At analogue, shifts are found to be about two or
three-orders of magnitude higher in absolute value than the lighter molecules 
considered herein. 

Another comment as to the accuracy of these estimates of rotational energy
shifts is in order: Herein, we have followed the most simple approach
\cite{quack:2000a} and
computed only the PV shift at the equilibrium structure as it would arise from
the PV gradient contribution. An improved treatment would include estimates of PV
induced shifts of vibrationally averaged rotational constants as well
as their influence on Coriolis coupling terms and centrifugal distortion 
constants.

\begin{table*}[]
\caption{Vibrationally averaged HF and LDA parity-violating potential
$E_{n_4,\mathrm{PV}}^{S}$ for energy levels $n_4$ in  $10^{-12}$ cm$^{-1}$
for the C-F stretching mode ($\nu_4$) of the (S)-enantiomer of
CHBrClF, CHClFI, CHBrFI and CHAtFI.}
\begin{center}
\begin{tabular}{ll
S[table-format=-4.4]
S[table-format=-4.4]
S[table-format=-4.4]
S[table-format=2.2]
c
S[table-format=-4.4]
S[table-format=-4.4]
S[table-format=-4.4]
S[table-format=-4.4]
S[table-format=2.2]}
\toprule
&  & \multicolumn{4}{c}{HF}&&\multicolumn{4}{c}{LDA}\\ 
\cline{3-6}\cline{8-11}
{Molecule}&    {$n_4$}           & 
{Full 1D\footnote{Reference
\cite{berger:2007}}}   & {Perturbed 1D} & {Perturbed 2D} & {2D effects
($\%$)\footnote{$\mathrm{2D\,effects} =
(\mathrm{Perturbed\,2D}-\mathrm{Perturbed\,1D})/\mathrm{Perturbed\,1D}$}} 
&& {Full 1D$^\mathrm{a}$}  & {Perturbed 1D} & {Perturbed 2D} &{2D
effects ($\%$)$^\mathrm{b}$}\\ 
\midrule
\multirow{ 5}{*}{ CHBrClF }&0& -1.4328   & -1.4326 & -1.4211 & -0.80 && 0.5621 & 0.5624 & 0.5802 & 3.17 \\
                           &1& -1.3929   & -1.3906 & -1.3562 & -2.47 && 0.6160 & 0.6192 & 0.6727 & 8.64 \\
                           &2& -1.3554   & -1.3485 & -1.2912 & -4.25 && 0.6668 & 0.6761 & 0.7652 & 13.18\\
                           &3& -1.3202   & -1.3065 & -1.2263 & -6.14 && 0.7144 & 0.7329 & 0.8577 & 17.03\\
           &$1\leftarrow0$   &  0.0399   &  0.0420 &  0.0649 & 54.53 && 0.0539 & 0.0568 & 0.0925 & 62.75\\
\\
\multirow{ 5}{*}{ CHClFI } &0& -13.463   & -13.461 & -13.445 & -0.12 && 5.5610 & 5.5645 & 5.6159 &  0.92\\
                        &1   & -12.953   & -12.938 & -12.890 & -0.37 && 5.9640 & 5.9800 & 6.1342 &  2.58\\
                        &2   & -12.459   & -12.415 & -12.336 & -0.64 && 6.3500 & 6.3955 & 6.6525 &  4.02\\
                        &3   & -11.979   & -11.892 & -11.781 & -0.93 && 6.7170 & 6.8110 & 7.1708 &  5.28\\
           &$1\leftarrow0$   &   0.510   &   0.523 &   0.555 &  6.07 && 0.4030 & 0.4155 & 0.5183 & 24.74\\
\\
\multirow{ 5}{*}{ CHBrFI } &0& -37.997   & -37.996 & -38.117 & 0.32 && 20.008 & 20.007 & 19.883 & -0.62\\
                        &1   & -37.035   & -37.017 & -37.380 & 0.98 && 21.311 & 21.306 & 20.932 & -1.75\\
                        &2   & -36.092   & -36.037 & -36.643 & 1.68 && 22.611 & 22.604 & 21.982 & -2.75\\
                        &3   & -35.169   & -35.058 & -35.906 & 2.42 && 23.904 & 23.903 & 23.032 & -3.65\\
           &$1\leftarrow0$   &   0.962   &   0.979 &   0.737 &-24.74&& 1.303  &  1.299 &  1.050 &-19.17\\
\\
\multirow{ 5}{*}{ CHAtFI } &0& -2343.3   & -2343.4 & -2320.3 & -0.99 && 938.77 & 937.98 & 966.88 &  3.08\\
                        &1   & -2401.2   & -2402.2 & -2332.9 & -2.88 && 887.70 & 883.10 & 969.80 &  9.82\\
                        &2   & -2458.1   & -2461.0 & -2345.5 & -4.69 && 840.46 & 828.22 & 972.72 & 17.45\\
                        &3   & -2513.8   & -2519.7 & -2358.0 & -6.42 && 797.14 & 773.34 & 975.64 & 26.16\\
           &$1\leftarrow0$   &   -57.9   &   -58.8 &   -12.6 &-78.62 && -51.07 & -54.88 &   2.92 & -105.3 \\
\bottomrule
\end{tabular}
\end{center}
\label{table:vibenergy}
\end{table*}

\subsection{Vibrational transitions for C-F stretching mode}
For the vibrational spectrum, the results from the separable
anharmonic adiabatic approximation (SAAA)\cite{quack:2000a} used in earlier 
work \cite{berger:2007} have been
compared to the one-dimensional (1D) perturbative approach (Eq.
$\ref{vib}$) of the present work.  
The influence of multi-mode coupling terms to the vibrational energy levels 
is also accounted for after adding the perturbative multi-mode contribution
(Eq. $\ref{vibD}$) to the single-mode results. 
Table \ref{table:vibenergy} presents the vibrationally averaged HF and LDA parity
violating potential energy levels up to 3$^\mathrm{rd}$ vibrational state for
the C-F stretching mode ($\nu_4$) of the \textit{S}-enantiomer of
CHBrClF, CHClFI, CHBrFI and CHAtFI molecules along with a comparison
to SAAA results.\cite{berger:2007} 

For the lowest transition (from $n_{4} = 0$ to $n_{4} = 1$) in CHBrClF, the full 1D and the perturbative
1D treatment match to about $<1~\%$, which is far below the accuracy
of the employed ZORA-cGKS and ZORA-cGHF methods and therefore
negligible. The transition after perturbative inclusion of multi-mode
effects up to 2D coupling terms, however, differs by about
50~\% (\textit{cf.} Table \ref{table:vibenergy}).  A similar influence of non-separable
anharmonic effects on PV has been reported for similar
molecules.\cite{quack:2003a} Nonetheless, the conclusion to be
drawn is that the effect of the 2D treatment of the parity conserving
potential is more important than the higher order terms of the 1D
parity-violating potential. When comparing the different electronic
structure methods, one finds significant differences, but the values
for vibrational frequency shifts in CHBrClF
are in the same order of magnitude, deviating less than 30$\%$, with
the same sign, the latter of which is not the case for the PV
energy shifts. When compared with absolute values, the 2D contributions are found to 
increase for the higher vibrational levels.
When these calculated PV frequency shifts (Table \ref{table:vibenergy}) 
are multiplied by two and then divided by the corresponding
vibrational transition frequencies, they produce the dimensionless 
parity-violating relative vibrational frequency splittings 
between the C-F stretching fundamental of S- and R-enantiomers.
The calculated relative vibrational frequency splittings corresponding to
the lowest transition for the C-F stretching vibration of CHBrClF molecule is found to be 
in good agreement with the earlier theoretical 
values.\cite{berger:2007,quack:2000,laerdahl:2000a,schwerdtfeger:2005,viglione:2000,schwerdtfeger:2002}  

For CHClFI, these parity-violating $1\leftarrow0$ transition frequency
shifts are found to be about one order of magnitude
higher than CHBrClF (See Table \ref{table:vibenergy}). Again, the shifts in vibrational energy 
levels from the full and perturbed 1D treatment match better 
with each other, whereas the values obtained after
inclusion of 2D contributions differ by a larger margin as
compared to perturbative 1D treatment.  Substitution with the heavier
element bromine to CHClFI in place of chlorine leads to an increase of
the PV transition frequency shifts compared to CHClFI again, which is not quite an order of
magnitude this time (\textit{cf}. Table \ref{table:vibenergy}). 
PV transition frequency shifts in the C-F stretching fundamental of 
CHBrFI due to full 1D and perturbed 1D are roughly twice to that of
the values for CHClFI, whereas perturbed 2D effects in CHBrFI
measures about 75$\%$ higher than the 2D contributions in CHClFI
molecule. Akin to CHBrClF, the difference between the full 1D and
perturbed 1D approach is again negligible, whereas 2D contributions
are significant. 

A similar observation is also noticed for CHAtFI, which 
shows the largest absolute value of the PV vibrational frequency difference
in the C-F stretching fundamental among the four
halogenated methane derivatives discussed herein. This large shift is
expected and mainly attributed to the presence of the 
heavier astatine nucleus. The full 1D and
perturbative 1D treatment do not differ much. When taking the 2D
terms into account, the values change significantly with the multi-mode contributions 
causing a reduction in absolute value compare to the perturbative 1D estimate
of the frequency shift. This is expected
to result from the fact, that the couplings of the C-F stretching mode
$\nu_4$ to the C-H bending mode $\nu_3$, where the frequencies are
close ($\Delta \omega \approx $ 1.5$\%$), and to the C-H stretching
mode $\nu_1$, where there is a factor of about three ($ \approx $ 2.9)
between the fundamental frequencies are very strong. In the perturbative approach
for LDA, even the sign and the order of magnitude of the effects for 
the fundamental transition (from $n_{4} = 0$ to $n_{4} = 1$) are altered.

Thus, the 2D perturbative terms are found to be important in all
cases, which increase gradually for higher vibrational levels. Thus,
the multi-mode effects due to the perturbative 2D approach increases the
fundamental vibrational frequency shift for the CHBrClF, CHClFI
molecules whereas the shifts in CHBrFI molecule are lower in absolute value as
compared to the perturbative 1D treatment counterparts. The heaviest At
derivative shows the maximum frequency shift after the inclusion of the
perturbative 2D effects at both HF and LDA level. 
 
\subsection{Multimode effects in CHBrClF and CHAtFI}
\begin{figure}[]
\includegraphics[width=\linewidth]{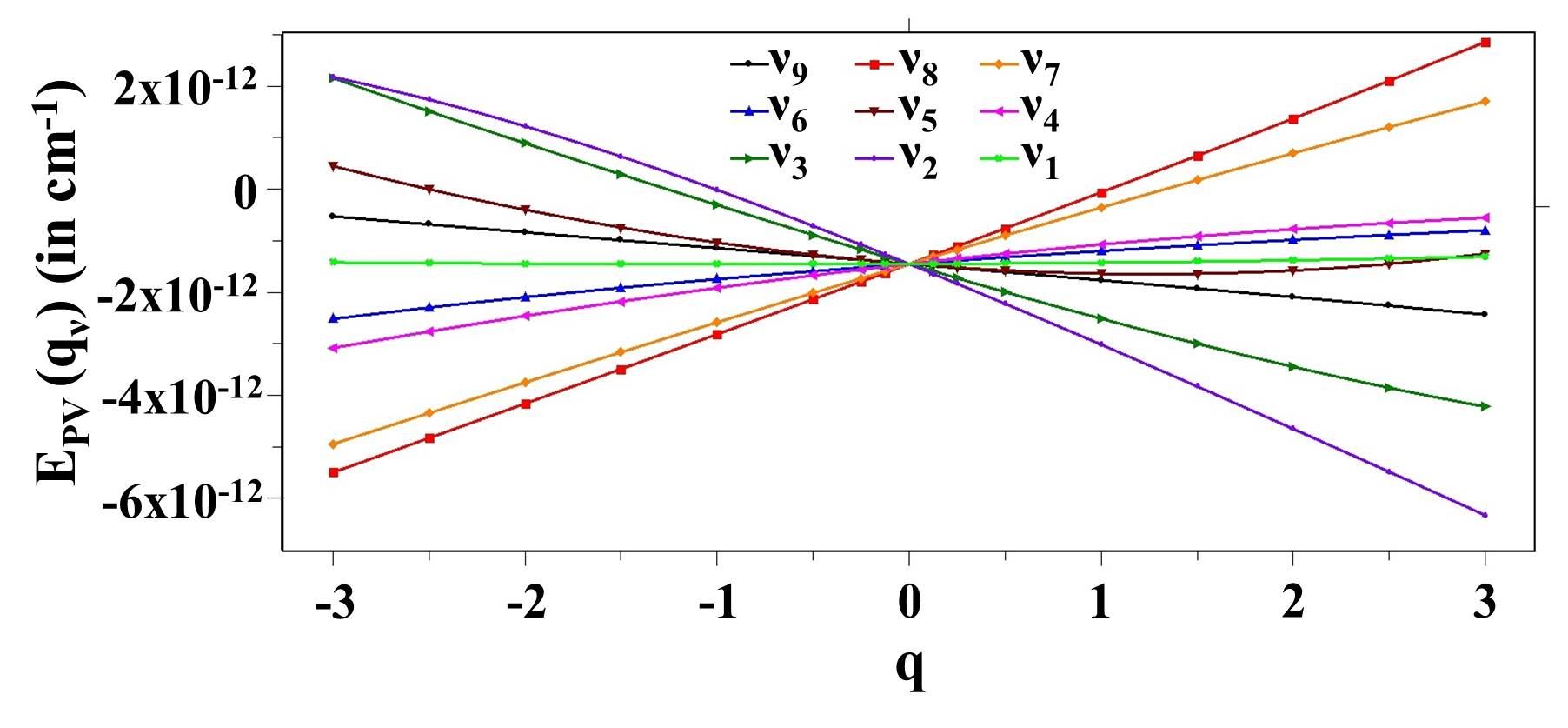}
\caption{HF parity-violating energy along the dimensionless reduced normal coordinates of CHBrClF for all the normal modes.}
\label{fig:chbrclf-hf-epv}
\end{figure}

\begin{figure}[]
\includegraphics[width=\linewidth]{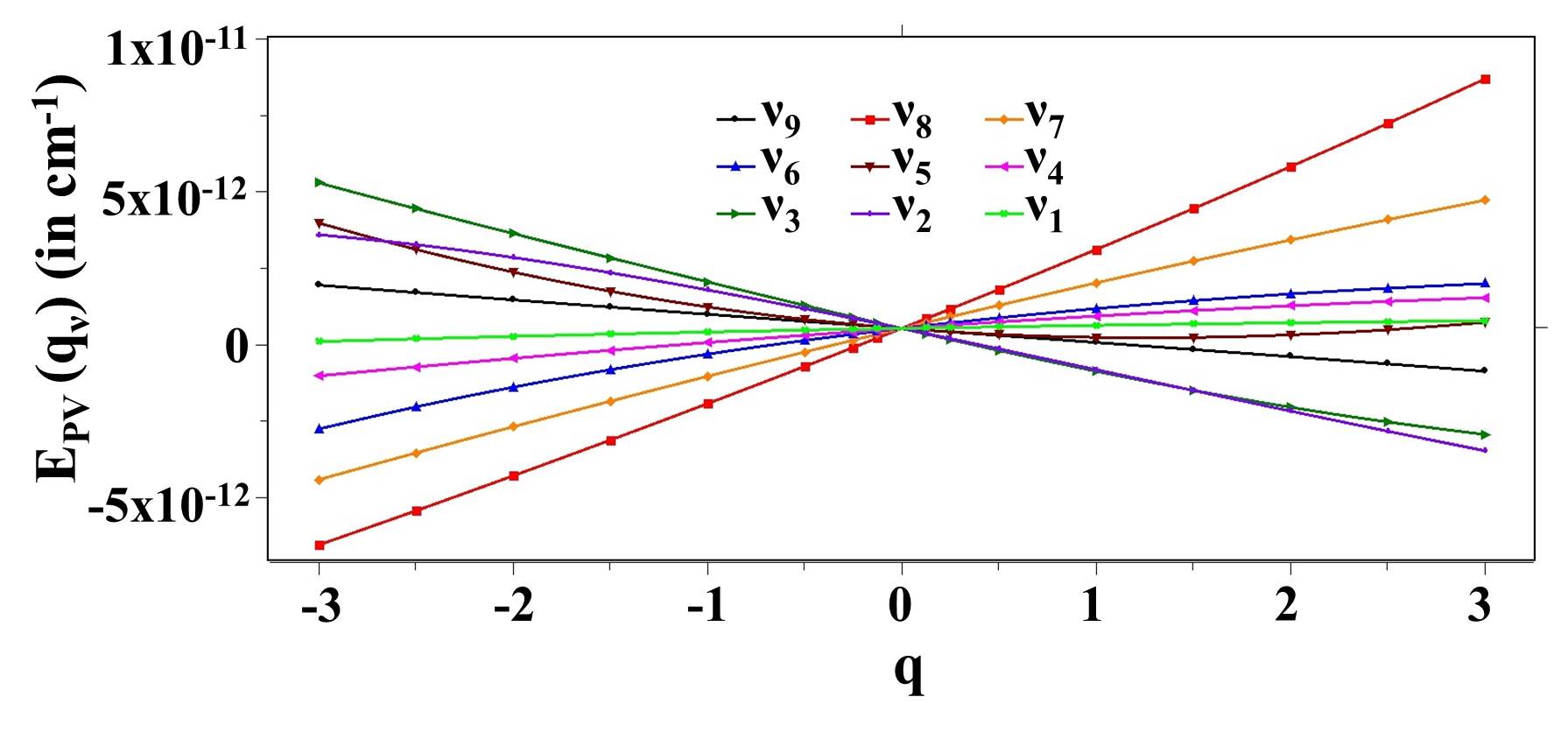}
\caption{LDA parity-violating energy along the dimensionless reduced normal coordinates of CHBrClF for all the normal modes.}
\label{fig:chbrclf-lda-epv}
\end{figure}

\begin{figure*}
\centering
\includegraphics[width=18cm]{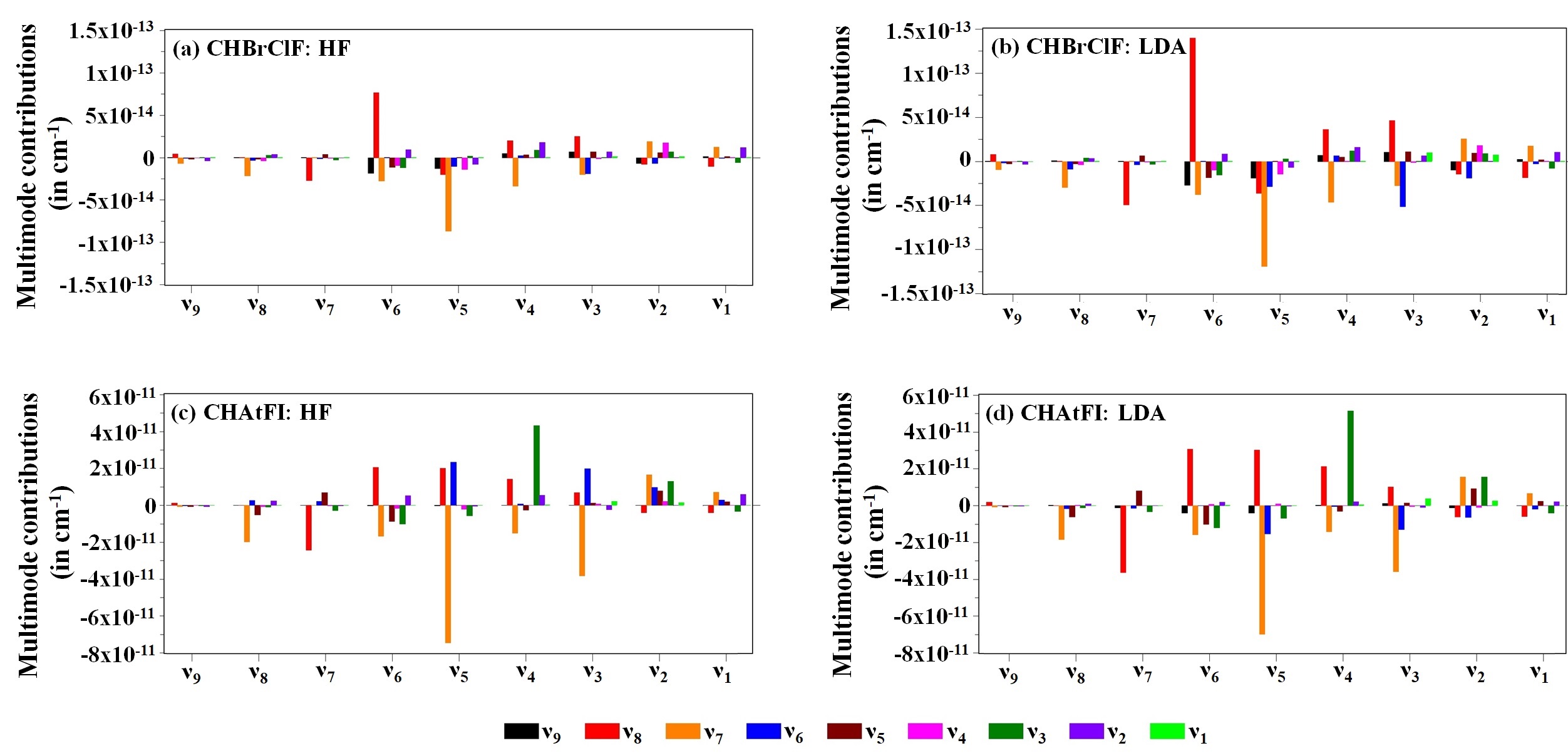}
\caption{Contributions (in cm$^{-1}$) to multi-mode effects from each of the
normal coordinates to other fundamental vibrational transitions ($1
\leftarrow 0$) in the ($\textit{S}$)-enantiomer of CHBrClF and CHAtFI molecules at HF and LDA level of theory.}
\label{FigCE}
\end{figure*}

\begin{figure}
\centering
\includegraphics[width=\linewidth]{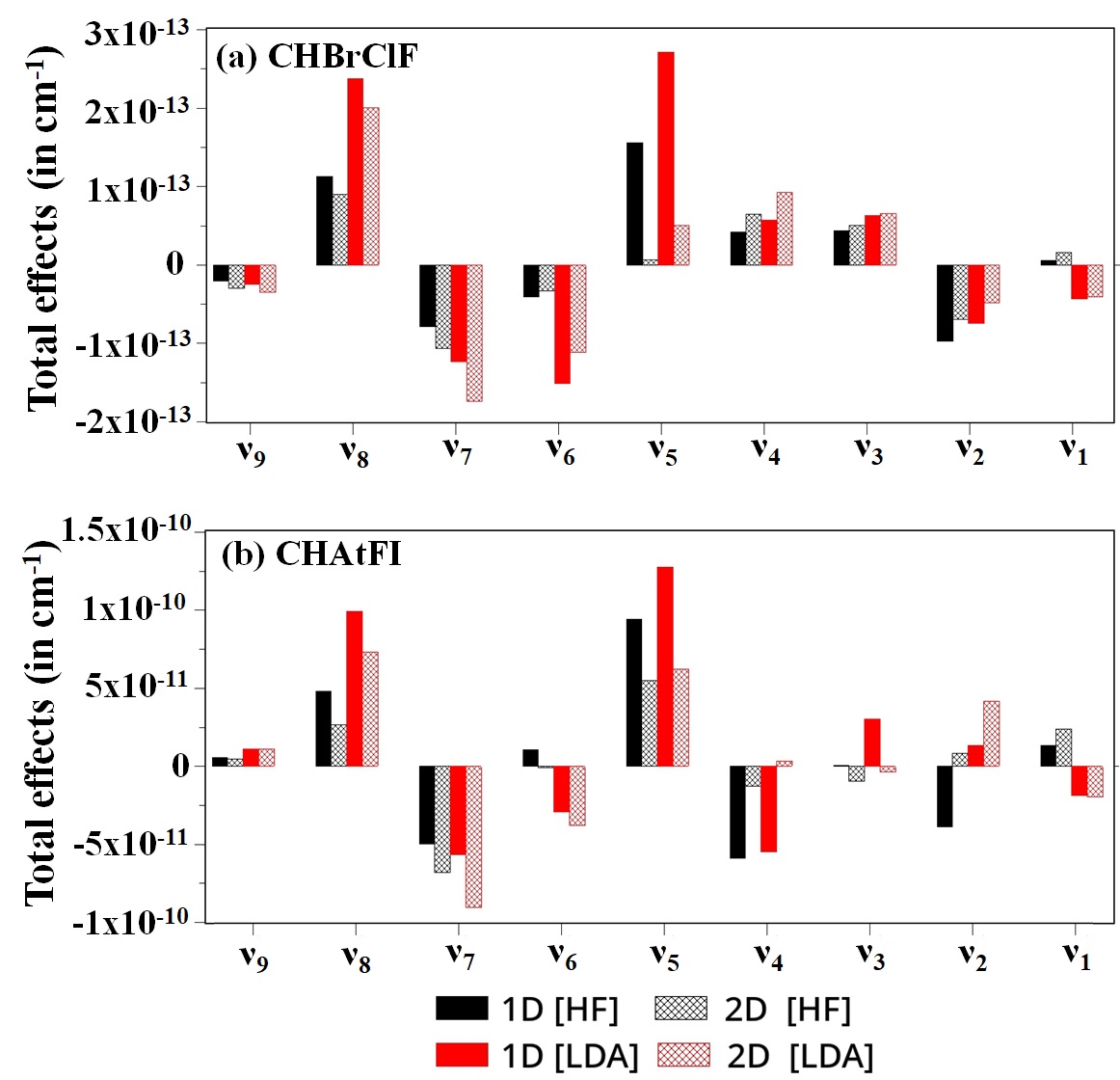}
\caption{1D and 2D perturbative treatment of the vibrationally
averaged HF and LDA parity-violating potential $E_{n ,PV}^{S}$ leading to
the shown PV frequency shifts (in cm$^{-1}$) for the
fundamental transition ($1\leftarrow 0$) in all the normal modes of the ($\textit{S}$)-enantiomer of CHBrClF and CHAtFI molecules.}
\label{FigTE}
\end{figure}

The influence of multi-mode effects\cite{quack:2003a,gaul:2020c,gaul:2020d} from
non-separable anharmonic effects have been estimated for CHBrClF and
the heaviest CHAtFI molecules after calculating
the PV energy gradients along all the normal modes. Figures
\ref{fig:chbrclf-hf-epv} and \ref{fig:chbrclf-lda-epv},
respectively, display the variation of the HF and LDA PV energies along
the dimensionless reduced normal coordinates of all the normal modes of
CHBrClF molecule. In both cases, normal coordinates $q_8$, $q_7$, $q_3$ and $q_2$
show steep slopes for the PV energy with respect to displacements along the respective dimensionless reduced normal coordinates.
This may be naively expected, because the bending motion can alter the chiral 
structure of the molecule to a larger extent than a stretching
vibration. In view of this, the vibrationally averaged PV potential 
has been computed along all the normal modes of CHBrClF and the
heavier CHAtFI analogue. Akin to the observation in CHBrClF molecule, 
the bending vibrational modes in CHAtFI exhibit larger deviation 
in the PV energies as compared to other normal modes
and hence, larger vibrational splittings and an 
enhancement of the perturbative 2D effects can be 
expected due to these modes.

Now, the vibrationally averaged HF and LDA PV gradients are utilised
for calculating the 1D (Eq. $\ref{vib}$) and 2D (Eq. $\ref{vib2D}$) perturbative
treatment for the vibrational energy levels up to 3$^\mathrm{rd}$
vibrational states. For this purpose, in addition to the PV energy gradient
$\vec \nabla E_\mathrm{PV}$, which is available along all the normal
modes, the second derivative ($a_{1}$ in Table \ref{table:TableFit} which is $\sim
\frac{\partial^{2}E_{PV}}{\partial q_r^{2}}$ in Eqs. $\ref{vib}$ and
$\ref{vib2D}$) of the PV potential is needed.  But it is not necessary
to compute a full profile, because with the availability of the analytic
gradient, the second derivative ($a_{1}$) can be computed from the
linear fit of only a few energy gradient points close to the
equilibrium structure ($q = -0.125$ and $q = 0.125$). This term is
derived from a linear fit from these two points with another at $q =
0$ ($\it{cf}.$ Tables S19-S20 in the Supporting Information).

The multi-mode contributions for the fundamental vibrational transition (from
$n = 0$ to $n = 1$) for all the vibrational modes are estimated from Eq. $\ref{vib2D}$. 
The $q_8$, $q_7$, $q_3$ and $q_2$ bending normal coordinates
(although the corresponding fundamentals give rise to the least intense peaks in the infrared spectrum)
contribute significantly to the multi-mode effects in the various fundamental
transitions.  Figure \ref{FigCE} displays the contributions towards
the multi-mode effects for each of the vibrational modes of CHBrClF and
CHAtFI molecules at HF and LDA level of theory. Contributions from
$\nu_8$ (in red) and $\nu_7$ (in orange) are the largest and they
contribute to almost every fundamental, whereas $q_6$ and $q_3$
contribute moderately in CHBrClF. The normal coordinate of the mode with
the most intense fundamental in
the vibrational spectrum, the C-F stretching fundamental ($\nu_4$), and other
normal coordinates hardly contribute to multi-mode effects in other fundamentals.
The observations are slightly different in CHAtFI.
The contributions from $q_8$ (At-C-F bending in red) and $q_7$
(F-C-I bending in orange) normal coordinates are significant. In addition to this,
the C-At stretching vibration  ($q_6$) has shown a substantial
contribution to the multi-mode effects. This is expected, as astatine is
the heaviest nucleus of the molecule and, due to the steep $Z$-scaling of
parity-violating effects, with $Z$ being the nuclear charge, should also
have a pronounced influence on the parity
violating shifts. The C-F stretching fundamental ($\nu_4$) gets a
large contributions from the H-C-F bending ($q_3$ in olive) normal coordinate,
most probably due to the strong coupling between themselves since
their frequencies are very close. Otherwise, the contributions from
other vibrations are not that significant.     

The total observed shifts associated with the fundamental 
transition (from $n = 0$ to $n = 1$) for all the vibrational modes of the
($\textit{S}$)-enantiomer of these two chiral derivatives 
from the 1D and 2D perturbative treatment are displayed in Figure
\ref{FigTE}. It should be noted that one needs to multiply these values  
by 2 and then divide by the vibrational frequencies of the corresponding
fundamentals to get the relative PV vibrational frequency splittings 
between the C–F stretching fundamental of the S- and 
R-enantiomers of CHBrClF and CHAtFI. The corresponding PV shifts
up to the 3$^\mathrm{rd}$ vibrational energy level as well as the
PV shifts in the fundamental transitions (from $n = 0$ to $n = 1$) 
are included in Tables S25-S26 of the Supporting Information.

For CHBrClF, the vibrational frequency shifts from the 1D
perturbative treatment evaluated at HF level are larger in magnitude 
for $\nu_8$, $\nu_6$, $\nu_5$ and $\nu_2$ than that of the $\nu_4$ C-F
stretching mode. The steeper slope of the curves due to the large
variation in the PV energies (which increases the numerical value of
$\vec \nabla E_\mathrm{PV}$) along $q_8$, $q_7$, $q_3$ and
$q_2$ in Figure \ref{fig:chbrclf-hf-epv} favours these large
vibrational shifts in the bending modes. On the other hand, the 2D
perturbation results in large shifts for most modes as compared to the C-F
stretching mode ($\nu_4$). This is expected since the parity-violating
multi-mode effects are significant in other bending modes and stretching
modes other than the C-F stretching vibration (See Figure
\ref{FigCE}). From the comparison of 1D and 2D effects in frequency
shifts within individual modes, it is found that the 2D effects
increase the frequency shifts only in $\nu_9$, $\nu_7$, $\nu_4$ and
$\nu_3$, whereas other fundamentals follow the opposite trend
(\textit{cf.} Figure \ref{FigTE}) where a cancellation of the 1D
contribution is caused due to the 2D treatment. Total calculated
PV shifts in the fundamentals transitions from both of the
perturbative 1D and 2D treatments align either towards positive or
negative directions. No alternation of the sign is observed even after
adding multi-mode contributions to 1D perturbative ones. Overall, the estimated 
vibrational shift in all the fundamental transitions matches 
to the same order of magnitude of previous theoretical 
values.\cite{quack:2000,viglione:2000}

The case for CHAtFI is slightly complicated. The 2D terms are
important for all modes where $\nu_8$, $\nu_7$ and $\nu_5$ fundamentals show
frequency shifts that are larger in absolute value as compared to the C-F stretching
($\nu_4$) mode (\textit{cf.} Figure \ref{FigTE}). Multimode effects
increase the frequency shifts by a factor of $\sim$ 1.5 in $\nu_7$
when compared to the perturbative 1D corrections. The correction due
to multi-mode effects for the most intense C-F stretching mode reduces the
absolute value
as compared to the perturbative 1D result, i.e.  the perturbative 2D
treatment decreases the energy gap between the ground ($n = 0$) and
1$^\mathrm{st}$ ($n = 1$) vibrational state.
Even, as mentioned above (see Table \ref{table:vibenergy}), at LDA level the sign is
altered (\textit{cf.} Figure \ref{FigTE}).  A substantial increase in
the $\nu_2$ fundamental at LDA level is seen after including the multi-mode
effects which is about 215$\%$ of the perturbative 1D value (See
Figure \ref{FigTE}). Although the fundamentals $\nu_8$ and $\nu_5$ have
relatively larger frequency shifts for the vibrational transition from
$n = 0$ to $n = 1$, the values decrease when multi-mode effects are
introduced.  Multimode contributions in $\nu_7$ increases the frequency
shifts in magnitude as compared to 1D effects.
It may be noted that the associated transition frequencies of these
fundamentals are not in the range
of the CO$_2$ laser used in the previous\cite{ziskind:2002}
experimental set up. But, in spite of their lower intensity in the
infrared spectrum, this might not be a severe limitation due to new
developments in contemporary laser technology with quantum cascade laser
being available in a broader frequency range.

We should emphasise, finally, that we reused in our present work the
equilibrium structures and harmonic vibrational force fields from a
previous study \cite{berger:2007} to allow for direct comparison of
results, Whereas a sophisticated CCSD(T) approach was used in
Ref.~\onlinecite{berger:2007}, the basis sets were only of limited quality, which
impacts not only on the equilibrium structures and harmonic force constants
obtained, but also on the quality of molecular properties as was noted in
our recent study \cite{gaul:2020b} of nuclear electric quadrupole coupling
constants in CHBrClF and CHClFI. For CHBrClF and its deuterated isotopomer,
improved anharmonic ab initio force fields have been reported for instance
in Ref.~\onlinecite{rauhut:2006}. In particular for the compound with the
radioactive halogen, CHAtFI, for which no experimental information is
available and for which very pronounced multi-mode effects on the C-F
stretching fundamental were computed herein, a study with an improved
description of the conventional parity-conserving effects is indicated.
This becomes even more important by virtue of the potential role of heavy
elemental chiral molecules in the search for dark matter
candidates~\cite{gaul:2020c,gaul:2020d} and recent advances in laser
spectroscopy of radioactive molecules with short-lived
nuclei.\cite{garciaruiz:2020}

\section{Concluding Remarks}
In the present article, the gradient of the molecular parity-violating
(PV) potential has been derived and implemented within a
quasirelativistic framework at the level of HF and LDA, and is applied
to rotational and vibrational spectroscopy of polyatomic molecules. A
systematic study of frequency shifts in the rotational and
vibrational spectra of chiral polyhalomethanes has been carried out.

A one-dimensional perturbative treatment has been compared to a
previously calculated full anharmonic one-dimensional approach and
found to be in good agreement. For the hypothetical astatine compound,
the predicted frequency splitting is in the right order of magnitude
to be measured when treating the problem in the one-dimensional
approximation. The results form the multi-mode effects in CHBrClF and
CHAtFI reflects the importance of nonseparable anharmonic effects and
suggests that the C–F stretching mode is not ideally suited for a
measurement of electroweak PV effects, since other bending and
stretching modes contribute significantly and can not be neglected.
In view of this, choosing a proper vibrational transition is equally
important as choosing a suitable molecule for conducting PV
experiments and calculating the PV effects on the vibrational
transition of chiral derivatives. Multimode effects 
are crucial for getting more insights about the
vibrational transitions as well as rotational spectra in chiral
compounds containing heavier atoms. 

The present analytic derivative approach within a quasi-relativistic
mean-field scheme leads to an important simplification for routine
calculation of PV frequency shift in rotational spectra of chiral molecules
and of multi-mode contributions to the PV frequency shifts in rovibrational
spectroscopy.
The implementation of the analytic $\vec \nabla E_\mathrm{PV}$ at
other DFT levels such as GGA (e.g. BLYP) and GGA-based hybrid-functionals 
(e.g. B3LYP) is currently underway, which will increase its applicability
to diverse chiral systems containing heavier nuclei, in particular
transition metals. Also, such a development helps in measuring the
functional dependencies and influence of electron correlation effects
for the theoretical estimation of the PV effects in the rotational and
vibrational transitions of chiral compounds. This will provide
valuable information for future experiments aiming at the detection of
molecular parity violation.

\begin{acknowledgments}
The authors are particularly thankful to Yunlong Xiao, Sophie Nahrwold,
Timur Isaev and Christoph van W{\"u}llen for stimulating discussions.
Financial support by the Deutsche Forschungsgemeinschaft (DFG, German
Research Foundation) -- Projektnummer 328961117 -- SFB 1319 ELCH and the
VolkswagenFoundation is gratefully acknowledged. The center for scientific
computing (CSC) Frankfurt is thanked for computer time.
\end{acknowledgments}

\appendix
\section{Variational Perturbation Theory\label{variational_pt}}
Following Refs.~\onlinecite{jayatilaka:1992,bast:2011a,helgaker:2012} the total energy $E(\vec{R}_0)=\Braket{\Psi|\hat{H}(\vec{R}_0)|\Psi}$
parametrically depends on the nuclear coordinates. 
Within a mean-field approach, the energy satisfies
\begin{equation}
\vec{\nabla}_{\vec{t}} E(\vec{R}_0)= \vec{\nabla}_{\vec{t}}
\Braket{\Psi(\vec{t})|\hat{H}(\vec{R}_0)|\Psi(\vec{t})}
=\vec{0}\,,
\end{equation}
where $\vec{t}$ is the vector of orbital rotations and
$\vec{\nabla}_{\vec{t}}$ is the gradient with respect to
$\vec{t}$. The variational energy of slightly displaced nuclei
$E(\vec{R}_0+\vec{\eta})$ can be expanded in a Taylor
series in the nuclear displacements $\vec{\eta}$ as (see
e.g.~Ref.~\onlinecite{sellers:1988}):
\begin{multline}
E(\vec{R}_0+\vec{\eta}) 
= E(\vec{R}_0) 
+ \left.\vec{\nabla}_{\vec{\eta}}
E(\vec{R}_0+\vec{\eta})\right|_{\vec{\eta}=\vec{0}} \cdot \vec{\eta}
\\
+\frac{1}{2}\vec{\eta}^\mathsf{T}\cdot 
\left[\left.\vec{\nabla}_{\vec{\eta}}\otimes\vec{\nabla}_{\vec{\eta}}E(\vec{R}_0+\vec{\eta})\right|_{\vec{\eta}=\vec{0}}\right]
\cdot \vec{\eta}
+\dots\,,
\label{eq: displacementPT}
\end{multline}
where $\vec{\nabla}_{\vec{\eta}}$ is the gradient with respect to the
nuclear displacement vector. 

Thus, the total variational energy is a function of
$\lambda_\mathrm{PV}$ as well:
\begin{multline}
E_\infty =
E(\vec{R}_0+\vec{\eta},\lambda_\mathrm{PV}) 
\\ = E(\vec{R}_0+\vec{\eta}) 
+ \left.\frac{\partial
E(\vec{R}_0+\vec{\eta},\lambda_\mathrm{PV})}{\partial\lambda_\mathrm{PV}}\right|_{\lambda_\mathrm{PV}=0} \lambda_\mathrm{PV}
\\
+\frac{1}{2}
\left.\frac{\partial^2
E(\vec{R}_0+\vec{\eta},\lambda_\mathrm{PV})}{\partial\lambda_\mathrm{PV}^2}\right|_{\lambda_\mathrm{PV}=0}
\lambda_\mathrm{PV}^2
+\dots\,.
\label{eq: parityviolationPT}
\end{multline}
In combination with Eq. (\ref{eq: displacementPT}) the parity
violating contribution $E_\mathrm{PV}(\vec{R}_0+\vec{\eta},\lambda_\mathrm{PV}) =
E(\vec{R}_0+\vec{\eta},\lambda_\mathrm{PV})-E(\vec{R}_0)$ to the gradient of the energy with respect to
nuclear displacements is:
\begin{multline}
\left.\vec{\nabla}_{\vec{\eta}}
E_\mathrm{PV}(\vec{R}_0+\vec{\eta},\lambda_\mathrm{PV})\right|_{\vec{\eta}=\vec{0}}
= \left. \frac{\partial
\vec{\nabla}_{\vec{\eta}}E(\vec{R}_0+\vec{\eta},
\lambda_\mathrm{PV})}{\partial\lambda_\mathrm{PV}}\right|_{\substack{\vec{\eta}=\vec{0}\\\lambda_\mathrm{PV}=0}} \lambda_\mathrm{PV}
\\
+\left.
\frac{1}{2} \frac{\partial^2
\vec{\nabla}_{\vec{\eta}}E(\vec{R}_0+\vec{\eta},
\lambda_\mathrm{PV})}{\partial\lambda_\mathrm{PV}^2}\right|_{\substack{\vec{\eta}=\vec{0}\\\lambda_\mathrm{PV}=0}} \lambda_\mathrm{PV}^2
+\dots
 \\
\approx \left. \frac{\partial\vec{\nabla}_{\vec{\eta}}
E(\vec{R}_0+\vec{\eta},
\lambda_\mathrm{PV})}{\partial\lambda_\mathrm{PV}}\right|_{\substack{\vec{\eta}=\vec{0}\\\lambda_\mathrm{PV}=0}} \lambda_\mathrm{PV}
\end{multline}

\section{Spin polarised density functionals\label{xcfun}}
We employ the approximation of non-collinear DFT. Then,  we can write
a exchange-correlation potential operator
$\hat{V}_\mathrm{XC}^{(\kappa)}$ as (see also
Ref.~\onlinecite{wullen:2007})
\begin{equation}
\hat{V}_\mathrm{XC}^{(\kappa)} = \hat{V}_\mathrm{XC, LDA}^{(\kappa)} +
\vec{\hat{V}}_\mathrm{XC, GGA}^{(\kappa)}\cdot\vec{\nabla}\,, 
\end{equation}
where $\vec{\hat{V}}_\mathrm{XC, GGA}^{(\kappa)} = \frac{\delta
F_\mathrm{XC, GGA}}{\delta
\vec{\nabla}\rho_\mathrm{e}^{(\kappa)}(\vec{r};\mathbf{D}^{(\kappa)})}$
and for the explicit definitions of $\vec{\hat{V}}_\mathrm{XC}$ see
e.g. Refs.~\onlinecite{wullen:2007,wullen:2010}.

We define the spin or number density
function $\rho_\mathrm{e}^{(\kappa)}$ as 
\begin{equation}
\rho^{(\kappa)}_\mathrm{e}(\vec{r};\mathbf{D}^{(\kappa)})=\sum_{\mu\nu}
\mathfrak{R} \left\{D^{(\kappa)}_{\mu\nu}\right\}\chi_\mu(\vec{r})
\chi_\nu(\vec{r})\,.
\end{equation}
Note, that this definition of the spin density coincides with what is
usually called in non-collinear DFT the magnetization density and
should not be confused with the length of the magnetization density
which is what is called spin density in non-collinear DFT.

We can write the exchange-correlation energy as
\begin{equation}
E_\mathrm{XC} = \sum\limits_\kappa \sum\limits_{\mu\nu}
\Re\\{D^{(\kappa)}_{\mu\nu}\\}
\Braket{\chi_\mu|\hat{V}_\mathrm{XC}^{(\kappa)}|\chi_\nu}
\end{equation}
and the gradient with respect to nuclear displacements is 
\begin{equation}
\vec{\nabla}_{\vec{\eta}} E_\mathrm{XC} = 2\sum\limits_\kappa \sum\limits_{\mu\nu}
\Re\\{D^{(\kappa)}_{\mu\nu}\\}
\Braket{\vec{\nabla}_{\vec{\eta}}\chi_\mu|\hat{V}_\mathrm{XC}^{(\kappa)}|\chi_\nu}\,.
\end{equation} 
In presence of a perturbing operator such as the parity-violating
potential we have to compute the first order perturbed
exchange-correlation potential operator as
\begin{multline}
\hat{V}'^{(\kappa)}_\mathrm{XC} = \sum_\lambda \frac{\delta
\hat{V}_\mathrm{XC}^{(\kappa)}}{\delta
\rho^{(\lambda)}_\mathrm{e}(\vec{r};\mathbf{D}^{(\lambda)})}\rho^{(\lambda)}_\mathrm{e}(\vec{r};\mathbf{D}'^{(\lambda)})
\\+ \sum_\lambda\frac{\delta
\hat{V}_\mathrm{XC}^{(\kappa)}}{\delta
\vec{\nabla}\rho^{(\lambda)}_\mathrm{e}(\vec{r};\mathbf{D}^{(\lambda)})}\vec{\nabla}\rho^{(\lambda)}_\mathrm{e}(\vec{r};\mathbf{D}'^{(\lambda)})\,,
\end{multline}
where $\rho^{(\kappa)}_\mathrm{e}(\vec{r};\mathbf{D}'^{(\kappa)})$ is the perturbed density function of first
order in $\lambda_\mathrm{PV}$.

\section{Nuclear displacement gradient of a first order
property\label{nucdisplace_property}}
We can express terms that are proportional to gradients of the density
matrix in terms of the Fock matrix
$\mathbf{F}^{(\kappa)}(\mathbf{D}^{\infty})=\mathbf{h}_{\mathrm{ZORA}}^{(\kappa)}
+\mathbf{h}_{\mathrm{PV}}^{(\kappa)} +
\mathbf{G}^{(\kappa)}(\mathbf{D}^{\infty})$
\begin{widetext}
\begin{equation}
\begin{aligned}
\vec{\nabla}_{\vec{\eta}}E_\infty &=
\mathfrak{R}\left\{\sum\limits_{\mu\nu}
\left[\sum\limits_{\kappa=0}^{3}
 D^{\infty,(\kappa)}_{\mu\nu}\left(\vec{\nabla}_{\vec{\eta}}h_{\mathrm{ZORA},\mu\nu}^{(\kappa)}\right)
+ \frac{1}{2}\sum\limits_{\kappa=0}^3 
D^{\infty,(\kappa)}_{\mu\nu} \vec{G}^{(\kappa)}_{\mathrm{grad},\mu\nu}(\mathbf{D}^{\infty})
+ \sum\limits_{\kappa=1}^{3} 
D^{\infty,(\kappa)}_{\mu\nu} \left(\vec{\nabla}_{\vec{\eta}}h_{\mathrm{PV},\mu\nu}^{(\kappa)}\right)
+\sum\limits_{\kappa=0}^{3}\left(\vec{\nabla}_{\vec{\eta}}D^{\infty,(\kappa)}_{\mu\nu}\right) F_{\mu\nu}^{(\kappa)}(\mathbf{D}^{\infty})
\right]\right\}\,,
\label{eq: totenergy_withfock}
\end{aligned}
\end{equation}
where we have defined the matrix $\vec{\mathbf{G}}_\mathrm{grad}^{(\kappa)}$ of 
contracted gradients of two-electron integrals with elements:
\begin{equation}
\vec{G}^{(\kappa)}_{\mathrm{grad},\mu\nu}(\mathbf{D})=\sum\limits_{\rho\sigma}\left[\delta_{k0}D^{(\kappa)}_{\rho\sigma}\vec{\nabla}_{\vec{\eta}}(\mu\nu|\rho\sigma)-a_{\mathrm{X}}\frac{1}{2}D^{(\kappa)}_{\rho\sigma}\vec{\nabla}_{\vec{\eta}}(\mu\sigma|\rho\nu)+2a_\mathrm{DFT}\Braket{\vec{\nabla}_{\vec{\eta}}\chi_\mu|V^{(\kappa)}_\mathrm{XC}(\mathbf{D})|\chi_\nu}\right].
\end{equation}
\end{widetext}
The canonical SCF equations are
\begin{equation}
\label{eq:canonicalhf}
\mathbf{F}\mathbf{C}=\mathbf{S}\mathbf{C}\bm{\varepsilon};\qquad
\mathbf{C}^\dagger\mathbf{S}\mathbf{C} = \bm{1}\,,
\end{equation}
where $\mathbf{S}=\bm{\sigma}^0\otimes\mathbf{S}_\mathrm{1c}$ is the overlap
matrix constructed from the one component overlap matrix
$\mathbf{S}_\mathrm{1c}$ of the basis functions and $\mathbf{F}=\sum_{\kappa=0}^3\bm{\sigma}^\kappa\otimes\mathbf{F}^{(\kappa)}(\mathbf{D})$. The matrix of
coefficients is chosen as 
\begin{equation}
\mathbf{C}= \begin{pmatrix}
\mathbf{C}^{(\alpha)}\\ \mathbf{C}^{(\beta)}
\end{pmatrix}\,.
\end{equation}
Exploiting the Hermiticity of $\mathbf{F}$ and $\bm{\varepsilon}$ we find from this
\begin{align}
(\vec{\nabla}_{\vec{\eta}}\mathbf{C}^\dagger)\mathbf{F}\mathbf{C} +
\mathbf{C}^\dagger\mathbf{F}(\vec{\nabla}_{\vec{\eta}}\mathbf{C})
=&(\vec{\nabla}_{\vec{\eta}}\mathbf{C}^\dagger)\mathbf{S}\mathbf{C}\bm{\varepsilon}
+
\bm{\varepsilon}\mathbf{C}^\dagger\mathbf{S}(\vec{\nabla}_{\vec{\eta}}\mathbf{C})
\end{align}
Furthermore, the gradient of the orthonormality condition in
Eq.~\ref{eq:canonicalhf} gives
\begin{align}
(\vec{\nabla}_{\vec{\eta}}\mathbf{C}^\dagger)\mathbf{S}\mathbf{C} 
+\mathbf{C}^\dagger\mathbf{S}(\vec{\nabla}_{\vec{\eta}}\mathbf{C}) 
=&
-\mathbf{C}^\dagger(\vec{\nabla}_{\vec{\eta}}\mathbf{S})\mathbf{C} 
\end{align}
Therewith, the contribution to the energy gradient from the Fock
matrix in Eq. (\ref{eq: totenergy_withfock}) can be expressed via 
the orthonormality condition as
\begin{widetext}
\begin{equation}
\begin{aligned}
\sum_{\mu\nu}\sum\limits_{\kappa=0}^{3}\left(\vec{\nabla}_{\vec{\eta}}D^{\infty,(\kappa)}_{\mu\nu}\right)
F_{\mu\nu}^{(\kappa)}(\mathbf{D}^{\infty})
=&
\sum_{\mu\nu}\sum_i n_i\sum\limits_{\kappa=0}^{3}
\left[(\vec{\nabla}_{\vec{\eta}}\vec{C}^\infty_{\mu
i})^\dagger\bm{\sigma}^{(\kappa)}
F^{(\kappa)}_{\mu\nu}(\mathbf{D}^{\infty})\vec{C}^\infty_{\nu i} +
(\vec{C}^\infty_{\mu i})^\dagger\bm{\sigma}^{(\kappa)}
F^{(\kappa)}_{\mu\nu}(\mathbf{D}^{\infty})(\vec{\nabla}_{\vec{\eta}}\vec{C}_{\nu
i}^\infty)\right]\\
=&
\sum_{\mu\nu}\sum_i n_i\varepsilon_i^\infty\left[
(\vec{\nabla}_{\vec{\eta}}\vec{C}^\infty_{\mu
i})^\dagger
S_{\mu\nu}\vec{C}^\infty_{\nu i}
+
(\vec{C}^\infty_{\mu i})^\dagger S_{\mu\nu}(\vec{\nabla}_{\vec{\eta}}\vec{C}_{\nu
i}^\infty)
\right]\\
=&
-\sum_{\mu\nu}\sum_i n_i\varepsilon_i^\infty(\vec{C}^\infty_{\mu
i})^\dagger(\vec{\nabla}_{\vec{\eta}}S_{\mu\nu})\vec{C}_{\nu
i}^\infty\\
=& -\sum\limits_{\mu\nu} W^{\infty,(0)}_{\mu\nu} \vec{\nabla}_{\vec{\eta}} S_{\mu \nu} 
\end{aligned}
\end{equation}
\end{widetext}
where we introduced the energy weighted density matrix
(EWDM) $\mathbf W$ as
\begin{equation}
\mathbf{W}^{(\kappa)} = \sum\limits_{i=1}^{N_\mathrm{orb}}n_i\varepsilon_i
\mathbf{D}^{(\kappa)}_i\,.
\end{equation}

\section{Linear response equations\label{linresponse}}
\subsection{First order perturbed density
matrix\label{first_order_pt_density}}
The infinite order coefficients can be expressed in terms of orbital
rotations, expressed via the anti-Hermitian matrix $\mathbf{T}_\infty$ as
\begin{equation}
\mathbf{C}^\infty = \mathbf{C}_0 \mathrm{e}^{\mathbf{T}_\infty}
\end{equation}
were $\mathbf{C}_0$ is the initial guess of
orthonormal orbitals and
$(\mathbf{C}^\infty)^\dagger=\mathrm{e}^{-\mathbf{T}_\infty}\mathbf{C}_0^\dagger$.
From the definition of the density matrix [Eq.~(\ref{eq:
density_matrix})] we see
$\mathbf{D}^\infty=\left(\mathbf{C}^\infty\mathbf{N}(\mathbf{C}^\infty)^\dagger\right)^\mathsf{T}=(\mathbf{C}^\infty)^*\mathbf{N}(\mathbf{C}^\infty)^\mathsf{T}$,
where $\mathbf{N}$ is a diagonal matrix containing the occupation
numbers $n_i$.
Thus, the perturbed density matrix of infinite order $\mathbf{D}^\infty$ can
be written in terms of orbital rotation matrix $\mathbf{T}_\infty$ as
\begin{equation}
\mathbf{D}^\infty =
\mathbf{C}_0^*\mathrm{e}^{-\mathbf{T}^\mathsf{T}_\infty}\mathbf{N}\mathrm{e}^{\mathbf{T}^\mathsf{T}_\infty}\mathbf{C}_0^\mathsf{T}
\,.
\label{eq: unitary_transform}
\end{equation}
Moreover, we can write $\mathbf{D}^\infty$ in
terms of the unperturbed orbital coefficients $\mathbf{C}$, received from
orbital rotations $\vec{t}_0$ in absence of the perturbation with
the corresponding rotation matrix $\mathbf{T}_0$ and an
the rotation matrix $\mathbf{T}$ as:
\begin{equation}
\mathbf{D}^\infty =
\mathbf{C}_0^*\mathrm{e}^{-\mathbf{T}^\mathsf{T}_0}\mathrm{e}^{-\mathbf{T}^\mathsf{T}}\mathbf{N}\mathrm{e}^{\mathbf{T}^\mathsf{T}}\mathrm{e}^{\mathbf{T}_0^\mathsf{T}}\mathbf{C}_0^\mathsf{T}
=
\mathbf{C}^*\mathrm{e}^{-\mathbf{T}^\mathsf{T}}\mathbf{N}\mathrm{e}^{\mathbf{T}^\mathsf{T}}\mathbf{C}^\mathsf{T}\,.
\label{eq: unitary_transform_steps}
\end{equation}
When assuming $\mathbf{N}$ to be successively sorted for occupied and
unoccupied orbitals we can write the transformation matrix
$\mathbf{T}$ in blocked form as
\begin{equation}
\mathbf{T} = \begin{pmatrix} 
\mathbf{T}_\mathrm{oo} & -\mathbf{T}_\mathrm{uo}^\dagger \\
\mathbf{T}_\mathrm{uo} & \mathbf{T}_\mathrm{uu} \\
\end{pmatrix}\,,
\end{equation} 
where, o means occupied and u means unoccupied.
In first order the perturbed density matrix can be written in terms of
the unoccupied-occupied block $\mathbf{T}_\mathrm{uo}$ of $\mathbf{T}$:
\begin{equation}
\begin{aligned}
\mathbf{D}' =&
\mathbf{C}^*\left(\mathbf{N}\mathbf{T}^\mathsf{T}-\mathbf{T}^\mathsf{T}\mathbf{N}\right)\mathbf{C}^\mathsf{T}\\
=&
\mathbf{C}^*\left(\mathbf{N}\mathbf{T}^\mathsf{T}+\mathbf{T}^*\mathbf{N}\right)\mathbf{C}^\mathsf{T}\\
=& \mathbf{C}^*\begin{pmatrix} 
\bm{0}& \mathbf{T}_\mathrm{uo}^\mathsf{T} \\
\mathbf{T}_\mathrm{uo}^* & \bm{0} \\
\end{pmatrix}\mathbf{C}^\mathsf{T}\,.
\end{aligned}
\label{eq: rotated_density}
\end{equation}
Thus, we arrive at
\begin{align}
D'^{(\kappa)}_{\mu\nu}(\mathbf{T}_\mathrm{uo}) &=
\sum\limits_{i}^{\mathrm{occ}}\sum\limits_{a}^{\mathrm{unocc}}\left[
\vec{C}^\dagger_{\mu a}\bm{\sigma}^{(\kappa)}\vec{C}_{\nu i} T_{ai}^*  
+
\vec{C}_{\nu i}^\dagger\bm{\sigma}^{(\kappa)}\vec{C}_{\mu a} T_{ai}
\right]
\,.
\end{align}

\subsection{Coupled perturbed HF/KS equations\label{cphf_ks}}
The coupled perturbed HF (CPHF) or coupled perturbed KS
(CPKS) equations read (see Ref.~\onlinecite{olsen:1989}).
\begin{equation}
\mathbf{F}\mathbf{C}'+\mathbf{F}'\mathbf{C}=\mathbf{S}'\mathbf{C}\bm{\varepsilon}+
\mathbf{S}\mathbf{C}'\bm{\varepsilon}+\mathbf{S}\mathbf{C}\bm{\varepsilon}'
\label{eq: cphf_ks}
\end{equation}
where, a prime denotes matrices linear in $\lambda_\mathrm{PV}$ and
unprimed matrices are unperturbed ($\lambda_\mathrm{PV}=0$). The
perturbed Fock matrix $\mathbf{F}'$ is for perturbation independent
basis functions as we have in the case of the perturbation due to the
PV potential:
\begin{equation}
F'_{\mu\nu} = \sum\limits_\kappa \bm{\sigma}^\kappa h_{\mathrm{PV},\mu\nu}^{(\kappa)}
+ \bm{\sigma}^\kappa G_{\mu\nu}^{(\kappa)}(\mathbf{D}'(\mathbf{T}_\mathrm{uo}))\,
\end{equation}
where, we introduced the self consistent transformation matrix of the
orbital coefficients $\mathbf{T}$ for which
$\mathbf{C}'=\mathbf{C}\mathbf{T}$. 
From Eq. (\ref{eq: rotated_density}) we see that only
the unoccupied-occupied rotations change the orbitals in first order,
such that we can focus on the of unoccupied-occupied block of
Eq (\ref{eq: cphf_ks}).  Thus, for perturbation independent basis
functions the CPHF/CPKS equations reduce to the coupled equations 
\begin{align}
\bm{\varepsilon}_\mathrm{u}\mathbf{T}_\mathrm{uo} +
\mathbf{C}_\mathrm{u}^\dagger\mathbf{F}'\mathbf{C}_\mathrm{o}&=\mathbf{T}_\mathrm{uo}\bm{\varepsilon}_\mathrm{o}\\
\bm{\varepsilon}_\mathrm{o}\mathbf{T}_\mathrm{ou} +
\mathbf{C}_\mathrm{o}^\dagger\mathbf{F}'\mathbf{C}_\mathrm{u}&=\mathbf{T}_\mathrm{ou}\bm{\varepsilon}_\mathrm{u},
\end{align}
with the unoccupied (u) and occupied (o) orbital sub-blocks of
$\mathbf{C}$ and $\bm{\varepsilon}$.
From this, one arrives at the linear response equations (see e.g.
Refs.~\onlinecite{salek:2002,saue:2003})
\begin{equation}
\sum\limits_{bj}
\begin{pmatrix}
\mathbf{A}_{bj} & \mathbf{B}_{bj}\\
\mathbf{B}_{bj}^* & \mathbf{A}_{bj}^* 
\end{pmatrix}
\begin{pmatrix}T_{bj}\\hT_{bj}^*\end{pmatrix} =
-\begin{pmatrix}\mathbf{H}^{\mathrm{uo}}_\mathrm{PV,MO}\\(\mathbf{H}^{\mathrm{uo}}_\mathrm{PV,MO})^*\end{pmatrix},
\end{equation}
where the index $b$ runs over unoccpuied orbitals and the index $j$
runs over occupied orbitals. The Hermiticity factor $h$ results from
the orthonormality condition which for perturbation independent basis
functions yields
$\mathbf{C}^\dagger\mathbf{S}\mathbf{C}\mathbf{T}=-\mathbf{T}^\dagger\mathbf{C}^\dagger\mathbf{S}\mathbf{C}\Leftrightarrow\mathbf{T}^\dagger=-\mathbf{T}$,
which is in accordance with Eq. (\ref{eq: unitary_transform}).  For
observables such as $\mathbf{H}_{\mathrm{PV}}$ we have $h=+1$.

\end{document}


\begin{center}
{\bf \huge Supporting Information for} \\
\vspace{40mm}
{\bf \LARGE Quasi-relativistic approach to analytical gradients of parity violating potentials} \\ 
\vspace{10mm}
{\large Sascha A Br\"{u}ck,$^{1}$ Nityananda Sahu,$^{2}$ Konstantin Gaul,$^{2}$ and Robert Berger$^{1,2,3,\ast}$} \\ 
\end{center}
\vspace{15mm}
{\it $^{1)}$Frankfurt Institute for Advanced Studies, Ruth-Moufang-Stra{\ss}e 1, 60438 Frankfurt am Main, Germany \\
$^{2)}$Fachbereich Chemie, Philipps--Universit{\"a}t Marburg, Hans-Meerwein-Stra{\ss}e 4, 35032 Marburg, Germany \\
$^{3)}$Clemens-Sch{\"o}pf-Institut, Technische Universit{\"a}t, Darmstadt, Alarich-Weiss-Stra{\ss}e 4, 64287, Darmstadt, Germany \\}

Herein, we report analytically determined Cartesian gradient components of
the parity violating potential as obtained on the Hartree-Fock (HF) and
local density approximation (LDA) level at the equilibrium structures of
the ($\textit{S}$)-enantiomer of the various methane derivatives (see
Tables~\ref{tab:gradCHBrClF}--\ref{tab:gradCHAtFI}). They refer to the
specific equilibrium structures and orientation as reported previously in
the Supplementary Material to R. Berger and J. L. Stuber, Mol. Phys.
{\bf105}, 41 (2007). Based on the calculations and verifications on the convergence 
of parity violating energies and parity violating energy gradients of few
planar molecules\footnote{We are particularly grateful to Dr. Yunlong Xiao,
who proposed this crucial test of the gradient to us.},
we here give the values up to four digits after the decimal point.
The parity violating energy at the equilibrium structures as obtained on the HF and LDA levels are given in table \ref{tab:Epv0}.
Molecular orbitals (MOs) have been
converged until the change of the SCF energy and relative change of spin-orbit
energy between two successive iterations
dropped below at least $10^{-6}$ $E_{\textrm{h}}$ and $10^{-12}$
respectively. MOs of CHClFI at the HF level, however, have been converged until
the relative change of the spin-orbit energy was on the level of about
$10^{-15}$ as the corresponding PV energy for the equilibrium structure did
not achieve the desired accuracy with the $10^{-12}$ criterion. In 
practice, the spin-orbit energy criterion was by far the more restrictive one. 
At the end of the process, the change in SCF energy typically remained below
$10^{-9}$ $E_{\textrm{h}}$.

In table \ref{tab:freq}, we repeat unscaled harmonic vibrational wavenumbers
of all normal modes as computed in R. Berger and J. L. Stuber, Mol. Phys.
{\bf105}, 41 (2007), but give for precision reasons one more digit. The
Cartesian displacement coordinates of each of these modes, corresponding to
a unit shift along the dimensionless reduced normal coordinates
$q_{i}$, are given in tables
\ref{tab:displacementsCHBrClF}--\ref{tab:displacementsCHAtFI}. The displacements 
refer to the same equilibrium structure and orientation as given in the
Supplementary Material to R. Berger 
and J. L. Stuber, Mol. Phys.  {\bf105}, 41 (2007), where the displacements
for $q_{4}$ were given. The various displacements are provided herein
to clarify also their phase used in determining cuts through the parity 
violating potentials along the normal modes.
Table \ref {tab:freqm} depicts the unscaled harmonic vibrational wavenumbers 
as computed from the anharmonic force field calculations
at the CCSD(T) level of theory for the equilibrium
structure of the ($\textit{S}$)-enantiomer of CHBrClF, CHClFI, CHBrFI
and CHAtFI molecules using the MOLPRO, which are utilized for 
calculating vibrationally averaged HF and LDA parity violating potential for energy levels.

In tables \ref{tab:cubicCHBrClF}--\ref{tab:cubicCHAtFI}, we give the
semi-diagonal cubic force constants as computed with the SURF module
of the program package Molpro. Signs have been adjusted where neccessary
to reflect the same phases of the normal modes as reported in tables
\ref{tab:displacementsCHBrClF}--\ref{tab:displacementsCHAtFI}.

In table \ref{tab:EpvCHBrClF}, HF and LDA level parity violating energies along 
the dimensionless reduced normal coordinates $q$ ranging from $-$3.00 to $+$3.00
for all (except C-F stretching mode $\nu_4$) the normal modes of the 
($\textit{S}$)$-$enantiomer of CHBrClF are given. See the Supplementary Material to R. Berger 
and J. L. Stuber, Mol. Phys.  {\bf105}, 41 (2007) for the parity violating energies 
of the C-F stretching mode. The values are reported up to four digits after the decimal point.

Tables \ref{tab:EpvGCHBrClF}--\ref{tab:EpvGCHAtFI} depict the HF and LDA level  
PV energy gradients along the dimensionless reduced normal 
coordinates ($q_{4}$ in the range from $-$3.00 to $+$3.00) corresponding to the C-F stretching mode ($\nu_4$) for 
the ($\textit{S}$)-enantiomers of CHBrClF, CHClFI, CHBrFI and CHAtFI respectively. On the other hand,
the HF and LDA level PV energy gradients along the dimensionless reduced normal coordinates
$q$ = $-$0.125, $+$0.00 and $+$0.125 corresponding to all (except C-F stretching mode $\nu_4$) 
the normal modes of the ($\textit{S}$)-enantiomer of CHBrClF and CHAtFI are given respectively 
in tables \ref{tab:3ptsCHBrClF} and \ref{tab:3ptsCHAtFI}. The values are reported up to four digits after the decimal point.

In table \ref{tab:LDAFit}, the fitting coefficients of the LDA PV energy
and the LDA PV energy gradients along the C-F stretching mode ($\nu_4$) of 
the methane derivatives are given. Whereas in table \ref{tab:LinearFit}, 
the second order derivative terms of the HF and PV LDA energy are given. These are 
obtained from a linear fit of the PV energy gradients at 
$q$ = $-$0.125, $+$0.00 and $+$0.125 along the C-F stretching mode. 

In tables \ref{tab:CHBrClFVibE} and \ref{tab:CHAtFIVibE}, the vibrationally averaged HF and
LDA parity violating potential $E_{v,\mathrm{PV}}^{S}$ up to four lowest vibrational energy levels 
as obtianed from the 1-dimensional and 2-dimensional perturbative treatment are given
for all (expect C-F stretching vibration $\nu_4$) the normal modes of ($\textit{S}$)-enantiomers
of CHBrClF and CHAtFI molecules. The frequency shift for the fundamental transition for all 
the vibrational modes are calculated. The percentage of the multimode effects in each of the energy 
levels are also included. 
\newpage

\begin{table*}[]
        \caption{Cartesian HF and LDA PV gradients [in Hartree/Bohr ($E_\mathrm h/a_0$)] for the equilibrium structure of the ($\textit{S}$)-enantiomer of CHBrClF.\label{tab:gradCHBrClF}}
\begin{center}
\setlength{\tabcolsep}{12pt}
\begin{tabular}{lS[round-precision=4,round-mode=places,table-format=-1.4e-2]S[round-precision=4,round-mode=places,table-format=-1.4e-2]S[round-precision=4,round-mode=places,table-format=-1.4e-2]} \toprule
\multicolumn{4}                                  {c}{\bf{HF}}                                                               \\  \midrule
 Atom & \multicolumn{1}{c}{$x$} & \multicolumn{1}{c}{$y$} & \multicolumn{1}{c}{$z$} \\ \midrule
C  &   -3.884190389248646E-017 &  2.124439914094466E-018 &  6.004386987268143E-018 \\ 
H  &    1.947783486242358E-017 & -1.905093374187981E-017 &  2.302633185615121E-019 \\
F  &    1.130578067429468E-017 &  4.491559395238545E-017 &  3.290274443732623E-018 \\
Cl &    3.011753944446303E-017 & -2.846179572343325E-017 & -1.333424231356668E-017 \\
Br &   -2.205925115207907E-017 &  4.726956127281288E-019 &  3.809317688101694E-018 \\ \midrule 
 \multicolumn{4}                                   {c}{\bf{LDA}}                                                             \\  \midrule
 Atom & \multicolumn{1}{c}{$x$} & \multicolumn{1}{c}{$y$} & \multicolumn{1}{c}{$z$} \\ \midrule
C  &   -3.627929346649333E-017 & -6.244648615926249E-018 &  3.001807106976615E-018 \\
H  &    1.592229649823852E-017 & -2.574686911530290E-017 & -2.350973360348403E-018 \\
F  &    1.874382867005071E-017 &  7.877526811456384E-017 &  4.087011747196073E-018 \\
Cl &    4.561792186825698E-017 & -4.316554773792785E-017 & -1.470944618550085E-017 \\
Br &   -4.400475360291593E-017 & -3.618202640199601E-018 &  9.971600748126677E-018 \\  \toprule
\end{tabular}
\end{center}
\end{table*}

\begin{table*}[]
\caption{Cartesian HF and LDA PV gradients [in Hartree/Bohr ($E_\mathrm h/a_0$)] for the equilibrium structure of the ($\textit{S}$)-enantiomer of CHClFI.\label{tab:gradCHClFI}}
\begin{center}
\setlength{\tabcolsep}{12pt}
\begin{tabular}{lS[round-precision=4,round-mode=places,table-format=-1.4e-2]S[round-precision=4,round-mode=places,table-format=-1.4e-2]S[round-precision=4,round-mode=places,table-format=-1.4e-2]} \toprule
\multicolumn{4}                                  {c}{\bf{HF}}                                                                \\  \midrule
 Atom & \multicolumn{1}{c}{$x$} & \multicolumn{1}{c}{$y$} & \multicolumn{1}{c}{$z$} \\ \midrule
C  &   -2.789320232752986E-016 &  7.297038903345494E-017 &  5.618260029519358E-017 \\
H  &    1.044178776916815E-016 & -6.949528537366308E-017 &  8.531200487455046E-018 \\
F  &    9.309981973436639E-017 &  2.227086239762475E-016 & -8.534979890258278E-018 \\
Cl &    1.867552317316197E-016 & -2.366589442969286E-016 & -4.734894101773326E-017 \\
I  &   -1.053409061166126E-016 &  1.047521671042447E-017 & -8.829879281990645E-018 \\  \toprule
 \multicolumn{4}                                   {c}{\bf{LDA}}                                                             \\  \midrule
 Atom & \multicolumn{1}{c}{$x$} & \multicolumn{1}{c}{$y$} & \multicolumn{1}{c}{$z$} \\ \midrule
C  &  -2.142687750494225E-016 &  1.006480973852376E-018 &  2.722338449763796E-017 \\
H  &   7.426830768051654E-017 & -1.072086746916684E-016 & -2.263423144859199E-017 \\
F  &   1.039455890855705E-016 &  3.357707849678507E-016 &  1.517650797051813E-017 \\
Cl &   2.067282836070985E-016 & -2.269100268361163E-016 & -5.423845212814412E-017 \\
I  &  -1.706734054743849E-016 & -2.658564387285980E-018 &  3.447279144373953E-017 \\ \toprule
\end{tabular}
\end{center}
\end{table*}

\begin{table*}[]
\caption{Cartesian HF and LDA PV gradients [in Hartree/Bohr ($E_\mathrm h/a_0$)] for the equilibrium structure of the ($\textit{S}$)-enantiomer of CHBrFI.\label{tab:gradCHBrFI}}
\begin{center}
\setlength{\tabcolsep}{12pt}
\begin{tabular}{lS[round-precision=4,round-mode=places,table-format=-1.4e-2]S[round-precision=4,round-mode=places,table-format=-1.4e-2]S[round-precision=4,round-mode=places,table-format=-1.4e-2]} \toprule
\multicolumn{4}                                  {c}{\bf{HF}}                                                                \\  \midrule
 Atom & \multicolumn{1}{c}{$x$} & \multicolumn{1}{c}{$y$} & \multicolumn{1}{c}{$z$} \\ \midrule
C  &   -4.980697176154770E-016 & -2.385247275083793E-016 &  3.122369572938899E-016 \\
H  &    1.639225608428647E-016 & -4.885341331700197E-016 & -3.543994294905995E-017 \\
F  &   -2.108515277196401E-017 &  1.336496171917163E-015 & -9.415491131957960E-017 \\
Br &    8.361084468061412E-016 & -4.780226805619512E-016 & -3.209124950727979E-016 \\
I  &   -4.808761373775227E-016 & -1.314146306683241E-016 &  1.382703921228268E-016 \\ \toprule
 \multicolumn{4}                                   {c}{\bf{LDA}}                                                             \\  \midrule
 Atom & \multicolumn{1}{c}{$x$} & \multicolumn{1}{c}{$y$} & \multicolumn{1}{c}{$z$} \\ \midrule
C  &   -4.669534027947073E-016 & -3.466249983607791E-016 &  2.499164084283264E-016 \\
H  &    4.657728633347492E-017 & -6.775364240129691E-016 & -9.371489737984194E-017 \\
F  &    8.114060080314221E-017 &  2.274876748936827E-015 & -9.492868233875715E-017 \\
Br &    1.345046449963887E-015 & -8.477162677525761E-016 & -3.279048922106019E-016 \\
I  &   -1.005810934485286E-015 & -4.029990587961998E-016 &  2.666320635311707E-016 \\ \toprule
\end{tabular}
\end{center}
\end{table*}

\begin{table*}[]
\caption{Cartesian HF and LDA PV gradients [in Hartree/Bohr ($E_\mathrm h/a_0$)] for the equilibrium structure of the ($\textit{S}$)-enantiomer of CHAtFI.\label{tab:gradCHAtFI}}
\begin{center}
\setlength{\tabcolsep}{12pt}
\begin{tabular}{lS[round-precision=4,round-mode=places,table-format=-1.4e-2]S[round-precision=4,round-mode=places,table-format=-1.4e-2]S[round-precision=4,round-mode=places,table-format=-1.4e-2]} \toprule
\multicolumn{4}                                  {c}{\bf{HF}}                                                                \\  \midrule
 Atom & \multicolumn{1}{c}{$x$} & \multicolumn{1}{c}{$y$} & \multicolumn{1}{c}{$z$} \\ \midrule
C &   -1.834598323949893E-014 & -5.524044848168744E-016 &  1.199448800132946E-014 \\
H &    7.904427010649049E-015 & -1.063001950556959E-014 &  5.650582490184070E-016 \\
F &   -2.270716317219369E-015 &  3.266085034796097E-014 & -5.236347133707917E-015 \\
I &    2.349982312678920E-014 & -1.375329507184285E-014 & -7.228306250107943E-015 \\
At&   -1.078755058654457E-014 & -7.725131286182207E-015 & -9.489286336793902E-017 \\ \toprule
\multicolumn{4}                                   {c}{\bf{LDA}}                                                             \\  \midrule
 Atom & \multicolumn{1}{c}{$x$} & \multicolumn{1}{c}{$y$} & \multicolumn{1}{c}{$z$} \\ \midrule
C &  -1.318874048725536E-014 & -2.442936018689857E-015 &  3.273387656118468E-015 \\
H &   1.640152216062033E-015 & -1.276635114239998E-014 & -6.445460628745692E-016 \\
F &   3.278905516406071E-015 &  4.339407723230330E-014 & -3.821602222105493E-015 \\
I &   2.661432242248070E-014 & -1.946318386720990E-014 & -3.526463797016486E-015 \\
At&  -1.834463966859251E-014 & -8.721606205524787E-015 &  4.719224425427727E-015 \\ \toprule
 \end{tabular}
\end{center}
\end{table*}

\begin{table*}[]
\caption{Parity violating energy $E_\mathrm{PV}$ [in cm$^{-1}$] at the
CCSD(T) optimized geometry (equilibrium structure) of the
($\textit{S}$)-enantiomer of chiral methane derivatives. The reported
values have been recomputed in this work, but agree with the values given
in the Supplementary Material to R. Berger and J. L. Stuber, Mol. Phys.
{\bf105}, 41 (2007), except for a minor deviation ($1.2\times10^{-14}$) on the 
HF level for CHClFI as explained in the main text.
\label{tab:Epv0}}
\begin{center}
\setlength{\tabcolsep}{12pt}
\begin{tabular}{lS[round-precision=4,round-mode=places,table-format=-1.4e-2]S[round-precision=4,round-mode=places,table-format=-1.4e-2]} \toprule
 Molecule   & \multicolumn{1}{c}{\bf{HF}} & \multicolumn{1}{c}{\bf{LDA}}  \\ \midrule
CHBrClF &  -1.4536049027844838e-12 &   5.340071043415603e-13  \\
CHClFI  &  -1.37227408675584e-11   &   5.3567892415276254e-12  \\
CHBrFI  &  -3.848578500417084e-11  &   1.9357655986870807e-11  \\
CHAtFI  &  -2.3140586435066584e-09 &   9.654197938068645e-10   \\
%
%
 \toprule
 \end{tabular}
\end{center}
\end{table*}

\begin{table*}[]
\caption{ Unscaled harmonic vibrational wavenumbers $\tilde{\omega }$ (in
cm$^{-1}$) as computed at the CCSD(T) level of theory for the equilibrium
structure of the ($\textit{S}$)-enantiomer of CHBrClF, CHClFI, CHBrFI
and CHAtFI molecules in R. Berger and J. L. Stuber, Mol. Phys. {\bf105},
41 (2007). Herein, we give for precision reasons one digit 
after the decimal point more than reported in Tables 1.2, 4.2, 3.2 and 6.2 of 
the Supplementary Material to R. Berger and J. L. Stuber, Mol. Phys. {\bf105}, 
41 (2007). The C–F stretching mode corresponds to mode $\nu_4$. 
These wavenumbers along with the corresponding harmonic force field data are utilized
for generating displaced geometries along one dimensional cuts.\label{tab:freq}}
\begin{center}
\setlength{\tabcolsep}{15pt}
\begin{tabular}{crrrr}\toprule 
 Mode& CHBrClF & CHClFI & CHBrFI & CHAtFI\\ \midrule
 $\nu_9$ &  224.23  &  193.51 &  143.36 &  100.34\\
 $\nu_8$ &  309.18  &  269.47 &  267.83 &  237.91\\
 $\nu_7$ &  422.59  &  412.95 &  326.37 &  280.21\\
 $\nu_6$ &  638.36  &  576.35 &  552.26 &  490.61\\
 $\nu_5$ &  786.56  &  776.20 &  653.72 &  603.87\\
 $\nu_4$ & 1122.11  & 1108.20 & 1101.67 & 1091.16\\
 $\nu_3$ & 1236.37  & 1210.33 & 1180.30 & 1105.56\\
 $\nu_2$ & 1334.45  & 1327.76 & 1319.34 & 1305.15\\
 $\nu_1$ & 3170.47  & 3162.06 & 3161.29 & 3154.01\\  \toprule
\end{tabular}
\end{center}
\end{table*}

\begin{table*}[]
\caption{ Unscaled harmonic vibrational wavenumbers $\tilde{\omega }$ (in
cm$^{-1}$) as computed from the anharmonic force field calculations
at the CCSD(T) level of theory for the equilibrium
structure of the ($\textit{S}$)-enantiomer of CHBrClF, CHClFI, CHBrFI
and CHAtFI molecules using the MOLPRO. These wavenumbers are utilized in 
calculating vibrationally averaged HF and LDA parity violating potential for energy levels.\label{tab:freqm}}
\begin{center}
\setlength{\tabcolsep}{15pt}
\begin{tabular}{crrrr}\toprule 
 Mode& CHBrClF & CHClFI & CHBrFI & CHAtFI\\ \midrule
 $\nu_9$ &  222.88  &  192.60 &  142.44 &  100.43\\
 $\nu_8$ &  308.45  &  269.30 &  267.40 &  237.82\\
 $\nu_7$ &  421.52  &  411.97 &  325.42 &  280.43\\
 $\nu_6$ &  637.64  &  575.76 &  552.38 &  490.46\\
 $\nu_5$ &  785.52  &  775.15 &  652.68 &  603.88\\
 $\nu_4$ & 1121.55  & 1107.69 & 1100.83 & 1090.74\\
 $\nu_3$ & 1235.71  & 1210.04 & 1177.67 & 1105.91\\
 $\nu_2$ & 1333.79  & 1327.28 & 1316.83 & 1305.43\\
 $\nu_1$ & 3169.72  & 3161.39 & 3159.79 & 3153.64\\  \toprule
\end{tabular}
\end{center}
\end{table*}

\begin{table*}[]
\caption{ Cartesian displacements (in \AA) corresponding to a unit shift along the dimensionless reduced normal coordinates for all the normal modes of the ($\textit{S}$)-enantiomer of CHBrClF.\label{tab:displacementsCHBrClF}}
\begin{center}
\setlength{\tabcolsep}{12pt}
\begin{tabular}{ccrrr} \toprule 
 Mode & Atom & \multicolumn{1}{c}{$x$} & \multicolumn{1}{c}{$y$} & \multicolumn{1}{c}{$z$} \\ \hline
 \multirow{ 5}{*}{ $\nu_9$ }
&  C &        $-$0.0236951470&        0.0121369995&       $-$0.0087080375\\
&  H &        $-$0.0322354154&        0.0063717963&       $-$0.0088260917\\
&  F &        $-$0.0235104244&        0.0069300442&       $-$0.0041925549\\
&  Cl&         0.0163653898&        0.0456299712&        0.0041736273\\
&  Br&         0.0024228979&       $-$0.0238138947&        0.0005967752 \\ \hline 
 \multirow{ 5}{*}{ $\nu_8$ }
&  C &         0.0215246864&        0.0049634827&       $-$0.0044364218\\
&  H &         0.0158073895&       $-$0.0131073106&       $-$0.0045836435\\
&  F &         0.0265608221&        0.0555400827&        0.0071490103\\
&  Cl&         0.0093571217&       $-$0.0075179212&        0.0000617805\\
&  Br&        $-$0.0140150985&       $-$0.0106265759&       $-$0.0010152722  \\ \hline 
\multirow{ 5}{*}{ $\nu_7$ }
&  C &        $-$0.0217507319&       $-$0.0088114409&        0.0205479682\\
&  H &        $-$0.0091707683&       $-$0.0158381686&        0.0210902277\\
&  F &        $-$0.0374172088&        0.0214384035&       $-$0.0116551581\\
&  Cl&         0.0217087598&       $-$0.0176323065&       $-$0.0011468561\\
&  Br&         0.0028128843&        0.0041940244&       $-$0.0000797945 \\ \hline 
\multirow{ 5}{*}{ $\nu_6$ }
&  C &        $-$0.0204677518&       $-$0.0287186861&       $-$0.0467130722\\
&  H &        $-$0.0455763639&       $-$0.0384435862&       $-$0.0483095150\\
&  F &         0.0028736501&        0.0060496129&        0.0078769417\\
&  Cl &         0.0027155810&       $-$0.0014139336&        0.0046862569\\
&  Br &         0.0017992081&        0.0040279521&        0.0037471908 \\ \hline 
\multirow{ 5}{*}{ $\nu_5$ }
&  C &        $-$0.0095604597&        0.0462889358&       $-$0.0209516218\\
&  H &        $-$0.0210491366&        0.0417292646&       $-$0.0220444023\\
&  F &        $-$0.0016544127&       $-$0.0056890009&        0.0031035543\\
&  Cl&         0.0050943299&       $-$0.0115518078&        0.0067081591\\
&  Br&        $-$0.0001364993&       $-$0.0010832420&       $-$0.0002521952 \\ \hline
\multirow{ 5}{*}{ $\nu_4$ }
&  C &        $-$0.0350799297&        0.0003784503&        0.0149144708\\
&  H &        $-$0.0265347062&        0.0091576250&        0.0156634792\\
&  F &         0.0225345433&        0.0001640155&       $-$0.0100171270\\
&  Cl&         0.0004680430&       $-$0.0002860938&        0.0001007929\\
&  Br&         0.0000407286&       $-$0.0000872085&       $-$0.0001010525 \\ \hline 
\multirow{ 5}{*}{ $\nu_3$ }
&  C &        $-$0.0000495143&       $-$0.0136417558&       $-$0.0024901127\\
&  H &         0.0125806925&        0.1559634666&       $-$0.0021091781\\
&  F &        $-$0.0009637683&        0.0019117634&        0.0010722579\\
&  Cl&         0.0007570909&       $-$0.0003366187&        0.0017648758\\
&  Br&        $-$0.0002565881&       $-$0.0002284905&       $-$0.0006345778 \\ \hline 
\multirow{ 5}{*}{ $\nu_2$ }
&  C &         0.0139161242&       $-$0.0012544171&        0.0045797156\\
&  H &        $-$0.1477580099&        0.0120477682&        0.0038286553\\
&  F &        $-$0.0000825568&        0.0000291054&       $-$0.0050377751\\
&  Cl&        $-$0.0002687167&       $-$0.0002247786&        0.0006683842\\
&  Br&        $-$0.0000901452&        0.0001294789&        0.0001713398 \\ \hline 
\multirow{ 5}{*}{ $\nu_1$ }
&  C &        $-$0.0000581460&       $-$0.0000528993&        0.0085152676\\
&  H &         0.0005730257&        0.0002040558&       $-$0.0984260016\\
&  F &         0.0000233032&        0.0000057069&       $-$0.0000485033\\
&  Cl&        $-$0.0000059327&        0.0000185525&       $-$0.0000314752\\
&  Br&        $-$0.0000014575&       $-$0.0000041567&       $-$0.0000122264 \\ \bottomrule 
\end{tabular}
\end{center}
\end{table*}

\begin{table*}[]
\caption{ Cartesian displacements (in \AA) corresponding to a unit shift along the dimensionless reduced normal coordinates for all the normal modes of the ($\textit{S}$)-enantiomer of CHClFI.\label{tab:displacementsCHClFI}}
\begin{center}
\setlength{\tabcolsep}{12pt}
\begin{tabular}{ccrrr} \toprule 
 Mode & Atom & \multicolumn{1}{c}{$x$} & \multicolumn{1}{c}{$y$} & \multicolumn{1}{c}{$z$} \\ \hline
 \multirow{ 5}{*}{ $\nu_9$ }
&  C  &       $-$0.0212404063&        0.0199492085&       $-$0.0090853486\\
&  H  &       $-$0.0295188582&        0.0109319755&       $-$0.0092036033\\
&  F  &       $-$0.0203089710&        0.0111972216&       $-$0.0033684127\\
&  Cl &        0.0197109129&        0.0522134767&        0.0061381660\\
&  I  &       $-$0.0001481094&       $-$0.0180376311&       $-$0.0002549267\\ \hline 
 \multirow{ 5}{*}{ $\nu_8$ }
&  C  &        0.0296818756&        0.0088539676&       $-$0.0042764455\\
&  H  &        0.0210641146&       $-$0.0074645969&       $-$0.0044089053\\
&  F  &        0.0353448555&        0.0582786261&        0.0107873970\\
&  Cl &        0.0055498284&       $-$0.0073900626&       $-$0.0003007127\\
&  I  &       $-$0.0097949130&       $-$0.0074665109&       $-$0.0010927226\\ \hline 
\multirow{ 5}{*}{ $\nu_7$ }
&  C  &       $-$0.0207574042&       $-$0.0090463514&        0.0201598914\\
&  H  &       $-$0.0080216300&       $-$0.0161584574&        0.0206862612\\
&  F  &       $-$0.0355673584&        0.0268612060&       $-$0.0111820587\\
&  Cl &        0.0228323360&       $-$0.0176434830&       $-$0.0012953910\\
&  I  &        0.0010596945&        0.0018242147&       $-$0.0000396167\\ \hline 
\multirow{ 5}{*}{ $\nu_6$ }
&  C  &       $-$0.0207018243&       $-$0.0299003889&       $-$0.0495556294\\
&  H  &       $-$0.0506795808&       $-$0.0489561842&       $-$0.0509950074\\
&  F  &        0.0028737576&        0.0071753347&        0.0088925595\\
&  Cl &        0.0029915407&       $-$0.0012020640&        0.0050328596\\
&  I  &        0.0011055186&        0.0024732711&        0.0023729172\\ \hline 
\multirow{ 5}{*}{ $\nu_5$ }
&  C  &       $-$0.0096126701&        0.0468249381&       $-$0.0208181974\\
&  H  &       $-$0.0211539443&        0.0370008191&       $-$0.0219739353\\
&  F  &       $-$0.0014882621&       $-$0.0056629015&        0.0030134026\\
&  Cl &        0.0050174077&       $-$0.0122486450&        0.0065786270\\
&  I  &       $-$0.0000827953&       $-$0.0004986800&       $-$0.0001208230\\ \hline 
\multirow{ 5}{*}{ $\nu_4$ }
&  C  &       $-$0.0357030151&       $-$0.0004263954&        0.0142219281\\
&  H  &       $-$0.0226701910&        0.0139784793&        0.0149046739\\
&  F  &        0.0229046432&        0.0004550924&       $-$0.0096654294\\
&  Cl &        0.0004879819&       $-$0.0002871787&        0.0001569501\\
&  I  &       $-$0.0000073432&       $-$0.0000596917&       $-$0.0000594575\\ \hline 
\multirow{ 5}{*}{ $\nu_3$ }
&  C  &       $-$0.0001613314&       $-$0.0133512694&       $-$0.0048511523\\
&  H  &        0.0195682941&        0.1564269145&       $-$0.0044208196\\
&  F  &       $-$0.0016253549&        0.0018734898&        0.0018635742\\
&  Cl &        0.0008615416&       $-$0.0004420261&        0.0019598576\\
&  I  &       $-$0.0001342274&       $-$0.0001384761&       $-$0.0003252128\\ \hline 
\multirow{ 5}{*}{ $\nu_2$ }
&  C  &        0.0131505071&       $-$0.0016887122&        0.0049986275\\
&  H  &       $-$0.1478201172&        0.0198353660&        0.0042420419\\
&  F  &        0.0002012968&        0.0000063869&       $-$0.0052420336\\
&  Cl &       $-$0.0002124418&       $-$0.0002212976&        0.0007432630\\
&  I  &       $-$0.0000411662&        0.0000621830&        0.0000736043\\ \hline 
\multirow{ 5}{*}{ $\nu_1$ }
&  C  &       $-$0.0000578731&       $-$0.0000753681&        0.0084976492\\
&  H  &        0.0005928256&        0.0002245937&       $-$0.0985860331\\
&  F  &        0.0000198004&        0.0000116463&       $-$0.0000410375\\
&  Cl &       $-$0.0000048365&        0.0000206554&       $-$0.0000298142\\
&  I  &       $-$0.0000008671&       $-$0.0000020921&       $-$0.0000062381\\ \bottomrule 
\end{tabular}
\end{center}
\end{table*}

\begin{table*}[]
\caption{ Cartesian displacements (in \AA) corresponding to a unit shift along the dimensionless reduced normal coordinates for all the normal modes of the ($\textit{S}$)-enantiomer of CHBrFI.\label{tab:displacementsCHBrFI}}
\begin{center}
\setlength{\tabcolsep}{12pt}
\begin{tabular}{ccrrr} \toprule 
 Mode & Atom & \multicolumn{1}{c}{$x$} & \multicolumn{1}{c}{$y$} & \multicolumn{1}{c}{$z$} \\ \hline
 \multirow{ 5}{*}{ $\nu_9$ }
& C  &        0.0252210102&       $-$0.0081164111&        0.0103067250\\
& H  &        0.0326015084&       $-$0.0051137294&        0.0103866888\\
& F  &        0.0239053551&       $-$0.0044202716&        0.0043240643\\
& Br &       $-$0.0076555727&       $-$0.0393145903&       $-$0.0026691476\\
& I  &       $-$0.0014618339&        0.0259193115&       $-$0.0000445572 \\ \hline 
 \multirow{ 5}{*}{ $\nu_8$ }
& C  &        0.0196740909&        0.0127328864&       $-$0.0037806426\\
& H  &        0.0126686453&       $-$0.0059926238&       $-$0.0038421466\\
& F  &        0.0240814368&        0.0659518640&        0.0081220619\\
& Br &        0.0060997564&       $-$0.0053594406&        0.0004150514\\
& I  &       $-$0.0093596895&       $-$0.0076971927&       $-$0.0010860596  \\ \hline 
\multirow{ 5}{*}{ $\nu_7$ }
& C  &        0.0366416247&       $-$0.0023428777&       $-$0.0140064313\\
& H  &        0.0225974872&        0.0049967733&       $-$0.0142981862\\
& F  &        0.0487085811&       $-$0.0236950216&        0.0143266957\\
& Br &       $-$0.0126376213&        0.0108898991&       $-$0.0007024870\\
& I  &       $-$0.0030773657&       $-$0.0030430686&       $-$0.0002699587  \\ \hline 
\multirow{ 5}{*}{ $\nu_6$ }
& C  &        0.0231049368&        0.0128727133&        0.0573958444\\
& H  &        0.0590454967&        0.0230939771&        0.0591514984\\
& F  &       $-$0.0043967409&       $-$0.0029855215&       $-$0.0106851815\\
& Br &       $-$0.0014211361&        0.0021842479&       $-$0.0030830013\\
& I  &       $-$0.0011117563&       $-$0.0023120845&       $-$0.0023802769  \\ \hline 
\multirow{ 5}{*}{ $\nu_5$ }
& C  &       $-$0.0047566857&        0.0588596220&       $-$0.0105205363\\
& H  &       $-$0.0098449476&        0.0569860348&       $-$0.0110406418\\
& F  &        0.0010673899&       $-$0.0107078916&        0.0016564301\\
& Br &        0.0012581799&       $-$0.0045481104&        0.0024967356\\
& I  &       $-$0.0004142612&       $-$0.0015869577&       $-$0.0007181555  \\ \hline 
\multirow{ 5}{*}{ $\nu_4$ }
& C  &       $-$0.0355805786&       $-$0.0008124577&        0.0147539179\\
& H  &       $-$0.0209277779&        0.0047231298&        0.0154297507\\
& F  &        0.0232806577&        0.0002803673&       $-$0.0097029087\\
& Br &        0.0000661550&        0.0000577165&       $-$0.0000422751\\
& I  &        0.0000042750&       $-$0.0000385502&       $-$0.0000387828  \\ \hline 
\multirow{ 5}{*}{ $\nu_3$ }
& C  &       $-$0.0002207449&       $-$0.0134182504&       $-$0.0022986074\\
& H  &        0.0071296253&        0.1609709288&       $-$0.0023310393\\
& F  &       $-$0.0006005751&        0.0018864492&        0.0007729801\\
& Br &        0.0003391174&       $-$0.0002580776&        0.0007731425\\
& I  &       $-$0.0001567302&       $-$0.0001314794&       $-$0.0003606602  \\ \hline 
\multirow{ 5}{*}{ $\nu_2$ }
& C  &        0.0127900952&       $-$0.0004634496&        0.0057629219\\
& H  &       $-$0.1494682381&        0.0078811713&        0.0048312972\\
& F  &        0.0003529074&       $-$0.0000195203&       $-$0.0055191868\\
& Br &       $-$0.0000617102&       $-$0.0001209125&        0.0002384229\\
& I  &       $-$0.0000368571&        0.0000593503&        0.0000946852  \\ \hline 
\multirow{ 5}{*}{ $\nu_1$ }
& C  &       $-$0.0000641177&       $-$0.0000210670&        0.0084897660\\
& H  &        0.0006689137&       $-$0.0000288032&       $-$0.0986075181\\
& F  &        0.0000222592&        0.0000059831&       $-$0.0000379301\\
& Br &       $-$0.0000020040&        0.0000072846&       $-$0.0000122954\\
& I  &       $-$0.0000013355&       $-$0.0000032051&       $-$0.0000063564  \\  \bottomrule 
\end{tabular}
\end{center}
\end{table*}

\begin{table*}[]
\caption{ Cartesian displacements (in \AA) corresponding to a unit shift along the dimensionless reduced normal coordinates for all the normal modes of the ($\textit{S}$)-enantiomer of CHAtFI.\label{tab:displacementsCHAtFI}}
\begin{center}
\setlength{\tabcolsep}{12pt}
\begin{tabular}{ccrrr} \toprule 
 Mode & Atom & \multicolumn{1}{c}{$x$} & \multicolumn{1}{c}{$y$} & \multicolumn{1}{c}{$z$} \\ \hline
 \multirow{ 5}{*}{ $\nu_9$ }
 &C  &    0.0266233549&       $-$0.0069932752&        0.0111689178 \\
 &H  &    0.0332310722&       $-$0.0051860255&        0.0111990872\\
 &F  &    0.0247007571&       $-$0.0049217068&        0.0046691440\\
 &I  &   $-$0.0038895916&       $-$0.0383979520&       $-$0.0006485078\\
 &At &   $-$0.0015651225&        0.0240746465&       $-$0.0007225401 \\ \hline 
 \multirow{ 5}{*}{ $\nu_8$ }
 &C  &   0.0226677518&        0.0169298855&       $-$0.0052820358\\
 &H  &   0.0146791658&       $-$0.0023822828&       $-$0.0053489364\\
 &F  &   0.0271647016&        0.0722494002&        0.0074145024\\
 &I  &   0.0036672971&       $-$0.0032090604&        0.0004430506\\
 &At &  $-$0.0060397724&       $-$0.0055534350&       $-$0.0006110459 \\ \hline 
 \multirow{ 5}{*}{ $\nu_7$ }
 &C  &  $-$0.0424737299&        0.0053931873&        0.0122135326\\
 &H  &  $-$0.0259596898&       $-$0.0026013704&        0.0123476460\\
 &F  &  $-$0.0537882108&        0.0283347896&       $-$0.0168441066\\
 &I  &   0.0091620633&       $-$0.0078444073&        0.0010118288\\
 &At &   0.0018813924&        0.0018812874&        0.0001552619  \\ \hline 
\multirow{ 5}{*}{ $\nu_6$ }
 &C  &   0.0242887442&        0.0114222854&        0.0616510669\\
 &H  &   0.0676705934&        0.0193911426&        0.0633567616\\
 &F  &  $-$0.0048330981&       $-$0.0022062847&       $-$0.0121365522\\
 &I  &  $-$0.0009825309&        0.0016584086&       $-$0.0021554098\\
 &At &  $-$0.0006817577&       $-$0.0015484146&       $-$0.0014266010  \\ \hline 
 \multirow{ 5}{*}{ $\nu_5$ }
 &C  &  $-$0.0036518357&        0.0620893140&       $-$0.0092640116\\
 &H  &  $-$0.0100282089&        0.0620934260&       $-$0.0097122176\\
 &F  &   0.0008238230&       $-$0.0116676241&        0.0017969964\\
 &I  &   0.0008113156&       $-$0.0026750283&        0.0014999561\\
 &At &  $-$0.0003080136&       $-$0.0011739895&       $-$0.0004930242  \\ \hline 
 \multirow{ 5}{*}{ $\nu_4$ }
 &C  &  $-$0.0344882683&       $-$0.0045274544&        0.0130794042\\
 &H  &  $-$0.0130998469&        0.0539402097&        0.0138560935\\
 &F  &   0.0223988699&        0.0007556002&       $-$0.0088266869\\
 &I  &   0.0000749075&       $-$0.0000201182&        0.0001162955\\
 &At &  $-$0.0000380356&       $-$0.0000563607&       $-$0.0000856355  \\ \hline 
 \multirow{ 5}{*}{ $\nu_3$ }
 &C  &  $-$0.0099191450&        0.0131203429&        0.0063835206\\
 &H  &  $-$0.0189718874&       $-$0.1571545648&        0.0062184214\\
 &F  &   0.0075394215&       $-$0.0018656000&       $-$0.0038404212\\
 &I  &  $-$0.0001926328&        0.0001460040&       $-$0.0004186315\\
 &At &   0.0000921880&        0.0000850321&        0.0002058081  \\ \hline 
\multirow{ 5}{*}{ $\nu_2$ }
 &C  &   0.0123507661&       $-$0.0011859941&        0.0066340299\\
 &H  &  $-$0.1496771513&        0.0140968938&        0.0056730542\\
 &F  &   0.0004938261&        0.0000931605&       $-$0.0057932150\\
 &I  &  $-$0.0000192636&       $-$0.0000531357&        0.0001359464\\
 &At &  $-$0.0000204687&        0.0000238003&        0.0000356421  \\ \hline 
 \multirow{ 5}{*}{ $\nu_1$ }
 &C  &  $-$0.0000694028&       $-$0.0000357173&        0.0084745120\\
 &H  &   0.0007047366&        0.0001657833&       $-$0.0987465230\\
 &F  &   0.0000305779&        0.0000061370&       $-$0.0000322123\\
 &I  &  $-$0.0000019069&        0.0000052481&       $-$0.0000064353\\
 &At &  $-$0.0000010303&       $-$0.0000024813&       $-$0.0000035540 \\ \bottomrule 
\end{tabular}
\end{center}
\end{table*}

\begin{landscape}
\begin{table*}[]
\footnotesize
\caption{\footnotesize Anharmonic potential parameters $\frac{\sqrt{h^3c^3\tilde{\omega}_i^2\tilde{\omega}_j/E_\mathrm{h}^3}}{N} \frac{\partial^3V_{BO}}{\partial
q_{i}^{2}\partial q_{j}}$ (in $10^{-10}$ $E_\mathrm h$ where $N = 2$ for $i \neq j$ and $N = 6$ for $i = j$) for all normal modes in the ($\textit{S}$)-enantiomer of CHBrClF molecule. Values are
determined by the SURF module of MOLPRO package.\label{tab:cubicCHBrClF}}
\begin{center}
\setlength{\tabcolsep}{8pt}
\begin{tabular}{c|c|SSSSSSSSS}
\midrule
& i$\downarrow $ j$\rightarrow $ & \multicolumn{1}{c}{9} & \multicolumn{1}{c}{8} & \multicolumn{1}{c}{7} & \multicolumn{1}{c}{6} & \multicolumn{1}{c}{5} & \multicolumn{1}{c}{4} & \multicolumn{1}{c}{3} & \multicolumn{1}{c}{2} & \multicolumn{1}{c}{1} \\ \midrule  
\multirow{9}{*}{$\frac{\sqrt{h^3c^3\tilde{\omega}_i^2\tilde{\omega}_j/E_\mathrm{h}^3}}{\mathrm N} \frac{\partial^3V_{BO}}{\partial q_{i}^{2}\partial q_{j}}$} 
&9& -6.13375   & -1.62456   &  5.36623   &  4.10039   & -12.4538   &  4.22933   &  0.50045   & -11.9076   &  7.06316 \\
&8&  0.71152   & -17.6720   &  23.0462   &  25.7329   & -17.7983   &  44.9737   &  15.6896   &  17.7276   & -10.3145 \\ 
&7&  0.42873   &  19.7941   &  29.2981   &  14.9578   &  53.7267   &  10.6989   & -18.7207   & -1.45219   & -1.31528 \\
&6& -55.4164   & -85.0839   &  60.0186   &  108.029   & -231.883   &  241.612   & -129.507   &  84.6826   & -77.9350 \\
&5& -47.6829   &  27.0592   &  235.059   &  215.617   &  339.614   &  430.106   &  27.0377   & -85.5338   & -36.1604 \\
&4&  24.8567   & -38.8926   &  130.756   & -66.9905   &  111.865   & -1249.16   &  171.079   &  289.211   & -102.387 \\
&3&  40.0876   & -54.2464   &  85.5867   &  612.840   &  268.704   &  58.6542   &  11.0916   &  124.468   & -8633.10 \\
&2& -42.2623   &  18.0845   & -86.3162   &  245.746   &  247.044   & -917.066   &  152.883   & -220.018   & -7319.62 \\
&1&  20.8170   &  56.0084   & -137.426   &  79.3191   &  116.667   & -3.22304   & -319.689   &  545.220   &  25965.9 \\  \midrule
\end{tabular}
\end{center}
\end{table*}

 \begin{table*}[]\footnotesize 
\caption{\footnotesize Anharmonic potential parameters $\frac{\sqrt{h^3c^3\tilde{\omega}_i^2\tilde{\omega}_j/E_\mathrm{h}^3}}{N} \frac{\partial^3V_{BO}}{\partial
q_{i}^{2}\partial q_{j}}$ (in $10^{-10}$ $E_\mathrm h$ where $N = 2$ for $i \neq j$ and $N = 6$ for $i = j$) for all normal modes in the ($\textit{S}$)-enantiomer of CHClFI molecule. Values are determined by
the SURF module of MOLPRO package.\label{tab:cubicCHClFI}}
\begin{center}
\setlength{\tabcolsep}{8pt}
\begin{tabular}{c|c|SSSSSSSSS}
\midrule
& i$\downarrow $ j$\rightarrow $ & \multicolumn{1}{c}{9} & \multicolumn{1}{c}{8} & \multicolumn{1}{c}{7} & \multicolumn{1}{c}{6} & \multicolumn{1}{c}{5} & \multicolumn{1}{c}{4} & \multicolumn{1}{c}{3} & \multicolumn{1}{c}{2} & \multicolumn{1}{c}{1} \\ \midrule  
\multirow{9}{*}{$\frac{\sqrt{h^3c^3\tilde{\omega}_i^2\tilde{\omega}_j/E_\mathrm{h}^3}}{\mathrm N}\frac{\partial^3V_{BO}}{\partial q_{i}^{2}\partial q_{j}}$}  
& 9 & -4.19337   & -1.70009   &  2.61317   &  5.11825   & -13.0533   &  5.23462   & -1.60110   & -10.7496   &  3.98759 \\
& 8 & -3.46222   & -12.6386   &  5.42075   &  19.4367   & -21.3291   &  40.8662   &  11.9068   &  17.2613   & -5.85553 \\
& 7 &  3.68053   &  15.5899   &  30.3290   &  10.9146   &  55.8242   &  14.3710   & -22.3170   &  3.34165   &  0.43056 \\
& 6 & -44.3359   & -60.0953   &  35.7051   &  74.1749   & -192.347   &  238.719   & -138.832   &  82.7481   & -174.690 \\
& 5 & -36.2867   &  7.93016   &  226.964   &  169.990   &  344.654   &  423.203   &  24.4129   & -89.5973   & -62.1807 \\
& 4 &  14.6965   & -27.2956   &  116.342   & -35.8603   &  83.2129   & -1221.51   &  275.218   &  258.731   & -185.571 \\
& 3 &  27.1614   & -54.9990   &  86.3211   &  537.408   &  242.523   &  52.8710   &  70.1948   &  68.1178   & -8661.77 \\
& 2 & -30.3118   &  20.4339   & -95.2055   &  201.466   &  279.917   & -876.640   &  269.044   & -230.763   & -7265.97 \\
& 1 &  18.3341   &  46.2788   & -133.782   &  61.7095   &  133.774   &  21.4595   & -401.942   &  542.422   &  25887.1 \\ \midrule
\end{tabular}
\end{center}
\end{table*}

\begin{table*}[]\footnotesize 
\caption{\footnotesize Anharmonic potential parameters $\frac{\sqrt{h^3c^3\tilde{\omega}_i^2\tilde{\omega}_j/E_\mathrm{h}^3}}{N} \frac{\partial^3V_{BO}}{\partial
q_{i}^{2}\partial q_{j}}$ (in $10^{-10}$ $E_\mathrm h$ where $N = 2$ for $i \neq j$ and $N = 6$ for $i = j$) for all normal modes in the ($\textit{S}$)-enantiomer of CHBrFI molecule. Values are determined by
the SURF module of MOLPRO package.\label{tab:cubicCHBrFI}}
\begin{center}
\setlength{\tabcolsep}{8pt}
\begin{tabular}{c|c|SSSSSSSSS}
\midrule
& i$\downarrow $ j$\rightarrow $ & \multicolumn{1}{c}{9} & \multicolumn{1}{c}{8} & \multicolumn{1}{c}{7} & \multicolumn{1}{c}{6} & \multicolumn{1}{c}{5} & \multicolumn{1}{c}{4} & \multicolumn{1}{c}{3} & \multicolumn{1}{c}{2} & \multicolumn{1}{c}{1} \\ \midrule  
\multirow{9}{*}{$\frac{\sqrt{h^3c^3\tilde{\omega}_i^2\tilde{\omega}_j/E_\mathrm{h}^3}}{\mathrm N}\frac{\partial^3V_{BO}}{\partial q_{i}^{2}\partial q_{j}}$} 
& 9 &  1.79341   & -0.60263   & -0.86106   & -0.07134   & -2.08984   &  3.09786   & -0.94865   & -4.76112   &  8.09375  \\
& 8 & -0.68138   & -10.1028   & -19.7331   & -15.5721   & -16.9851   &  47.6269   &  10.6298   &  19.4276   & -13.6836  \\
& 7 &  3.91013   &  17.4723   & -15.9871   & -14.7202   &  30.7159   &  9.94616   & -16.8515   & -5.03895   & -0.54352  \\
& 6 &  27.8037   & -31.2720   & -32.4707   & -32.3795   & -93.4651   &  243.544   & -57.4575   &  94.4397   & -188.794  \\
& 5 &  33.1357   & -9.14426   & -138.918   & -305.388   &  101.118   &  333.219   &  29.3854   & -18.0517   & -21.3783  \\
& 4 & -9.80643   & -15.9821   & -74.7237   &  52.1214   & -44.9397   & -1225.40   &  116.039   &  219.153   & -144.128  \\
& 3 & -20.2874   & -34.8657   & -115.646   & -540.682   &  131.077   &  66.2428   &  52.3268   &  67.2387   & -9110.10  \\
& 2 &  23.9787   &  16.8476   &  59.4691   & -265.576   &  100.108   & -893.103   &  115.747   & -249.058   & -7326.72  \\
& 1 & -8.35726   &  26.8477   &  107.242   & -137.057   &  71.1484   &  48.3217   & -32.1477   &  597.936   &  25802.9  \\ \midrule
\end{tabular}
\end{center}
\end{table*}

\begin{table*}[]\footnotesize 
\caption{\footnotesize Anharmonic potential parameters $\frac{\sqrt{h^3c^3\tilde{\omega}_i^2\tilde{\omega}_j/E_\mathrm{h}^3}}{N} \frac{\partial^3V_{BO}}{\partial
q_{i}^{2}\partial q_{j}}$ (in $10^{-10}$ $E_\mathrm h$ where $N = 2$ for $i \neq j$ and $N = 6$ for $i = j$) for all normal modes in the ($\textit{S}$)-enantiomer of CHAtFI molecule. Values are determined by
the SURF module of MOLPRO package.\label{tab:cubicCHAtFI}}
\begin{center}
\setlength{\tabcolsep}{8pt}
\begin{tabular}{c|c|SSSSSSSSS}
\midrule
& i$\downarrow $ j$\rightarrow $ & \multicolumn{1}{c}{9} & \multicolumn{1}{c}{8} & \multicolumn{1}{c}{7} & \multicolumn{1}{c}{6} & \multicolumn{1}{c}{5} & \multicolumn{1}{c}{4} & \multicolumn{1}{c}{3} & \multicolumn{1}{c}{2} & \multicolumn{1}{c}{1} \\ \midrule 
\multirow{9}{*}{$\frac{\sqrt{h^3c^3\tilde{\omega}_i^2\tilde{\omega}_j/E_\mathrm{h}^3}}{\mathrm N}\frac{\partial^3V_{BO}}{\partial q_{i}^{2}\partial q_{j}}$}
& 9 &  0.59835   & -0.22248   &  0.17400   &  0.17674   & -0.91244   &  1.81629   &  0.42816   & -2.48089   &  7.30931 \\
& 8 & -0.32623   & -7.96470   &  10.8695   & -11.2747   & -17.1353   &  46.6723   &  5.11632   &  20.6620   & -11.3335 \\
& 7 &  2.90929   &  12.4169   &  10.3253   & -10.7379   &  26.1788   &  9.33768   &  17.9418   & -1.44316   & -6.44476 \\
& 6 &  17.0482   & -18.4112   &  19.0939   & -16.0849   & -58.0260   &  223.427   &  114.345   &  90.3820   & -256.771 \\
& 5 &  20.5969   & -22.2024   &  104.276   & -254.497   &  49.6887   &  324.090   &  78.0286   & -10.8920   & -16.8507 \\
& 4 & -2.79086   & -28.2270   &  37.9584   & -15.8978   & -39.7440   & -1035.41   & -1080.63   &  207.679   & -1197.41 \\
& 3 & -11.7552   & -13.7079   &  98.0477   & -397.161   &  18.6263   & -239.018   & -79.0787   & -89.2830   & -8502.01 \\
& 2 &  13.7907   &  9.70384   & -50.2351   & -230.491   &  141.044   & -757.859   & -392.472   & -291.813   & -7329.56 \\
& 1 & -1.01056   &  22.9703   & -52.0535   & -156.455   &  83.2103   & -19.4454   &  245.595   &  644.071   &  25733.2 \\ \midrule
\end{tabular}
\end{center}
\end{table*}
\end{landscape}

\begin{landscape}
\begin{table*}[]
\caption{HF and LDA level parity violating energy ($E_\mathrm{PV}$ in $10^{-12}$ cm$^{-1}$) along the dimensionless reduced normal coordinates $q$ ranging from $-$3.00 to $+$3.00 for all (except C-F stretching mode $\nu_4$) the normal modes for the ($\textit{S}$)-enantiomer of CHBrClF.\label{tab:EpvCHBrClF}}
\begin{center}
\setlength{\tabcolsep}{12pt}
\begin{tabular}{l|S[round-precision=4,round-mode=places]S[round-precision=4,round-mode=places]S[round-precision=4,round-mode=places]S[round-precision=4,round-mode=places]S[round-precision=4,round-mode=places]S[round-precision=4,round-mode=places]S[round-precision=4,round-mode=places]S[round-precision=4,round-mode=places]} \toprule 
\multicolumn{9}{c}{\bf{HF}$^{a}$} \\ \hline
\multicolumn{1}{c}{$q$ $\downarrow$} & \multicolumn{1}{|c}{$\nu_9$}  & \multicolumn{1}{c}{$\nu_8$} & \multicolumn{1}{c}{$\nu_7$}  & \multicolumn{1}{c}{$\nu_6$} & \multicolumn{1}{c}{$\nu_5$}  & \multicolumn{1}{c}{$\nu_3$} & \multicolumn{1}{c}{$\nu_2$}  & \multicolumn{1}{c}{$\nu_1$}\\ \midrule
$-$3.00   & -0.52868948 &  -5.50006588 & -4.94984417   &  -2.52001446 &   0.44910684 &   2.15780669 &  2.17790747 &  -1.42390859 \\ 
$-$2.50   & -0.68382083 &  -4.83215169 & -4.34785443   &  -2.29878134 &  -0.00162608 &   1.51351201 &  1.73975811 &  -1.43748735 \\
$-$2.00   & -0.83790805 &  -4.16306159 & -3.75313671   &  -2.09787753 &  -0.39932233 &   0.89293248 &  1.22155906 &  -1.44812930 \\
$-$1.50   & -0.99139897 &  -3.49177258 & -3.16596805   &  -1.91486744 &  -0.74386417 &   0.28616619 &  0.63305265 &  -1.45544387 \\
$-$1.00   & -1.14478446 &  -2.81718944 & -2.58668019   &  -1.74769601 &  -1.03529839 &  -0.30701425 & -0.01561332 &  -1.45940385 \\
$-$0.50   & -1.29859003 &  -2.13813961 & -2.01566381   &  -1.59451538 &  -1.27223776 &  -0.88823805 & -0.71432313 &  -1.45873371 \\
$-$0.25   & -1.37582273 &  -1.79655163 & -1.73339597   &  -1.52226419 &  -1.36995853 &  -1.17289121 & -1.07938561 &  -1.45653203 \\
$-$0.125  & -1.41455285 &  -1.62517091 & -1.59309909   &  -1.48742738 &  -1.41350940 &  -1.31391859 & -1.26533246 &  -1.45522862 \\
$+$0.125  & -1.49228629 &  -1.28113018 & -1.31422001   &  -1.41925262 &  -1.48971352 &  -1.59108769 & -1.64336737 &  -1.45095434 \\
$+$0.25   & -1.53130803 &  -1.10842679 & -1.17565517   &  -1.38709133 &  -1.52222421 &  -1.72817327 & -1.83519280 &  -1.44873693 \\
$+$0.50   & -1.60970861 &  -0.76154312 & -0.90031974   &  -1.32274550 &  -1.57685375 &  -1.99731805 & -2.22383875 &  -1.44295400 \\
$+$1.00   & -1.76820516 &  -0.06122138 & -0.35709894   &  -1.20199871 &  -1.64050264 &  -2.51440937 & -3.01798639 &  -1.42631125 \\
$+$1.50   & -1.92947973 &  0.64913216  &   0.17562556  &  -1.08952693 &  -1.64215250 &  -2.99951761 & -3.82957389 &  -1.40643646 \\
$+$2.00   & -2.09416412 &  1.37116284  &   0.69710189  &  -0.98507963 &  -1.57943494 &  -3.44776332 & -4.65404793 &  -1.38031833 \\
$+$2.50   & -2.26289919 &  2.10663494  &   1.20648059  &  -0.88839155 &  -1.45043659 &  -3.85662326 & -5.48861402 &  -1.34893993 \\
$+$3.00   & -2.43633165 &  2.85743516  &   1.70280227  &  -0.79956340 &  -1.25335928 &  -4.22516388 & -6.33229166 &  -1.31345622 \\
\toprule 
\multicolumn{9}{c}{\bf{LDA}$^{b}$} \\ \hline
\multicolumn{1}{c}{$q$ $\downarrow$} & \multicolumn{1}{|c}{$\nu_9$}  & \multicolumn{1}{c}{$\nu_8$} & \multicolumn{1}{c}{$\nu_7$}  & \multicolumn{1}{c}{$\nu_6$} & \multicolumn{1}{c}{$\nu_5$}  & \multicolumn{1}{c}{$\nu_3$} & \multicolumn{1}{c}{$\nu_2$}  & \multicolumn{1}{c}{$\nu_1$}\\ \midrule
$-$3.00   &   1.94960529  & -6.55060116   & -4.41536400   & -2.75718608   & 3.98219071  &  5.29675317  &   3.60083510  &  0.10239197   \\ 
$-$2.50   &   1.70691456  & -5.43183650   & -3.53946797   & -2.03539074   & 3.12090647  &  4.45043120  &   3.27429107  &  0.19116310   \\ 
$-$2.00   &   1.46731733  & -4.28878980   & -2.68443213   & -1.39112117   & 2.37872483  &  3.63807256  &   2.85438813  &  0.27239741   \\ 
$-$1.50   &   1.23064074  & -3.12138931   & -1.84975925   & -0.81793954   & 1.75204681  &  2.84295640  &   2.35551721  &  0.34670269   \\ 
$-$1.00   &   0.99658546  & -1.92914072   & -1.03516579   & -0.30975036   & 1.23771157  &  2.05860398  &   1.79271411  &  0.41469513   \\ 
$-$0.50   &   0.76458154  & -0.71111870   & -0.24055273   &   0.13913967  & 0.83272978  &  1.28652394  &   1.18084682  &  0.47695838   \\ 
$-$0.25   &   0.64915858  & -0.09204296   &   0.14924414  &   0.34301281  & 0.67028844  &  0.90719330  &   0.86096016  &  0.50610396   \\ 
$-$0.125  &   0.59155296  &   0.22009488  &   0.34225496  &   0.44008053  & 0.59890275  &  0.71974120  &   0.69826507  &  0.52020566   \\ 
$+$0.125  &   0.47649346  &   0.84970694  &   0.72447975  &   0.62484911  & 0.47548478  &  0.35016539  &   0.36836364  &  0.54749900   \\ 
$+$0.25   &   0.41901492  &   1.16723957  &   0.91368196  &   0.71269182  & 0.42342885  &  0.16842762  &   0.20153675  &  0.56070213   \\ 
$+$0.50   &   0.30409459  &   1.80791254  &   1.28821815  &   0.87963433  & 0.33808993  &-0.18793952   & -0.13496421   &  0.58624555   \\ 
$+$1.00   &   0.07402149  &   3.11263678  &   2.02165807  &   1.18032137  & 0.24098405  &-0.86643890   & -0.81467986   &  0.63399819   \\ 
$+$1.50   & -0.15712354   &   4.45050402  &   2.73356566  &   1.43980169  & 0.23788419  &-1.48863053   & -1.49535304   &  0.67745131   \\ 
$+$2.00   & -0.39029728   &   5.82411660  &   3.42296668  &   1.66127978  & 0.32297667  &-2.04371977   & -2.16870095   &  0.71671928   \\ 
$+$2.50   & -0.62649258   &   7.23635291  &   4.08856471  &   1.84746523  & 0.48933755  &-2.52539476   & -2.82777274   &  0.75176759   \\ 
$+$3.00   & -0.86672669   &   8.69033538  &   4.72873202  &   2.00059507  & 0.72870943  &-2.93449941   & -3.46674790   &  0.78250735   \\ 
\toprule
\end{tabular}
\end{center}
\footnotesize{$^a$ See Fig. 5 in the manuscript.} \\
\footnotesize{$^b$ See Fig. 6 in the manuscript.} \\
\end{table*}
\end{landscape}

\begin{table*}[]
\caption{HF and LDA level PV energy gradients (in $10^{-13}$ $\mathrm{cm}^{-1}$) along the dimensionless reduced normal coordinates ($q_{4}$ in the range from $-$3.00 to $+$3.00) corresponding to the C-F stretching mode ($\nu_4$) for the ($\textit{S}$)-enantiomer of CHBrClF.\label{tab:EpvGCHBrClF}}
\begin{center}
\setlength{\tabcolsep}{12pt}
\begin{tabular}{lS[round-precision=4,round-mode=places]S[round-precision=4,round-mode=places]} \toprule 
\multicolumn{1}{c} {$q_{4}$} & \multicolumn{1}{c}{\bf{HF}$^{c}$}  & \multicolumn{1}{c}{\bf{LDA}$^{d}$} \\ \midrule
$-$3.00    &6.5347549649578119     &	5.9820024610504738  \\
$-$2.50    &6.2155859160882766     &	5.7207070321386548  \\
$-$2.00    &5.8549081664446065     &	5.4571278173089667  \\
$-$1.50    &5.4637260789037641     &	5.1880802922755659  \\
$-$1.00    &5.0524427490903015     &	4.9114943770209709  \\
$-$0.50    &4.6303725845656077     &	4.6259456738990592  \\
$-$0.25    &4.4178318242624753     &	4.4794500015028526  \\
$-$0.125   &4.3115763336111507     &	4.4052143321332711  \\
$+$0.00    &4.2054637567388996     &	4.3302960514519083  \\
$+$0.125   &4.0995862725523970     &	4.2546763963665609  \\
$+$0.25    &3.9940312170536575     &	4.1783352152510791  \\
$+$0.50    &3.7842087281862644     &	4.0234002210676960  \\
$+$1.00    &3.3716549806182974     &	3.7037881393022956  \\
$+$1.50    &2.9715140093769364     &	3.3694711845378326  \\
$+$2.00    &2.5863114952625486     &	3.0179418918671761  \\
$+$2.50    &2.2175566940061782     &	2.6462530417598307  \\
$+$3.00    &1.8659119638954356     &	2.2511546605585006  \\ \toprule
\end{tabular}
\end{center}
\footnotesize{$^c$ See the black curve of Fig. 1 in the manuscript.} \\
\footnotesize{$^d$ See the red curve of Fig. 1 in the manuscript.} \\
\end{table*}

\begin{table*}[]
\caption{HF and LDA level PV energy gradients (in $10^{-12}$ $\mathrm{cm}^{-1}$) along the dimensionless reduced normal coordinates ($q_{4}$ in the range from $-$3.00 to $+$3.00) corresponding to the C-F stretching mode ($\nu_4$) for the ($\textit{S}$)-enantiomer of CHClFI.\label{tab:EpvGCHClFI}}
\begin{center}
\setlength{\tabcolsep}{12pt}
\begin{tabular}{lS[round-precision=4,round-mode=places]S[round-precision=4,round-mode=places]} \toprule 
\multicolumn{1}{c} {$q_{4}$} & \multicolumn{1}{c}{\bf{HF}$^e$}  & \multicolumn{1}{c}{\bf{LDA}$^f$} \\ \midrule
 $-$3.00  & 5.9614469976375759    & 4.0531420368112015   \\ 
 $-$2.50  & 5.6970551778892118    & 3.8461091548743730   \\
 $-$2.00  & 5.4058295465557536    & 3.6511813020420861   \\
 $-$1.50  & 5.0971571324329642    & 3.4646965384029860   \\
 $-$1.00  & 4.7792226796957984    & 3.2837099990811554   \\
 $-$0.50  & 4.4587150565089919    & 3.1057268553547191   \\
 $-$0.25  & 4.2991520748972768    & 3.0171596135774187   \\
 $-$0.125 & 4.2197743436384827    & 2.9728658508185001   \\
 $+$0.00  & 4.1407385625079802    & 2.9285209733532998   \\
 $+$0.125 & 4.0620888356795510    & 2.8840920276451734   \\
 $+$0.25  & 3.9838699500224987    & 2.8357848973262145   \\
 $+$0.50  & 3.8288620742094660    & 2.7499631352289784   \\
 $+$1.00  & 3.5252789147656367    & 2.5678501570274150   \\
 $+$1.50  & 3.2309995809923472    & 2.3798233920294954   \\
 $+$2.00  & 2.9460820690352090    & 2.1834070754307735   \\
 $+$2.50  & 2.6698567855849895    & 1.9760675365366495   \\
 $+$3.00  & 2.4011371174504178    & 1.7553564242039853   \\  \toprule
\end{tabular}
\end{center}
\footnotesize{$^e$ See the black curve of Fig. 2 in the manuscript.} \\
\footnotesize{$^f$ See the red curve of Fig. 2 in the manuscript.} \\
\end{table*}

\begin{table*}[]
\caption{HF and LDA level PV energy gradients (in $10^{-12}$ $\mathrm{cm}^{-1}$) along the dimensionless reduced normal coordinates ($q_{4}$ in the range from  $-$3.00 to $+$3.00) corresponding to the C-F stretching mode ($\nu_4$) for the ($\textit{S}$)-enantiomer of CHBrFI.\label{tab:EpvGCHBrFI}}
\begin{center}
\setlength{\tabcolsep}{12pt}
\begin{tabular}{lS[round-precision=4,round-mode=places]S[round-precision=4,round-mode=places]} \toprule 
\multicolumn{1}{c} {$q_{4}$} & \multicolumn{1}{c}{\bf{HF}$^g$}  & \multicolumn{1}{c}{\bf{LDA}$^h$} \\ \midrule
$-$3.00 & 	10.131912699924995  & 10.331374647181686	 \\
$-$2.50 & 	9.6181841099927215  & 9.7015791343619727	 \\ 
$-$2.00 & 	9.1003932872940781  & 9.1613120034464838	 \\
$-$1.50 & 	8.5834404711466671  & 8.6982327544438491	 \\ 
$-$1.00 & 	8.0720326833067697  & 8.3007757936820156	 \\ 
$-$0.50 & 	7.5702407601367554  & 7.9578025889580457	 \\ 
$-$0.25 & 	7.3239867650675403  & 7.8033301430422475	 \\ 
$-$0.125 & 	7.2021671178497350  & 7.7297501316752998	 \\ 
$+$0.00  & 	7.0812605904766038  & 7.6583805916548703	 \\ 
$+$0.125 & 	6.9612982189896718  & 7.5890478022601428	 \\ 
$+$0.25  & 	6.8423085260378836  & 7.5215777595985488	 \\ 
$+$0.50  & 	6.6073179493717286  & 7.3915235626267638	 \\ 
$+$1.00  & 	6.1496894799117213  & 7.1459712193078899	 \\ 
$+$1.50  & 	5.7087838781752003  & 6.9102763301548311	 \\ 
$+$2.00  & 	5.2842590761514417  & 6.6731747367482389	 \\ 
$+$2.50  & 	4.8751560812978936  & 6.4239530487233263	 \\ 
$+$3.00  & 	4.4800132633932624  & 6.1527607189191038    \\ \toprule
\end{tabular}
\end{center}
\footnotesize{$^g$ See the black curve of Fig. 3 in the manuscript.} \\
\footnotesize{$^h$ See the red curve of Fig. 3 in the manuscript.} \\
\end{table*}
   
\begin{table*}[]
\caption{HF and LDA level PV energy gradients (in $10^{-10}$ $\mathrm{cm}^{-1}$) along the dimensionless reduced normal coordinates ($q_{4}$ in the range from $-$3.00 to $+$3.00) corresponding to the C-F stretching mode ($\nu_4$) for the ($\textit{S}$)-enantiomer of CHAtFI.\label{tab:EpvGCHAtFI}}
\begin{center}
\setlength{\tabcolsep}{12pt}
\begin{tabular}{lS[round-precision=4,round-mode=places]S[round-precision=4,round-mode=places]} \toprule 
\multicolumn{1}{c} {$q_{4}$} & \multicolumn{1}{c}{\bf{HF}$^i$}  & \multicolumn{1}{c}{\bf{LDA}$^j$} \\ \midrule
$-$3.00  &	7.3066299041193390  &	5.5571238801075143  \\
$-$2.50  &	5.7823116493004585  &	4.1578710810718935  \\
$-$2.00  &	4.4399748488120609  &	2.9386209265425308  \\
$-$1.50  &	3.2640848412177974  &	1.8924968364886441  \\
$-$1.00  &	2.2406949902555159  &	1.0123827727900745  \\
$-$0.50  &	1.3571243940330058  &	0.29079346481432903  \\
$-$0.25  &	0.96409795637428648 &	-0.013020401470651449  \\
$-$0.125 &	0.77916888052694585 &	-0.15114387101547641  \\
$+$0.00  &	0.60174403225843597 &	-0.28028827551529776  \\
$+$0.125 &	0.43165644278532525 &	-0.40054205595154380  \\
$+$0.25  &	0.26874260587708278 &	-0.51211477343042091  \\
$+$0.50  & -0.036177703088979229 &	-0.70981198328817002  \\
$+$1.00  & -0.56660465230210638  &	-1.0082888982748790  \\
$+$1.50  & -0.99879961485846877  &	-1.1876685266831115  \\
$+$2.00  & -1.3414463231791823  &	-1.2612545433721233  \\
$+$2.50  & -1.6027528174236466  &	-1.2436763400161933  \\
$+$3.00  &	-1.7904445391314834 &	-1.1508788414375567  \\
 \toprule
\end{tabular}
\end{center}
\footnotesize{$^i$ See the black curve of Fig. 4 in the manuscript.} \\
\footnotesize{$^j$ See the red curve of Fig. 4 in the manuscript.} \\
\end{table*}

\begin{table*}[]
\caption{HF and LDA level PV energy gradients (in $10^{-12}$ $\mathrm{cm}^{-1}$) along the dimensionless reduced normal coordinates $q$ = $-$0.125, $+$0.00 and $+$0.125 corresponding to all (except C-F stretching mode $\nu_4$)
the normal modes for the ($\textit{S}$)-enantiomer of CHBrClF.\label{tab:3ptsCHBrClF}}
\begin{center}
\setlength{\tabcolsep}{12pt}
\begin{tabular}{c|S[round-precision=4,round-mode=places]S[round-precision=4,round-mode=places]S[round-precision=4,round-mode=places]|S[round-precision=4,round-mode=places]S[round-precision=4,round-mode=places]S[round-precision=4,round-mode=places]} \toprule
level$\rightarrow$ & \multicolumn{3}{c|}{\bf{HF}} & \multicolumn{3}{c}{\bf{LDA}}                                                                    \\ \midrule
\multicolumn{1}{c} {$q$ $\rightarrow$}     & \multicolumn{1}{|c}{$-$0.125}  & \multicolumn{1}{c}{$+$0.00} & \multicolumn{1}{c}{$+$0.125} & \multicolumn{1}{|c}{$-$0.125} & \multicolumn{1}{c}{$+$0.00} & \multicolumn{1}{c}{$+$0.125} \\ \hline
$\nu_9$     & -0.3101818803701181  & -0.3109251729693801 & -0.31173700145109262 & -0.46061732311294402 & -0.46025799558377092 &  -0.45993916359652936 \\      
$\nu_8$     &  1.3727055386237214  &  1.3761534383152636 &  1.3797574720955024  &  2.5042311860199227  &  2.5185486863786154  &   2.5329955211297544  \\  
$\nu_7$     &  1.1201310931510338  &  1.1154957214517502 &  1.1108943604113841  &  1.5391186894855914  &  1.5289076083049610  &   1.5188124216621566  \\ 
$\nu_6$     &  0.27637717438349575 &  0.2709145977078968 &  0.26574538842311995 &  0.76386741828729461 &  0.73890496237865350 &   0.71467137110283054 \\     
$\nu_5$     & -0.33401747955677753 & -0.30520396271241852& -0.27637665593669769 & -0.54519993432991139 & -0.49335577004930628 &  -0.44234810333737756 \\ 
$\nu_3$     & -1.1209684915655805  & -1.1107820121936249 & -1.0997772210876061  & -1.4930077122081551  & -1.4785625297371090  &  -1.4625618869027345  \\ 
$\nu_2$     & -1.4961721688399879  & -1.5123532801971334 & -1.5275097851898123  & -1.3081232296503599  & -1.3197873892417577  &  -1.3301366087133664  \\  
$\nu_1$     &  0.01321848254556634 &  0.0157362547831661 &  0.01829284311554885 &  0.11163393880566476 &  0.10920490734018662 &   0.10685071062475385 \\   \toprule
\end{tabular}
\end{center}
\end{table*} 

\begin{table*}[]
\caption{HF and LDA level PV energy gradients (in $10^{-10}$ $\mathrm{cm}^{-1}$)
along the dimensionless reduced normal coordinates $q$ = $-$0.125, $+$0.00 and $+$0.125 corresponding to all (except C-F stretching mode $\nu_4$)
the normal modes for the ($\textit{S}$)-enantiomer of CHAtFI.\label{tab:3ptsCHAtFI}}
\begin{center}
\setlength{\tabcolsep}{12pt}
\begin{tabular}{c|S[round-precision=4,round-mode=places]S[round-precision=4,round-mode=places]S[round-precision=4,round-mode=places]|S[round-precision=4,round-mode=places]S[round-precision=4,round-mode=places]S[round-precision=4,round-mode=places]} \toprule
level$\rightarrow$ & \multicolumn{3}{c|}{\bf{HF}} & \multicolumn{3}{c}{\bf{LDA}}                                                                    \\ \midrule
\multicolumn{1}{c} {$q$ $\rightarrow$}     & \multicolumn{1}{|c}{$-$0.125}  & \multicolumn{1}{c}{$+$0.00} & \multicolumn{1}{c}{$+$0.125} & \multicolumn{1}{|c}{$-$0.125} & \multicolumn{1}{c}{$+$0.00} & \multicolumn{1}{c}{$+$0.125} \\ \hline
$\nu_9$     & 0.0046045984203452742  & 0.019054618332549601 & 0.03386243424003106 & 0.49399397277524388 & 0.52685185807177415  & 0.56031998525262690 \\
$\nu_8$     & 8.9103043146412160     & 8.8930348554815381   & 8.8765361845771050  & 13.276024323953419  & 13.317970595189941   & 13.360100962023558  \\
$\nu_7$     & 9.0001209279513964     & 8.9957995448713935   & 8.9905768667380051  & 8.4480845923550051  & 8.4192903944028511   & 8.3895579053121500  \\
$\nu_6$     & 2.6553122935118172     & 2.6698587399425054   & 2.6808800125490034  &-1.6812619829491660  &-1.7443752053913400   &-1.8103674525464174  \\
$\nu_5$     &-4.9943624579523335     &-4.8033113188864628   &-4.6127849917247127  &-5.9296028883703828  &-5.6624472997593264   &-5.3986462996493962  \\
$\nu_3$     & 7.1157003502838150     & 7.1016364279843260   & 7.0713154492562940  & 8.4073740789586756  & 8.4663041666106706   & 8.5023297578583935  \\
$\nu_2$     &-5.9393585710479019     &-5.9911487355586499   &-6.0372572895114051  &-2.2852043528879902  &-2.2363750382711028   &-2.1833048241876956  \\
$\nu_1$     & 0.16262726837086366    & 0.21239595490868658  & 0.26230097089363799 & 0.39809337540053273 & 0.38078873822824817  & 0.36379198398830222 \\   \toprule
\end{tabular}
\end{center}
\end{table*}

\begin{table*}[]
\caption{Fitting coefficients of the LDA PV energy ($E_\mathrm{PV}$)$^k$ and the LDA PV energy gradients ($\vec \nabla E_\mathrm{PV}$)$^l$ along the 
C-F stretching mode ($\nu_4$) of the chiral halogenated methane derivatives in $10^{-12}$ cm$^{-1}$.\label{tab:LDAFit}}
\begin{center}
\begin{tabular}{lS[table-format=+4.5(2)]S[table-format=+4.5(2)]S[table-format=+4.5(4)]l} 
\toprule 
	molecules &                       &    \multicolumn{1}{c}{$E_\mathrm{PV}$ $^{m}$}&      \multicolumn{1}{c}{$\vec \nabla E_\mathrm{PV}$} &  \\ 
\midrule
\multirow{ 5}{*}{CHBrClF}                         &b$_{0}$      &    +0.5343(2)    &                          &                  \\
                                                  &b$_{1}$      &    +0.4333(2)    &         +0.43314(5)      &         a$_{0}$  \\
                                                  &b$_{2}$      &    -0.03053(4)   &         -0.03004(3)      &      a$_{1}$/2   \\
                                                  &b$_{3}$      &    -0.00078(3)   &         -0.000792(4)     &      a$_{2}$/3   \\
                                                  \\
\multirow{ 5}{*}{CHClFI}                          &b$_{0}$      &    +5.35672(5)   &                          &                 \\
                                                  &b$_{1}$      &    +2.92880(4)   &         +2.9281(3)       &        a$_{0}$  \\
                                                  &b$_{2}$      &    -0.17741(3)   &         -0.17728(2)      &      a$_{1}$/2  \\
                                                  &b$_{3}$      &    -0.000912(7)  &         -0.00089(3)     &      a$_{2}$/3  \\
                                                  &b$_{4}$      &    -0.000774(4)  &         -0.00079(1)     &      a$_{3}$/4  \\ 
                                                  \\
\multirow{ 5}{*}{CHBrFI}                          &b$_{0}$      &    +19.35769(5)  &                          &                 \\
                                                  &b$_{1}$      &    +7.65824(5)   &         +7.6584(1)       &        a$_{0}$  \\
                                                  &b$_{2}$      &    -0.28138(4)   &         -0.2813(1)      &      a$_{1}$/2  \\
                                                  &b$_{3}$      &    +0.021587(7)  &         +0.0216(10)     &      a$_{2}$/3  \\
                                                  &b$_{4}$      &    -0.003719(4)  &         +0.00372(5)     &      a$_{3}$/4  \\  
                                                  \\
\multirow{ 5}{*}{CHAtFI}                         &b$_{0}$      &    +965.44(7)     &                          &                 \\
                                                  &b$_{1}$      &    -27.63(7)     &         -27.6(2)         &        a$_{0}$  \\
                                                  &b$_{2}$      &    -49.84(5)     &         -49.8(1)         &      a$_{1}$/2  \\
                                                  &b$_{3}$      &    +9.253(9)     &         +9.22(2)         &      a$_{2}$/3  \\
                                                  &b$_{4}$      &    -0.335(6)     &         -0.338(9)        &      a$_{3}$/4  \\ 
                                                 \bottomrule 
\end{tabular}
\end{center}
\footnotesize{$^k$ See Eq. 65 in the manuscript.}\\
\footnotesize{$^l$ See Eq. 64 in the manuscript.} \\
\footnotesize{$^m$ From Supplementary Material to R. Berger and J. L. Stuber, Mol. Phys. {\bf105}, 41 (2007).}
\end{table*}

\begin{table*}[]
\caption{Calculated values of second derivative coefficients [a$_1$ terms in $10^{-12}$ cm$^{-1}$] from a linear fit of the PV energy gradients 
at $q$ = $-$0.125, $+$0.00 and $+$0.125 along the C-F stretching mode ($\nu_4$) of the chiral methane derivatives.\label{tab:LinearFit}}
\begin{center}
\setlength{\tabcolsep}{12pt}
\begin{tabular}{cSS}  \toprule 
molecule &          \multicolumn{1}{c}{\bf{HF}}                         &     \multicolumn{1}{c}{\bf{LDA}}                             \\ \hline  
CHBrClF  & -0.08480(5)  &  -0.060(2)  \\ 
CHClFI   & -0.6307(9)   &  -0.3551(2) \\ 
CHBrFI   & -0.963(2)    &  -0.563(5)  \\ 
CHAtFI   & -139.0(2)   & -100.0(2)   \\ \bottomrule 
\end{tabular}
\end{center}
\end{table*} 

\begin{table*}[]
\caption{Vibrationally averaged HF and LDA parity violating potential  $E_{v,\mathrm{PV}}^{S}$ for energy levels $n$ in  $10^{-12}$ cm$^{-1}$ for all (expect C-F stretching vibration $\nu_4$) the normal modes of ($\textit{S}$)-enantiomer of CHBrClF.\label{tab:CHBrClFVibE}}
\begin{center}
\begin{tabular}{llS[table-format=-4.4]S[table-format=-4.4]S[table-format=2.2]cS[table-format=-4.4]S[table-format=-4.4]S[table-format=-4.4]S[table-format=2.2]}
\toprule
&  & \multicolumn{3}{c}{\bf{HF}}&&\multicolumn{3}{c}{\bf{LDA}}\\ 
\cline{3-5}\cline{7-9}
{Mode}&    {$n$}           &  {Perturbed 1D} & {Perturbed 2D} & {2D effects ($\%$)$^{aa}$} && {Perturbed 1D} & {Perturbed 2D} &{2D effects ($\%$)$^{aa}$}\\ 
\multirow{ 5}{*}{ $\nu_9$ }                     & 0   & -1.4639   & -1.4686  &  0.32  & & 0.5218 & 0.5166 & -1.00 \\
                                                & 1   & -1.4844   & -1.4984  &  0.95  & & 0.4974 & 0.4819 & -3.11 \\
                                                & 2   & -1.5049   & -1.5283  &  1.56  & & 0.4730 & 0.4472 & -5.45 \\
                                                & 3   & -1.5254   & -1.5582  &  2.15  & & 0.4486 & 0.4125 & -8.05 \\
                                     &$1\leftarrow0$  & -0.0205   & -0.0299  & 45.67  & &-0.0244 & -0.0347& 42.24 \\\\    
\multirow{ 5}{*}{ $\nu_8$ }                     & 0   & -1.3973   & -1.4088  & 0.82   & & 0.6529 & 0.6345 & -2.83 \\
                                                & 1   & -1.2847   & -1.3191  & 2.68   & & 0.8908 & 0.8354 & -6.22 \\
                                                & 2   & -1.1720   & -1.2294  & 4.89   & & 1.1286 & 1.0363 & -8.18 \\
                                                & 3   & -1.0594   & -1.1397  & 7.58   & & 1.3665 & 1.2372 & -9.46 \\
                                     &$1\leftarrow0$  &  0.1126   &  0.0897  &-20.36  & & 0.2379 & 0.2009 & -15.54 \\\\ 
\multirow{ 5}{*}{ $\nu_7$ }                     &  0   & -1.4932  & -1.5069  & 0.92   & & 0.4721 & 0.4469 & -5.34 \\
                                                &  1   & -1.5723  & -1.6136  & 2.62   & & 0.3484 & 0.2727 & -21.72 \\
                                                &  2   & -1.6514  & -1.7202  & 4.16   & & 0.2247 & 0.0985 & -56.15 \\
                                                &  3   & -1.7305  & -1.8268  & 5.56   & & 0.1009 & -0.0757& -175.01 \\
                                     &$1\leftarrow0$   & -0.0791  & -0.1066  & 34.77  & &-0.1237 & -0.1742 & 40.78 \\\\
\multirow{ 5}{*}{ $\nu_6$ }                     &  0   & -1.4739  & -1.4702  & -0.25  & & 0.4585 & 0.4784 & 4.34 \\
                                                &  1   & -1.5144  & -1.5033  & -0.74  & & 0.3075 & 0.3671 & 19.41 \\
                                                &  2   & -1.5550  & -1.5364  & -1.20  & & 0.1564 & 0.2559 & 63.57 \\
                                                &  3   & -1.5956  & -1.5695  & -1.63  & & 0.0054 & 0.1446 & 2571 \\
                                     &$1\leftarrow0$  &  -0.0406  & -0.0331  & -18.36 & &-0.1510 & -0.1112& -26.34 \\\\
\multirow{ 5}{*}{ $\nu_5$ }                     &  0   & -1.3757  & -1.4504  & 5.43   & & 0.6697 & 0.5591 & -16.50 \\
                                                &  1   & -1.2198  & -1.4440  & 18.38  & & 0.9409 & 0.6094 & -35.23 \\
                                                &  2   & -1.0640  & -1.4377  & 35.12  & & 1.2122 & 0.6597 & -45.58 \\
                                                &  3   & -0.9081  & -1.4313  & 57.61  & & 1.4835 & 0.7100 & -52.14 \\
                                     &$1\leftarrow0$  &  0.1559   &  0.0064  & -95.91 & & 0.2713 & 0.0503 & -81.46 \\\\
\multirow{ 5}{*}{ $\nu_3$ }                     &  0   & -1.4316  & -1.4281  & -0.25  & & 0.5655 & 0.5669 & 0.25 \\
                                                &  1   & -1.3877  & -1.3771  & -0.77  & & 0.6284 & 0.6326 & 0.66\\
                                                &  2   & -1.3438  & -1.3261  & -1.32  & & 0.6914 & 0.6984 & 1.00 \\
                                                &  3   & -1.2999  & -1.2750  & -1.91  & & 0.7544 & 0.7641 & 1.29 \\
                                     &$1\leftarrow0$  &   0.0439  &  0.0510  & 16.11  & & 0.0630 & 0.0657 & 4.41 \\\\
\multirow{ 5}{*}{ $\nu_2$ }                     &  0   & -1.5023  & -1.4882  & -0.94  & & 0.4969 & 0.5100 & 2.64 \\
                                                &  1   & -1.5996  & -1.5573  & -2.65  & & 0.4226 & 0.4619 & 9.31\\
                                                &  2   & -1.6970  & -1.6264  & -4.16  & & 0.3483 & 0.4139 & 18.84 \\
                                                &  3   & -1.7943  & -1.6955  & -5.51  & & 0.2740 & 0.3659 & 33.52 \\
                                     &$1\leftarrow0$   & -0.0973  & -0.0691  & -29.02 & &-0.0743 & -0.0480 & -35.33 \\\\ 
\multirow{ 5}{*}{ $\nu_1$ }                     &  0   & -1.4510  & -1.4458  & -0.36  & & 0.5123 & 0.5137 & 0.28 \\
                                                &  1   & -1.4457  & -1.4302  & -1.07  & & 0.4688 & 0.4730 & 0.92 \\
                                                &  2   & -1.4405  & -1.4146  & -1.79  & & 0.4252 & 0.4324 & 1.68 \\
                                                &  3   & -1.4352  & -1.3991  & -2.52  & & 0.3817 & 0.3918 & 2.63 \\
                                     &$1\leftarrow0$   &  0.0053  &  0.0156  & 196.35 & & -0.0435 & -0.0406 & -6.58 \\\\
\bottomrule
\end{tabular}
\end{center}
\footnotesize{$^{aa}$ 2D effects = (Perturbed 2D-Perturbed 1D)/Perturbed 1D}
\end{table*}

\begin{table*}[]
\caption{Vibrationally averaged HF and LDA parity violating potential  $E_{v,\mathrm{PV}}^{S}$ for energy levels $n$ in  $10^{-12}$ cm$^{-1}$ for all (expect C-F stretching vibration $\nu_4$) the normal modes of ($\textit{S}$)-enantiomer of CHAtFI.\label{tab:CHAtFIVibE}}
\begin{center}
\begin{tabular}{llS[table-format=-4.4]S[table-format=-4.4]S[table-format=2.2]cS[table-format=-4.4]S[table-format=-4.4]S[table-format=-4.4]S[table-format=2.2]}
\toprule
&  & \multicolumn{3}{c}{\bf{HF}}&&\multicolumn{3}{c}{\bf{LDA}}\\ 
\cline{3-5}\cline{7-9}
{Mode}&    {$n$}           &  {Perturbed 1D} & {Perturbed 2D} & {2D effects ($\%$)$^{aa}$} && {Perturbed 1D} & {Perturbed 2D} &{2D effects ($\%$)$^{aa}$}\\ 
\multirow{ 5}{*}{ $\nu_9$ }        &  0   & -2311.2  & -2311.8  & 0.03 &   &  971.00  & 970.95 & -0.005 \\
                                   &  1   & -2305.4  & -2307.3  & 0.08 &   &  982.15  & 982.01 & -0.01\\
                                   &  2   & -2299.6  & -2302.8  & 0.14 &   &  993.30  & 993.07 & -0.02\\
                                   &  3   & -2293.8  & -2298.3  & 0.20 &   &  1004.5  & 1004.1 & -0.03 \\
                        &$1\leftarrow0$   &     5.8  &     4.5  &-22.16 &      & 11.15    & 11.06  & -0.84 \\ \\    
\multirow{ 5}{*}{ $\nu_8$ }        &  0   & -2289.9   & -2300.7  & 0.47 &  &  1015.0  & 1002.0 & -1.28 \\
                                   &  1   & -2241.7   & -2274.1  & 1.44 &  &  1114.1  & 1075.0 & -3.51 \\
                                   &  2   & -2193.5   & -2247.4  & 2.46 &  &  1213.3  & 1148.1 & -5.37 \\
                                   &  3   & -2145.3   & -2220.7  & 3.52 &  &  1312.4  & 1221.2 & -6.95 \\
                        &$1\leftarrow0$  & 48.2       & 26.7 & -44.71       && 99.1    & 73.1  & -26.30 \\ \\
\multirow{ 5}{*}{ $\nu_7$ }        &  0   & -2338.9   & -2348.1  & 0.39 &   & 937.22  & 920.14 & -1.82 \\
                                   &  1   & -2388.5   & -2416.1  & 1.15 &   & 880.83  & 829.58 & -5.82\\
                                   &  2   & -2438.2   & -2484.2  & 1.89 &   & 824.43  & 739.01 & -10.36 \\
                                   &  3   & -2487.9   & -2552.2  & 2.59 &   & 768.04  & 648.45 & -15.57 \\
                        &$1\leftarrow0$  & -49.7   &-68.0   & 37.02   & &-56.39   & -90.56  & 60.59 \\ \\
\multirow{ 5}{*}{ $\nu_6$ }        &  0   & -2308.8   & -2314.5  & 0.25 &   & 950.73  & 946.54 & -0.44\\
                                   &  1   & -2298.2   & -2315.5  & 0.75 &   & 921.34  & 908.79 & -1.36\\
                                   &  2   & -2287.6   & -2316.5  & 1.26 &   & 891.95  & 871.03 & -2.35 \\
                                   &  3   & -2277.1   & -2317.5  & 1.77 &   & 862.57  & 833.28 & -3.40 \\
                        &$1\leftarrow0$  & 10.6   & -1.0   & -109.1 &    &-29.39   & -37.75  & 28.48\\ \\
\multirow{ 5}{*}{ $\nu_5$ }        &  0   & -2266.9   & -2286.6  & 0.87 &   & 1029.1  & 996.65 & -3.16 \\
                                   &  1   & -2172.5   & -2231.7  & 2.72 &   & 1156.6  & 1059.1 & -8.43\\
                                   &  2   & -2078.2   & -2176.8  & 4.75 &   & 1284.0  & 1121.6 & -12.65 \\
                                   &  3   & -1983.8   & -2122.0  & 6.96 &   & 1411.5  & 1184.0 & -16.11 \\
                        &$1\leftarrow0$  & 94.3    &54.9 & -41.82    &  &  127.4   & 62.5  & -50.99 \\ \\
\multirow{ 5}{*}{ $\nu_3$ }        &  0   & -2313.8   & -2318.7  & 0.21 &   & 980.49  & 963.72 & -1.71 \\
                                   &  1   & -2313.4   & -2328.0  & 0.63 &   & 1010.6  & 960.32 & -4.98\\
                                   &  2   & -2312.9   & -2337.3   & 1.06&   & 1040.8  & 956.92 & -8.06 \\
                                   &  3   & -2312.4   & -2346.6   & 1.48&   & 1070.9  & 953.52 & -10.96 \\
                        &$1\leftarrow0$  & 0.5    &-9.3 & -2075    &    & 30.1  & -3.40  & -111.3 \\ \\
\multirow{ 5}{*}{ $\nu_2$ }        &  0   & -2333.5   & -2310.0  & -1.01&    &  972.02 & 986.25 & 1.46 \\
                                   &  1   & -2372.3   & -2301.7  & -2.97&    &  985.23 & 1027.9 & 4.33 \\
                                   &  2   & -2411.1   & -2293.5  & -4.88&    &  998.43 & 1069.6 & 7.13 \\
                                   &  3   & -2449.9   & -2285.3  & -6.72&    &  1011.6 & 1111.2 & 9.85 \\
                         &$1\leftarrow0$  & -38.8    & 8.2 & -121.2   &     & 13.20  & 41.7   & 215.5  \\ \\
\multirow{ 5}{*}{ $\nu_1$ }        &  0   & -2307.4   & -2302.2  & -0.22&    & 956.05  & 955.73 & -0.03  \\
                                   &  1   & -2294.1   & -2278.6  & -0.67&    & 937.31  & 936.34 & -0.10\\
                                   &  2   & -2280.8   & -2255.0  & -1.13&    & 918.58  & 916.95 & -0.18\\
                                   &  3   & -2267.5   & -2231.4  & -1.59&    & 899.84  & 897.56 & -0.25 \\
                         &$1\leftarrow0$  & 13.3    & 23.6 & 77.49   && -18.74 & -19.39  & 3.47 \\ \toprule
\end{tabular}
\end{center}
\footnotesize{$^{aa}$ 2D effects = (Perturbed 2D-Perturbed 1D)/Perturbed 1D}
\end{table*}